## Proceedings of the 1st International Workshop

# Radiopure Scintillators for EURECA
# RPScint'2008

## 9 - 10 September 2008
## Institute for Nuclear Research, Kyiv, Ukraine

http://lpd.kinr.kiev.ua/rps08/

# Contents

# Preface

The searches for non-baryonic dark matter and neutrinoless double beta decay are among the most active areas of nuclear and astro-particle physics. Scintillation materials are widely used in these investigations, as conventional scintillators and also as cryogenic phonon scintillation detectors, yielding important scientific results. High light output, presence of certain elements, and a low level of radioactive contamination are the most important requirements of the experiments to crystal scintillators. Present and future dark matter and double beta decay projects call for extremely low (ideally zero) background of the detectors. Thus, reduction of the radioactive contamination of scintillating materials is an issue of major importance. The first workshop to discuss R&D of radiopure scintillators for low-count rate experiments, and in particular for the EURECA cryogenic dark matter experiment, RPSCINT 2008 was organized in Kyiv (Ukraine) on 9$^{th}$ and 10$^{th}$ September 2008. The idea was to bring together physicists, chemists, crystal scintillator experts and manufacturers to discuss the requirements of low-count rate experiments, in particular the required radiopurity and scintillation properties; selection and screening of input materials; purification of materials; raw compound preparation; crystal growing, annealing and handling; test of crystals; search for and development of new scintillating materials. Some contributions to the RPSCINT 2008 workshop are presented in these proceedings.

Fedor Danevich and Hans Kraus

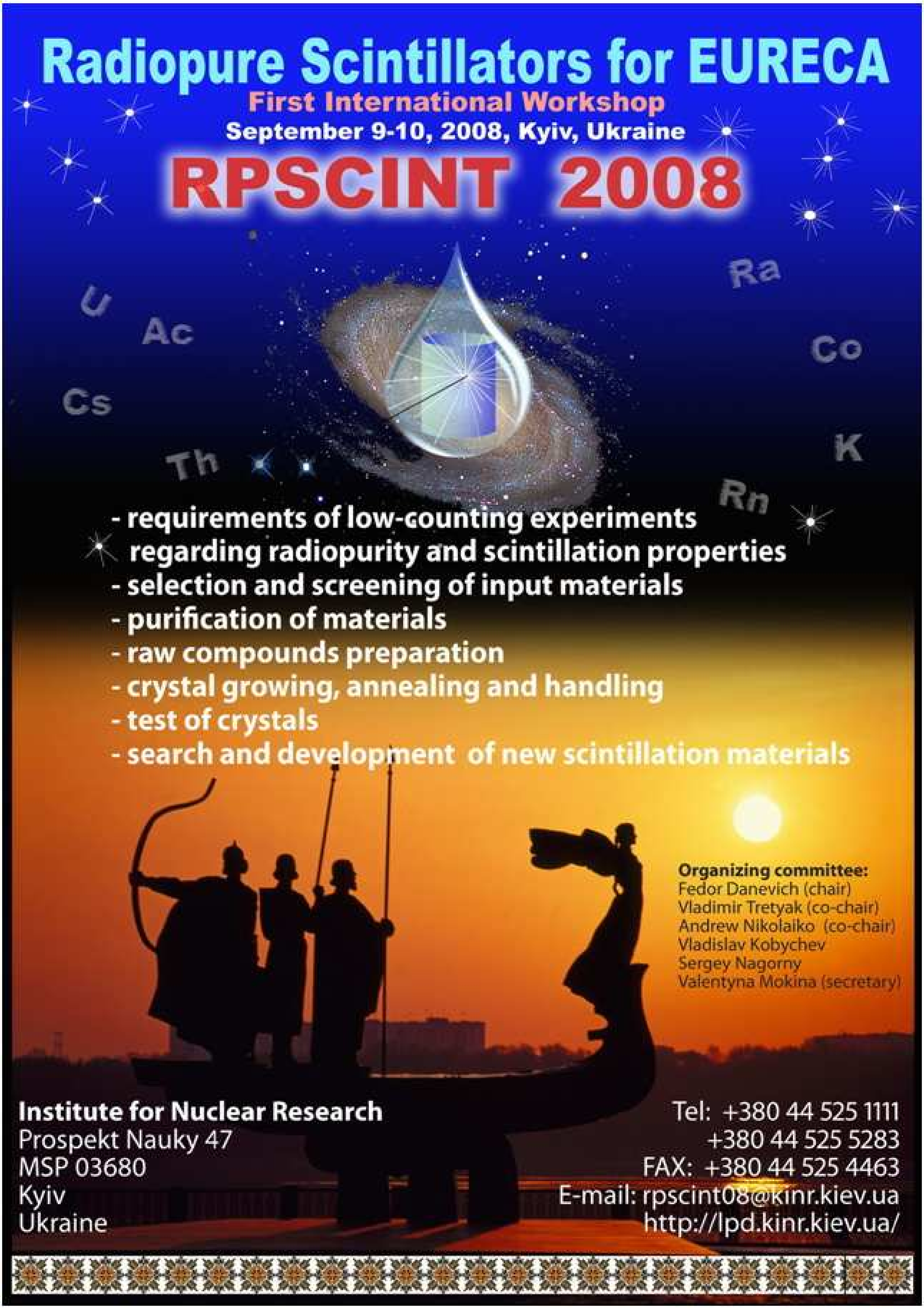
Radiopure Scintillators for EURECA
First International Workshop
September 9-10, 2008, Kyiv, Ukraine
RPSCINT 2008
U
Ac
Cs
Th
Ra
Co
K
Rn
- requirements of low-counting experiments
regarding radiopurity and scintillation properties
- selection and screening of input materials
- purification of materials
- raw compounds preparation
- crystal growing, annealing and handling
- test of crystals
- search and development of new scintillation materials
Organizing committee:
Fedor Danevich (chair)
Vladimir Tretyak (co-chair)
Andrew Nikolaiko (co-chair)
Vladislav Kobychev
Sergey Nagorny
Valentyna Mokina (secretary)
Institute for Nuclear Research
Prospekt Nauky 47
MSP 03680
Kyiv
Ukraine
Tel: +380 44 525 1111
+380 44 525 5283
FAX: +380 44 525 4463
E-mail: rpscint08@kinr.kiev.ua
http://lpd.kinr.kiev.ua/

# RPScint'2008 Program

## Tuesday, 9 September 2008

09:30 Registration
09:55 F. Danevich Welcome

**EURECA and related projects**

10:00 H. Kraus EURECA – an overview
10:30 F. Danevich R&D of crystal scintillators with the level of radiopurity required by EURECA
11:00 Coffee
11:30 P. de Marcillac Our – short – experience at IAS and within ROSEBUD with radioactive contaminations in scintillating bolometers: uses & needs
12:00 M.-A. Verdier Cryogenic scintillators for dark matter, status of the SciCryo project
12:30 Lunch
13:20 Workshop photo

**Double beta decay**

13:30 S. Pirro Radiopure scintillators for double beta decay searches
14:00 P.K. Raina Tin as a candidate for low background experimentation: Some issues in double beta decay

**Scintillator R&D**

14:30 M. Korzhik Tungstate and molybdate single-crystal scintillators development
15:00 L. Nagornaya Research and development for alkali-earth tungstate and molybdate crystal scintillators to search for rare processes
15:30 Coffee

**Scintillator characterisation**

16:00 V. Mokina Characterisation of scintillation crystals for cryogenic experimental search for rare events
16:20 V. Mikhailik Development of techniques for characterisation of scintillation materials for cryogenic applications at Oxford
16:40 D. Spassky Luminescence study of molybdates with cations of Li, Zn and Mg
17:00 V. Degoda Roentgen fluorescence of scintillation materials in wide temperature region
17:30 All Discussion
18:45 Workshop Dinner

## Wednesday, 10 September 2008

**Important issues of low radioactivity**

09:30 F. Danevich Radioactive contamination of crystal scintillators
10:00 D. Grigoriev Incidental radioactive background in BGO crystals
10:30 A. Nikolaiko Radioactive contamination of $CaWO_4$ scintillators
11:00 Coffee
11:30 G. Stryganyuk Effect of impurity segregation on the properties of single crystal scintillators
11:50 D. Poda Investigation of radiopure $ZnWO_4$
12:10 V. Kobychev Geant4-based simulator for response of scintillation detectors with typical geometries
12:30 Lunch

**Purification and production**

13:30 A. Dossovitski Raw materials for the production of low-background scintillation materials
14:00 V. Shlegel Growth of scintillation oxide crystals by the Low Thermal Gradient Czochralski technique (LTG Cz)
14:30 A. Shcherban Production of high-purity metals

14:50 D. Solopikhin Purification of cadmium and lead for low-background scintillators
15:10 R. Boiko Purification of calcium and molybdenum for $CaMoO_4$ crystals growing
15:30 Coffee
16:00 All Discussion on way forward
16:45 H. Kraus Summary remarks
17:00 End

# EURECA – setting the scene for scintillators

**H. Kraus[1*], E. Armengaud[7], M. Bauer[4], I. Bavykina[2], A. Benoit[13], A. Bento[2], J. Blümer[5,6], L. Bornschein[5], A. Broniatowski[10], G. Burghart[9], P. Camus[13], A. Chantelauze[6], M. Chapellier[8], G. Chardin[10], C. Ciemniak[3], C. Coppi[3], N. Coron[12], O. Crauste[10], F.A. Danevich[17], E. Daw[16], X. Defay[10], M. De Jésus[11], P. de Marcillac[12], G. Deuter[4], J. Domange[7], P. Di Stefano[11], G. Drexlin[5], L. Dumoulin[10], K. Eitel[6], F. von Feilitzsch[3], D. Filosofov[14], P. Gandit[13], E. Garcia[15], J. Gascon[11], G. Gerbier[7], J. Gironnet[12], H. Godfrin[13], S. Grohmann[6], M. Gros[7], M. Hannewald[7], D. Hauff[2], F. Haug[9], S. Henry[1], P. Huff[2], J. Imber[1], S. Ingleby[1], C. Isaila[3], J. Jochum[4], A. Juillard[10], M. Kiefer[2], M. Kimmerle[4], H. Kluck[6], V.V. Kobychev[17], V. Kozlov[6], V.M. Kudovbenko[17], V.A. Kudryavtsev[16], T. Lachenmaier[3], J.-C. Lanfranchi[3], R.F. Lang[2], P. Loaiza[18], A. Lubashevsky[14], M. Malek[1], S. Marnieros[10], R. McGowan[1], V. Mikhailik[1], A. Monfardini[13], X.-F. Navick[7], T. Niinikoski[9], A.S. Nikolaiko[17], L. Oberauer[3], E. Olivieri[10], Y. Ortigoza[15], E. Pantic[2], P. Pari[8], B. Paul[7], G. Perinic[9], F. Petricca[2], S. Pfister[3], C. Pobes[15], D.V. Poda[17], R.B. Podviyanuk[17], O.G. Polischuk[17], W. Potzel[3], F. Pröbst[2], J. Puimedon[15], M. Robinson[16], S. Roth[3], K. Rottler[4], S. Rozov[14], C. Sailer[4], A. Salinas[15], V. Sanglard[11], M.L. Sarsa[15], K. Schäffner[2], S. Scholl[4], S. Scorza[11], W. Seidel[2], S. Semikh[14], A. Smolnikov[14], M. Stern[11], L. Stodolsky[2], M. Teshima[2], V. Tomasello[16], A. Torrento[9], L. Torres[15], V.I. Tretyak[17], I. Usherov[4], M.A. Verdier[11], J.A. Villar[15], J. Wolf[5], E. Yakushev[14]**

*[1] University of Oxford, Department of Physics, Keble Road, Oxford OX1 3RH, UK*
*[2] Max-Planck-Insitut für Physik, Föhringer Ring 6, 80805 Munich, Germany*
*[3] Technische Universität München, Physik Department E15, 85748 Garching, Germany*
*[4] Eberhard Karls Universität Tübingen, Auf der Morgenstelle 14, 72076 Tübingen, Germany*
*[5] Institut für Experimentelle Kernphysik, Universität Karlsruhe (TH), Gaedestrasse 1, 76128 Karlsruhe, Germany*
*[6] Forschungszentrum Karlsruhe, Institut für Kernphysik, Postfach 3640, 76021 Karlsruhe, Germany*
*[7] CEA, Centre d'Etudes Saclay, IRFU, 91191 Gif-Sur-Yvette Cedex, France*
*[8] CEA, Centre d'Etudes Saclay, IRAMIS, 91191Gif-Sur-Yvette Cedex, France*
*[9] CERN, 1211 Geneva 23, Switzerland*
*[10] CSNSM, Université Paris-Sud and CNRS/IN2P3, 91405 Orsay, France*
*[11] Université de Lyon, F-69622, Lyon, France; Université de Lyon 1, Villeurbanne; CNRS/IN2P3, Institut de Physique Nucléaire de Lyon*
*[12] Institut d'Astrophysique Spatiale, UMR-8617 CNRS / Univ Paris Sud, Bat. 121, 91405 Orsay Cedex, France*
*[13] CNRS-Neel, 25 Avenue des Martyrs, 38042 Grenoble cédex 9, France*
*[14] Joint Institute for Nuclear Research, DLNP, 141980 Dubna, Moscow Region, Russia*
*[15] Laboratorio de Fisica Nuclear y Astropartículas, Facultad de Ciencias, Universidad de Zaragoza, C/ Pedro Cerbuna 12, 50009 Zaragoza, Spain*
*[16] Department of Physics and Astronomy, The University of Sheffield, Hicks Building, Hounsfield Road, Sheffield, S3 7RH, UK*
*[17] Institute for Nuclear Research, MSP 03680 Kyiv, Ukraine*
*[18] Laboratoire Souterrain de Modane, 90, Rue Polset, 73500 Modane, France*

EURECA (European Underground Rare Event Calorimeter Array) will be an astro-particle physics facility aiming to directly detect galactic dark matter. The Laboratoire Souterrain de Modane has been selected as host laboratory. The EURECA collaboration concentrates effort on cryogenic detector research in Europe into a single facility by bringing together colleagues from CRESST, EDELWEISS, ROSEBUD and additional new member institutes. EURECA will use a target mass of up to one ton for exploring WIMP-nucleon scalar scattering cross sections in the region of $10^{-9}$ – $10^{-10}$ picobarn. A major advantage of EURECA is the planned use of more than just one target material (multi target experiment for WIMP identification).

[*] Corresponding author. *E-mail address:* h.kraus@physics.ox.ac.uk

## 1. Motivation

Experimental data on the cosmic microwave background, combined with other astronomical and astrophysical data, give to high precision values for the fundamental parameters in our cosmological model [1]. Much of the matter density of the Universe seems to comprise non-luminous, non-baryonic particles [2]. Supersymmetry provides weakly interacting massive particles (WIMPs) as appealing and well-motivated candidates for this dark matter [3]. The WIMP-nucleon cross section appears to be at or below the electroweak scale and the expected event rates are correspondingly low. Thus, the identification of WIMP interaction in a detector could be challenging, owing to the rate of WIMP interactions being very small compared with the event rates expected from cosmic radiation and from the background radioactivity of present-day high-purity detectors. In addition, the recoil energies produced by elastic WIMP-nucleus scattering are very small, in the range of a few keV to a few tens of keV.

In order to address the experimental challenges mentioned above, a new generation of cryogenic detectors has been developed, exhibiting powerful background discrimination in combination with unprecedented energy threshold and resolution [4 – 7]. These detectors allow high-precision identification of nuclear recoils (caused by WIMP and also neutron interactions) by eliminating electron recoils due to radioactivity. Such detectors are installed in the EDELWEISS-II and CRESST-II dark matter search experiments, providing valuable R&D, expertise and experience for EURECA.

EURECA aims for a target sensitivity a factor >100 better than is currently projected by the 2$^{nd}$ phase of the above experiments. Although a discovery at WIMP-nucleon cross sections above $10^{-8}$ picobarn is not unlikely, the range covered by EURECA (extending to $10^{-10}$ picobarn) is currently the most favoured [8]. At the sensitivity limit, this translates to only few events per ton per year in typical targets, requiring an ultra-low background environment, excellent event type discrimination, neutron moderation and muon vetos.

## 2. Cryogenic detector technology

EURECA's detectors will evolve from those presently used in the CRESST, EDELWEISS and ROSEBUD experiments. The detectors are low-temperature calorimeters, operating in the millikelvin temperature range; and they use complementary techniques for the discrimination of nuclear and electron recoil events. EDELWEISS uses detectors based on charge-phonon discrimination [4], where the thermal signal induced by energy deposition in a germanium detector crystal is measured with a high-impedance thermistor attached to its surface. Simultaneously, the ionization signal is read out via electrodes on the crystal surface. The ratio between measured ionisation and heat signals provides an efficient method for the identification of the event type. CRESST and ROSEBUD use detectors based on scintillation-phonon discrimination [5, 6]. CRESST currently has several $CaWO_4$ absorbers and one $ZnWO_4$ crystal installed. The thermal signal is measured with a superconducting transition edge sensor (TES) on the crystal surface. Simultaneously, scintillation is detected by thin calorimeters again using TES sensors, but optimized for detection of scintillation.

The aim of EURECA R&D is to explore concepts and designs, based on the existing technologies, appropriate to a large-scale experiment. The exploitation phases of EDELWEISS-II and CRESST-II are aligned in time scale with the R&D for EURECA and the design of the experiment. This should allow us to select the optimum detector technology for EURECA.

A significant advantage of cryogenic detectors is their modularity. Once a design for an individual module has proven to be successful, the same design can be replicated in many copies of that module. This allows mass production, assembly, commissioning and quality control shared out among suppliers and some of the tasks can be carried out in parallel. The detectors will of course have to be cooled to millikelvin temperature for operation, requiring a suitable dilution refrigerator unit. Accommodating a larger target requires (only) a larger vacuum container and increased cooling power of the cryostat, neither of the two having direct impact on detector operation.

Furthermore, a modular approach is vital to achieving large detector masses. There are likely to be limitations in large-scale detectors due to radioactive backgrounds present in the target materials. With individual sub-kilogram solid targets, modules with abnormally high backgrounds can be isolated and replaced.

A further important feature of EURECA is its multi-material target. Having several targets is highly desirable for establishing a true WIMP signal by testing for the correct A-scaling of the WIMP-nucleon scattering cross section. Further strong motivation for equipping EURECA with a range of target materials is provided by kinematic considerations, as the mass of the WIMP is unknown. A natural initial choice for EURECA is to use germanium and tungstate targets, given the expertise of the collaboration. Additional absorbers are being researched and optimized [9 – 14].

Arranging the detectors in a large array of smaller absorbers has the additional advantage of allowing testing for a uniform rate within the target, and for providing an additional dark matter signature by requiring single interactions only for a dark matter candidate event. This should allow identification of residual neutron background through coincidences.

## 3. Scintillator targets

A key feature of EURECA is the operation of different target materials and complementary discrimination technologies within a common low-background volume. The ability to test the scaling of the WIMP-nucleon scattering cross section with atomic mass number will be an important tool for confirming a signal as being positive evidence for WIMP interactions. Scintillating crystals offer a wide selection of target nuclei, making them appealing absorbers for dark matter searches. The phonon-scintillation discrimination technique therefore adds desired complementarity to germanium targets with phonon-ionization discrimination. The application of scintillators in rare event searches already has a long history and the optimization of the most important scintillator parameters has thus been the subject of intense research. Rare event searches have an important advantage over main stream scintillator applications, namely that speed (short time constants) is not a priority. What really matters is high light yield and low intrinsic radioactivity. To achieve the sensitivity levels required for probing currently favoured dark matter models, scintillators with a light yield in excess of ~15,000 photons/MeV and intrinsic radioactivity below 0.1 mBq/kg are needed. Additional requirements for the materials are imposed through their operation at millikelvin temperatures. Scintillators suitable for cryogenic use have to have low specific heat and a surface compatible with being instrumented with a thermometer sensor glued to it [15 – 17].

Despite of a large number of materials that are known to be good scintillators, a deeper analysis of their performance against the above criteria shows that the majority do not meet the full set of requirements. Indeed, the light yield of classical, doped scintillators, such as NaI(Tl), CsI(Tl), CsI(Na) decreases substantially with temperature, thereby spoiling the merits of these materials for cryogenic applications. The same concern applies to the family of rare-earth-doped scintillators ($CaF_2$(Eu), $YAlO_3$(Ce), $Lu_2SiO_5$(Ce), $Y_3Al_5O_{12}$(Ce), $LaCl_3$(Ce), etc). In addition, these materials exhibit a very high level of intrinsic radioactive background inherent to the rare-earth host matrix or the dopant. This makes them totally unacceptable for use in rare event searches.

Given the constraints, our interest focuses on self-activated scintillators. These materials exhibit high light yield at low temperature; they are usually fairly stable and affordable. However, the suitability of each particular compound has to be analyzed individually. For example, the light yield of pure halide scintillators ($BaF_2$, $CaF_2$, CsI) is constant at low temperatures and can be considered reasonably good. However the radio purity of $CaF_2$ and $BaF_2$ requires some attention, while the hygroscopic nature of CsI causes practical and technological difficulties for detector production and handling. Oxide scintillators are fairly robust and stable, but some of the widely used scintillators, such as $Bi_4Ge_3O_{12}$ and $PbWO_4$, are problematic due to natural Bi and Pb inevitably containing radioactive isotopes at significant levels of abundance, i.e. they are members of the U-Th radioactive decay chains. In addition, $Bi_4Ge_3O_{12}$ is usually polluted by $^{207}$Bi, though a radio-pure version is available commercially. $CdWO_4$ contains the β-active $^{113}$Cd at a level of 0.56 Bq/kg. Production of these materials with the required level of purity is certainly a path to pursue and such attempts are being undertaken by scintillator manufacturers.

Among the candidate compounds that currently satisfy the selection criteria $ZnWO_4$, $CaWO_4$ and $CaMoO_4$ are materials of our particular interest. They prove to have a stable, high light yield at low temperature [10, 18] and they can be produced with low intrinsic radioactivity [19, 20]. The possibility of manufacturing these crystals with sufficiently large sizes [21] underpins the good prospect of these materials in dark matter search experiments. In addition, $Al_2O_3$ is considered for exploring scenarios in which a very low detection threshold is advantageous [13, 22].

A major task faced by EURECA will be mass production of the detector modules, which implies moving away from prototyping, which, by its nature, has a low rate of detector production. Regarding the supply of scintillating absorbers in large quantities, we are working with suppliers on the reduction of radioactive impurities and at the same time aim to increase the light yield and size of the scintillators. This avenue shows great promise for the development of substantially improved materials. In this systematic approach we put great emphasis on reproducibility, reliability and quality control.

Nevertheless significant progress in the area of crystal development is required before samples of ultimately high performance can be produced. We will have to improve considerably the radiopurity of the bulk detector materials. To perform this, careful selection of materials is needed, which requires a large number of samples to be tested with sensitivities of at least two orders of magnitude better than the present ones. An important aspect is the screening of raw detector materials before they enter the production process. This has to be done to a level compatible with sensitivities of measuring only a few events per year.

The success of scintillating bolometers in the large scale cryogenic dark matter experiment EURECA relies hugely on progress in material production. Owing to the extreme requirements on the quality of the absorbers and their properties, the complementary expertise from different areas of physics, chemistry and crystal production is a vital component of this development. Given the scale and the complexity of this project it is also clear that centralisation of research on one site is impossible and that the optimum solution is a strong and effective international collaboration. The research is genuinely interdisciplinary and it should be based upon the joint efforts of the leading experts, aiming to develop the ultimate scintillating absorber for EURECA.

## References

1. D.N. Spergel et al., Astrophys. J. Suppl. 170 (2007) 377.
2. L. Bergström, Rep. Prog. Phys. 63 (2000) 793.
3. B.W.L. Lee, S. Weinberg , Phys. Rev. Lett. 39 (1977) 165.
4. V. Sanglard et al., Phys. Rev. D 71 (2005) 122002.
5. G. Angloher et al., Astropart. Phys. 23 (2005) 325.
6. G. Angloher et al., to be published in Astropart. Phys., (arXiv:0809.1829).
7. D.S. Akerib et al., Phys. Rev. Lett. 96 (2006) 011302.
8. J. Ellis et al., Phys. Rev. D 71 (2005) 095007.
9. H. Kraus et al., Phys. Lett. B 610 (2005) 37.
10. V.B. Mikhailik, H. Kraus, J. Phys. D: Appl. Phys. 39 (2006) 1181.
11. F.A. Danevich et al., Phys. Stat. Sol. A 205 (2007) 335.
12. I. Bavykina et al., IEEE Trans. Nucl. Sci. 55 (2008) 1449.
13. P.C.F. Di Stefano et al., J. Low Temp. Phys. 151 (2008) 902.
14. A. Calleja et al., J. Low Temp. Phys. 151 (2008) 848.
15. J.-C. Lanfranchi et al., Nucl. Instr. Meth. A 520 (2004) 135.
16. S. Roth et al., arxiv:0810.0423 [astro-ph], to be published in the proceedings of CryoScint08, Optical Materials, Elsevier (doi:10.1016/j.optmat.2008.09.013).

17. M. Kiefer et al., arxiv:0809.4975 [astro-ph], to be published in the proceedings of CryoScint08, Optical Materials, Elsevier (doi:10.1016/j.optmat.2008.09.019).
18. V.B. Mikhailik et al., Phys. Rev. B 75 (2007) 184308.
19. A.N. Annenkov et al., Nucl. Instrum. Meth. A 584 (2008) 334.
20. H. Kraus et al., Nucl. Instr. Meth. A 600 (2009) 594.
21. L.L. Nagornaya et al., IEEE Nucl. Sci. Symp. 2008, p. 3266.
22. J. Amaré et al., Appl. Phys. Lett. 87 (2005) 264102.

# Our short experience at IAS and within ROSEBUD with radioactive contaminations in scintillating bolometers: uses and needs

**N. Coron[a], E. García[b], J. Gironnet[a], J. Leblanc[a], P. de Marcillac[a*], M. Martínez[b], Y. Ortigoza[b], A. Ortiz de Solórzano[b], C. Pobes[b], J. Puimedón[b], T. Redon[a], M.L. Sarsa[b], L. Torres[a,b], J.A. Villar[b]**

[a] *Institut d'Astrophysique Spatiale, Bât. 121, 91405 Orsay Cedex, France*
[b] *Laboratorio de Física Nuclear y Astropartículas, Facultad de Ciencias, Universidad de Zaragoza, C/ Pedro Cerbuna 12, 50009 Zaragoza, Spain*

Internal radioactive contamination in scintillating bolometers aiming to detect dark matter, which should be absolutely avoided in the ultimate stage of experiments, is a very valuable tool in their definition stage. The goal of this presentation is to report on our past experiences with scintillating bolometers, a mixed "heat and light" detection technique, both at sea level and underground. Focus is given to the last materials tested within the ROSEBUD collaboration in 2007: sapphire, BGO and LiF. An original use of delayed coincidences in the decays from the natural radioactive chains is also presented with the example of a $SrF_2$ crystal: it highlights position dependence in the light signal which worsens the resolution of this channel.

## 1. Introduction

The ROSEBUD collaboration is a joint effort between the IAS at Orsay (France) and the University of Zaragoza (Spain) to develop cryogenic detectors able to detect the hypothetical dark matter particles. The experiment is operated underground at the Laboratorio Subterráneo de Canfranc (LSC) but most of the developments and characterization of the prototypes are made at sea level in Orsay, for convenience, as well as for economical reasons: a unique, light weight dilution refrigerator is shared between the two sites. This requires relatively fast detectors with time constants less than some 10 ms in order not to blind the detectors with the cosmic rays at surface.

Scintillating bolometers with typical masses of ~50 g are able to efficiently discriminate between alphas, gammas and nuclear recoils above some tens of keV, at 20 mK, the base temperature of the refrigerator. Particles are discriminated through their ionization power. The technique uses the information provided by both heat and scintillation signals in the target, the latter being detected by an auxiliary light absorbing bolometer, made from a thin disk of Ge and optically coupled to the heat detector in a light reflecting cavity (see Fig. 1 for details on the double bolometer configuration).

The power of the technique relies on the high energy resolution power of the heat channel, found usually in every "good" single bolometer (below 2%), together with an independent measurement of the light emission, signing the nature of the incident particle. In order to explore this new technique, importance has been given first to the test of different materials (from known 300 K scintillators – as BGO, $CaWO_4$,… – to materials known to have excellent thermal properties at low temperature – as sapphire – independently of their radioactive content. As a result, rather high radioactive levels were encountered in the materials tested so far; however, associated events were used to gain a deeper comprehension of the detectors. Some intrinsic contaminations, as $^{207}$Bi in BGO, should be reduced for the next generation of detectors.

[*] Corresponding author. *E-mail address:* pierre.demarcillac@ias.u-psud.fr

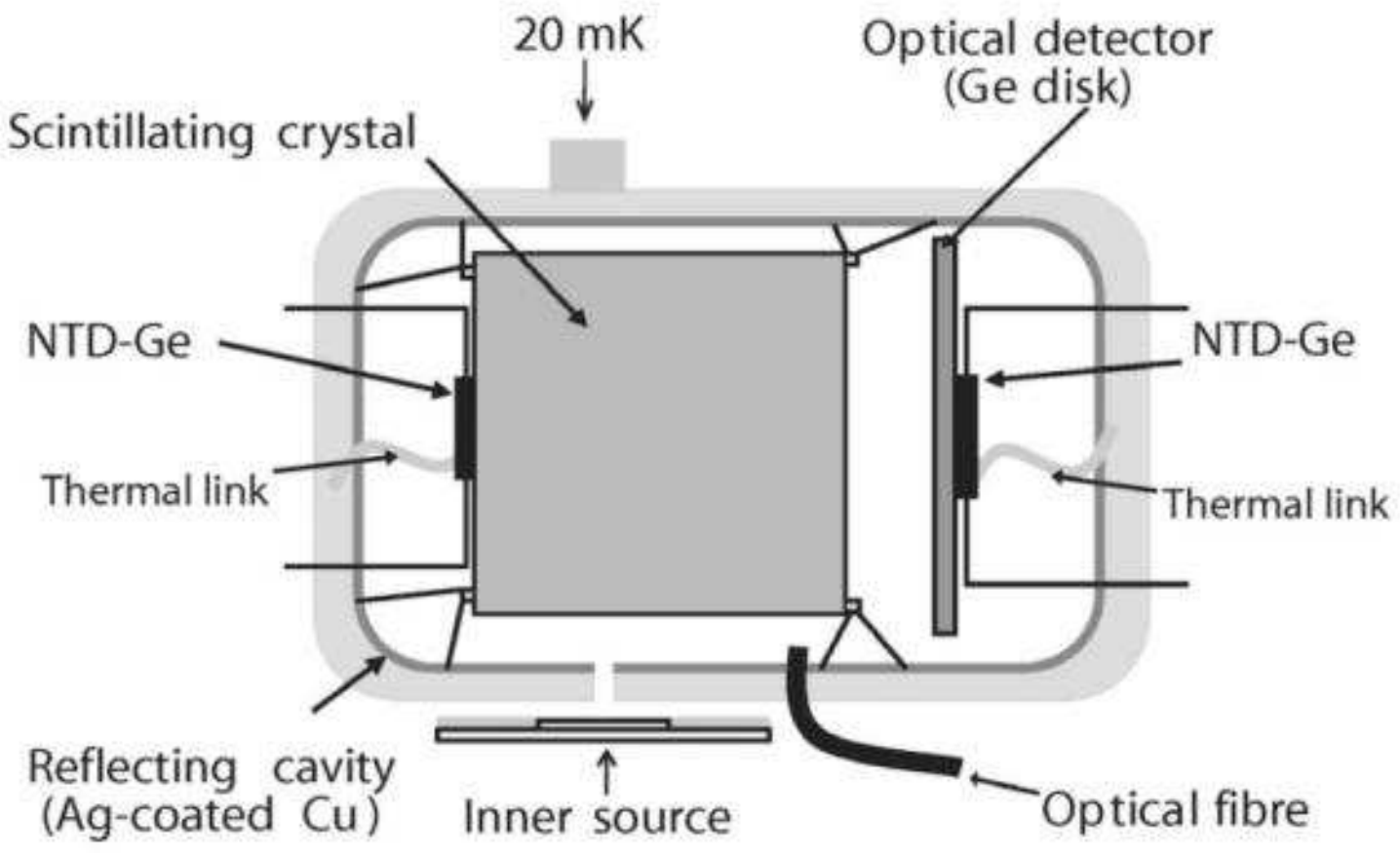

Fig. 1. Schematic view of the double bolometer configuration. Neutron Transmutation Doped Germanium (NTD-Ge) thermistors are used to read both signals (light and heat).

## 2. The ROSEBUD run in 2007: a complementary set of scintillating bolometers

Scintillating bolometers offer a wide choice of targets. This property is very welcome to face the uncertainties associated with the dark matter particles which are looked for, as its ability to couple to nuclear spin or, if not, the scaling of the cross section with the nuclear content. A set of three double bolometers with targets (from top to bottom) made of 46 g BGO ($Bi_4Ge_3O_{12}$), 33 g LiF and 50 g sapphire ($Al_2O_3$), each optically coupled to its own optical Ge bolometer, was mounted under the 20 mK mixing chamber of the refrigerator. A complete characterization – thermal responsivities, light yields, discrimination powers – was performed at Orsay before going underground.

Four underground runs, each lasting two weeks, were undertaken in 2007 in the ROSEBUD installation at LSC. Shielding was improved from the February to the May run (increase of lead shielding and removal of radon by nitrogen flushing) reducing the background as can be seen in Fig. 2 and 3. The last run was dedicated to neutron calibration with an external $^{252}$Cf source. The results obtained have been analyzed [1, 2], but a complete interpretation of the non-scintillating events seen in the so-called “recoil branch” in the light versus heat discrimination plots is still underway: it will probably need complementary measurements.

A different task was assigned to each bolometer in the experiment: the sapphire one (a low Z material with exceptional thermal properties at low temperature) was recording the low energy events at the keV level, the BGO detector (having 66% Bismuth content in weight) tracked the gamma background profiting from a high efficiency, while the LiF detector attempted to detect the residual neutrons using the $^6$Li neutron capture reaction. The questions concerning the suitability of these detectors for dark matter detection are only discussed here in the context of their internal radioactive contamination and of analysis of external backgrounds, which is the main concern of this RPScint’2008 workshop. A previous campaign in 2000 allowed us to quantify the radioactivity content of a 54 g $CaWO_4$ detector, which was found to be strongly contaminated in the U-Th chains [3].

### 2.1. Radio-purity in sapphire

Sapphire is a material with one of the highest melting temperature (~2050 °C). We might expect an important segregation of impurities during the growth of the crystal. The sample used in ROSEBUD was grown from the melt according to the Kyropoulos technique, “somewhere in Russia”. No line is seen in the sapphire background within the small exposure time, at low energy ($E < 200$ keV).

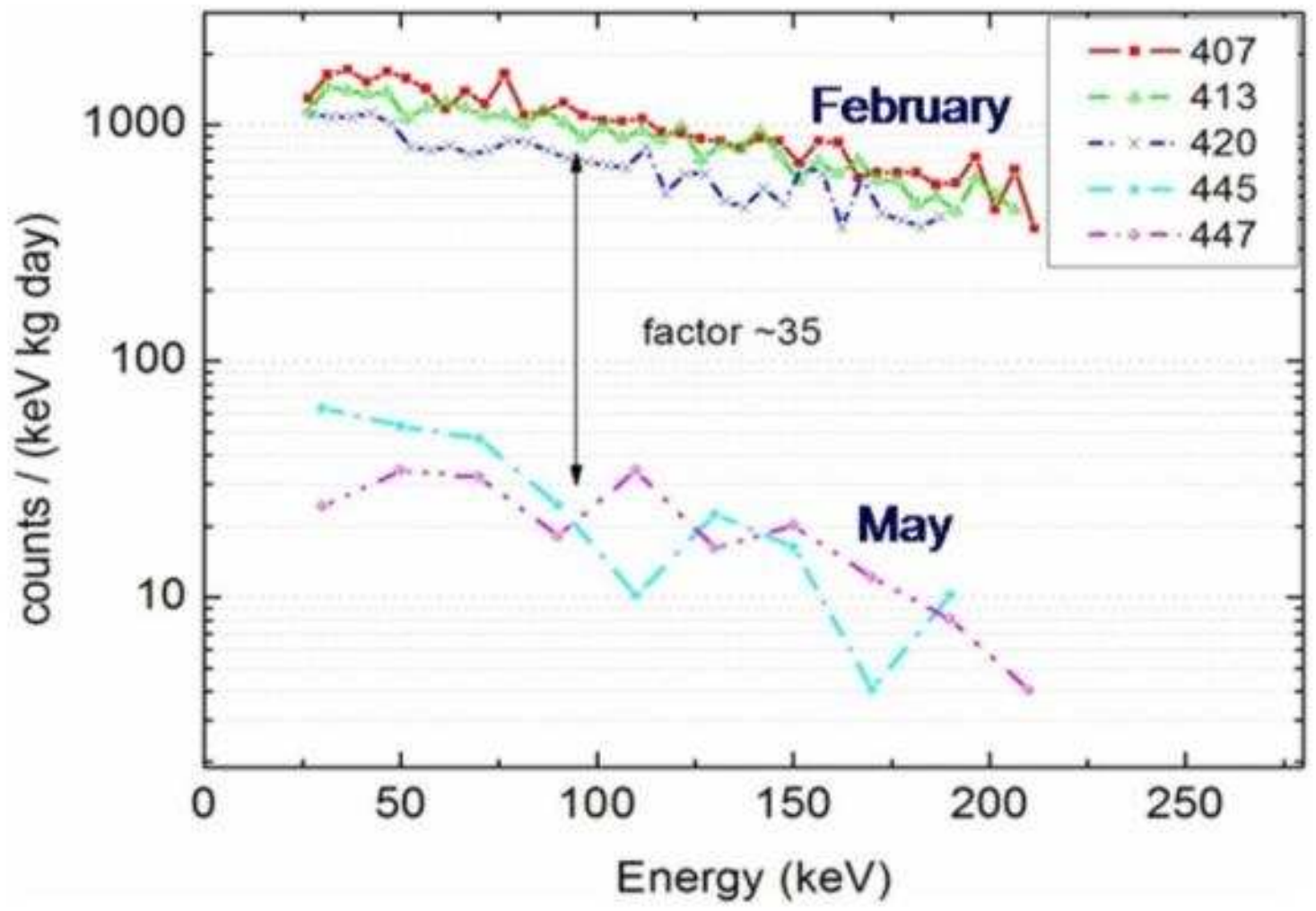

Fig. 2. Improvement of the gamma background in the 50 g sapphire detector at Canfranc between February and May 2007.

Little can be said at higher energy because of the dynamics chosen in the acquisition but one should remind that sapphire is a material difficult to calibrate with high energy gammas. Analysis of internal alpha contamination in the sapphire itself was not addressed in the particular detector tested within ROSEBUD in 2007, which was known to be unintentionally contaminated after a long exposure to a $^{236}$Pu source.

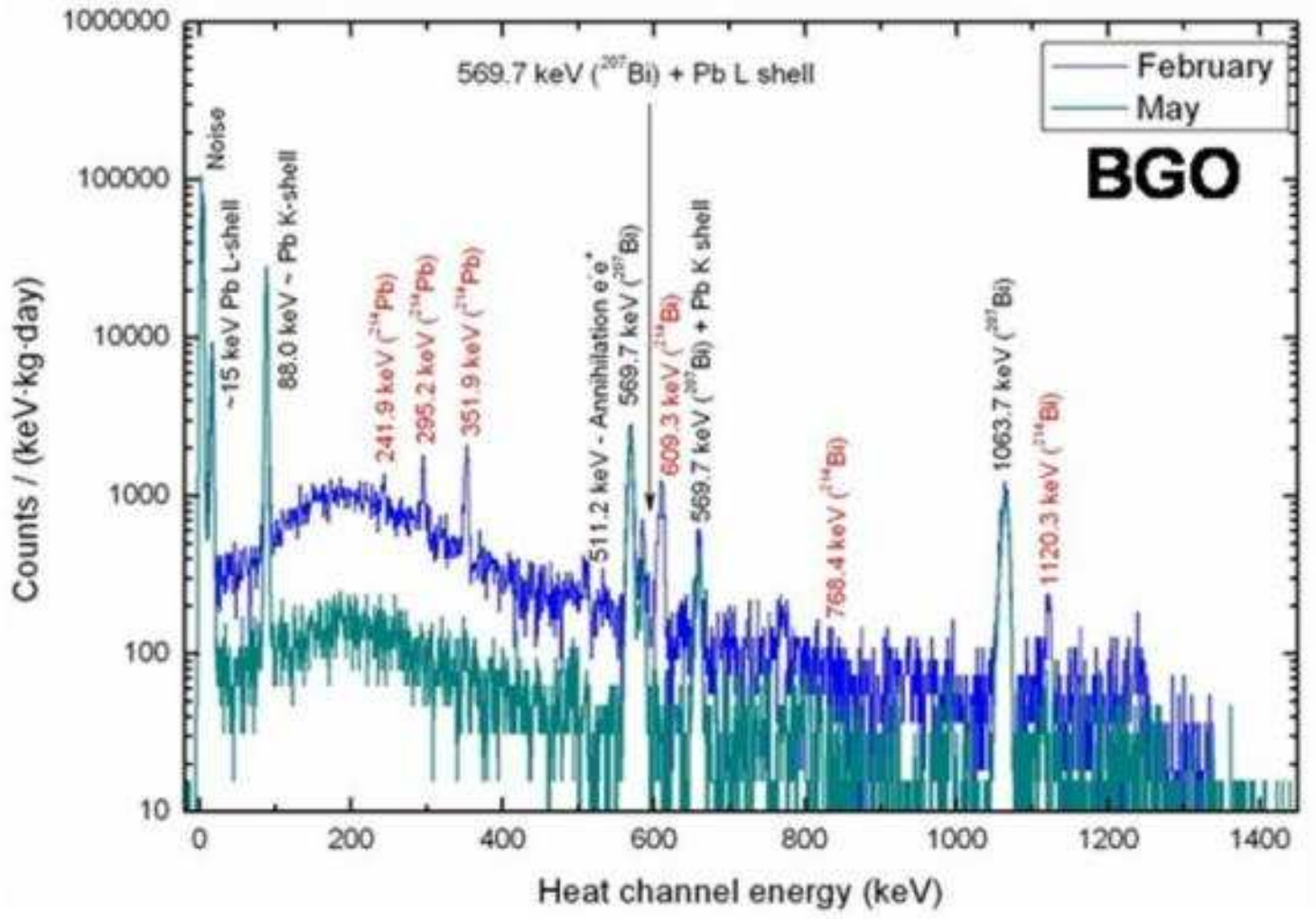

Fig. 3. Improvement of the gamma background in the 46 g BGO detector at Canfranc between February and May 2007. Radon lines disappear in May thanks to nitrogen flushing through the set-up. The lines from EC in $^{207}$Bi are dominating this spectrum, with a special mention to the 88 keV line (X-ray and Auger electrons cascade following capture in the $^{207}$Pb K-shell).

### 2.2. Radio-purity in BGO

The 46 g BGO detector presented a 6 keV energy threshold. It was known to be heavily contaminated by $^{207}$Bi from previous measurements at Orsay [4]. Using the event rate on the 88 keV line and published branching ratios the internal contamination in $^{207}$Bi was estimated at a level of 3300 mBq/kg of BGO (i.e. 5000 mBq/kg of the original bismuth material).

Special attention has been given to the 1063.7 keV gamma events seen in the BGO gamma background spectrum. The 1633.3 keV excited state of $^{207}$Pb that feeds this line has a relatively long lifetime ($T_{½}$~0.81 s) while the X rays and Auger electrons cascade emitted when filling the vacancies created by capture in the K, L, M… shells is fully absorbed. The delayed coincidence (cascade + gamma) could be identified in the data and will be used to improve the accuracy of the K/(L+M+…) EC ratios with respect to published ones.

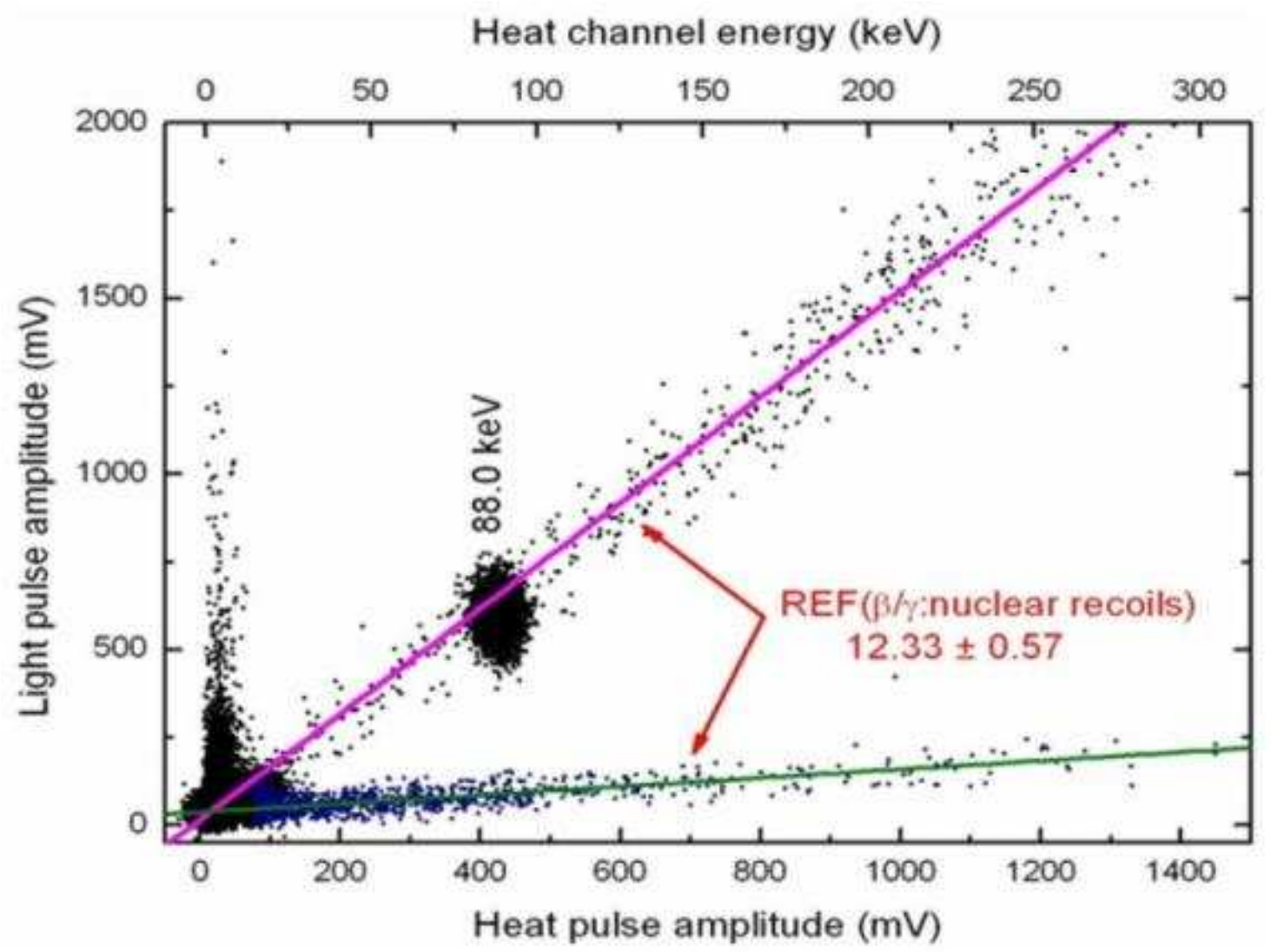

Fig. 4. Light versus heat discrimination plot in the 46 g BGO detector under a $^{252}$Cf neutron irradiation. The 88 keV line following EC at the $^{207}$Pb K-shell dominates and is well separated from the nuclear recoil events while the 15 keV lines (L-shell) could limit the detector discrimination threshold.

It is however obvious from the “heat and light” discrimination plots registered (see Fig. 4) that such a high contamination level in $^{207}$Bi could limit the use of these detectors for dark matter research. As we were aware of this serious drawback of BGO detectors, we made some years ago a compilation of $^{207}$Bi content in bismuth containing materials.

Table 1. $^{207}$Bi in bismuth compounds.

| Source | Activity mBq/kg Bi | Reference and comments |
|---|---|---|
| BGO Crismatec (radio-pure quality) | 7 ± 2 | LSC (1999); J. Puimedón, *private communication* |
| | 9 ± 3 | LSM (1999); C. Goldbach (also quoted in mBq/kg BGO: $^{40}$K<7; $^{212}$Pb=6±4; $^{208}$Tl=3±2), *private communication* |
| BGO Harshaw | 175 | LSM (<1996?); P. Hubert, *private communication* |
| BGO Crismatec | 750 | |
| PbS concentrate | 2000 | K. Fukuda et al. (1995) [5] |
| BGO Crismatec | 4400 | Y. Satoh et al. (1993) [6] |
| Bi ingot (1992) | 10000 | |
| Bi ingot (1992) | 2900 | |
| Bi ingot (1992) | 2000 | |
| Bi ingot (1967) | <200 | |

The BGO detector tested by ROSEBUD in 2007 has a $^{207}$Bi content similar to the one tested by Satoh et al in 1993: both crystals were bought from the same company. The origin of such high level of $^{207}$Bi in most material tested can be understood if we recall that Bismuth is mostly extracted as a byproduct in Pb metallurgy: the reactions $^{207}$Pb(p,n) and $^{206}$Pb(p,γ) may occur in lead with protons from the cosmic rays. A second source of $^{207}$Bi is anthropogenic: it has been detected in sediments in association with nuclear tests in the 1960-1970's. In principle it could be produced as well in bismuth ores through the $^{209}$Bi(n,3n) but fast neutrons are needed. The radio-pure quality BGO developed by the Crismatec company comes, probably, from selected bismuth materials extracted in leadless environments. New detectors were mounted with crystals issued from this radio-pure quality BGO, and they have been tested at sea level in Orsay. The background spectrum was published some years ago [4] and shows lines hardly seen above the continuous background in the heat channel that were attributed to $^{207}$Bi, giving a reduction factor in $^{207}$Bi content better than 15. However, they are probably due to $^{214}$Bi (see D. Grigoriev et al., in these Proceedings) suggesting that the reduction factor is much closer to the expected one (~500-1000).

### 2.3. Radio-purity in LiF

The scintillating LiF detector aimed to detect environmental neutrons using the neutron capture reaction on $^{6}$Li (n+$^{6}$Li→α+t; Q=4.78 MeV) as shown in Fig. 5. Alpha contaminations in LiF bolometers may be a relevant background source for the estimate of the neutron flux and a light yield improvement would be highly desirable for a better discrimination.

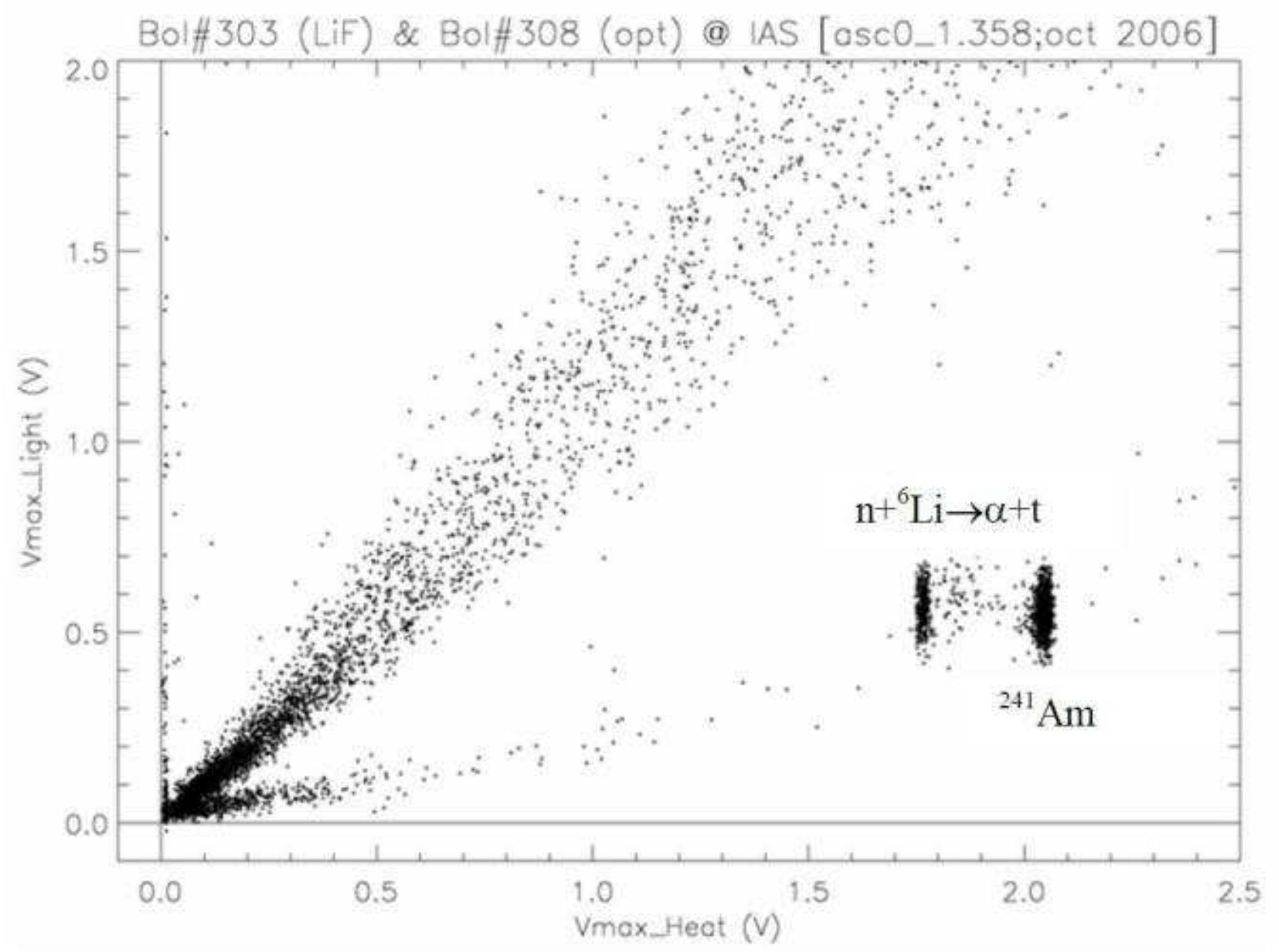

Fig. 5. Light versus heat discrimination plot during a background night at IAS (sea level) as recorded by the 33 g natural LiF detector. A $^{241}$Am source has been included in the set-up. While it is obvious that an alpha emits less light than an alpha + tritium pair releasing the same energy, a light yield improvement would be highly desirable for a better separation. Thermal and fast neutrons from the ambient background are detected: the latter detection underlines the need to perform these developments underground to avoid the nuclear recoils following fast neutron scattering in the bolometer targets.

Within the short exposure time during the 2007 runs at LSC, we could hardly detect any significant alpha contamination (see Fig. 6). A calibration with thermal neutrons from a $^{252}$Cf source suggests a slight internal $^{210}$Po contamination (~mBq/kg) at the level of one count per night, identified at 5.4 MeV. One should recall that bolometers, thanks to their high energy resolution power, can discriminate between internal and external contamination from alpha emitters: the 33 g natural LiF target used at Canfranc showed a better than 40 keV FWHM energy resolution, which is sufficient to resolve external decays (alpha only) from internal ones (alpha + recoil) that are separated ~100 keV.

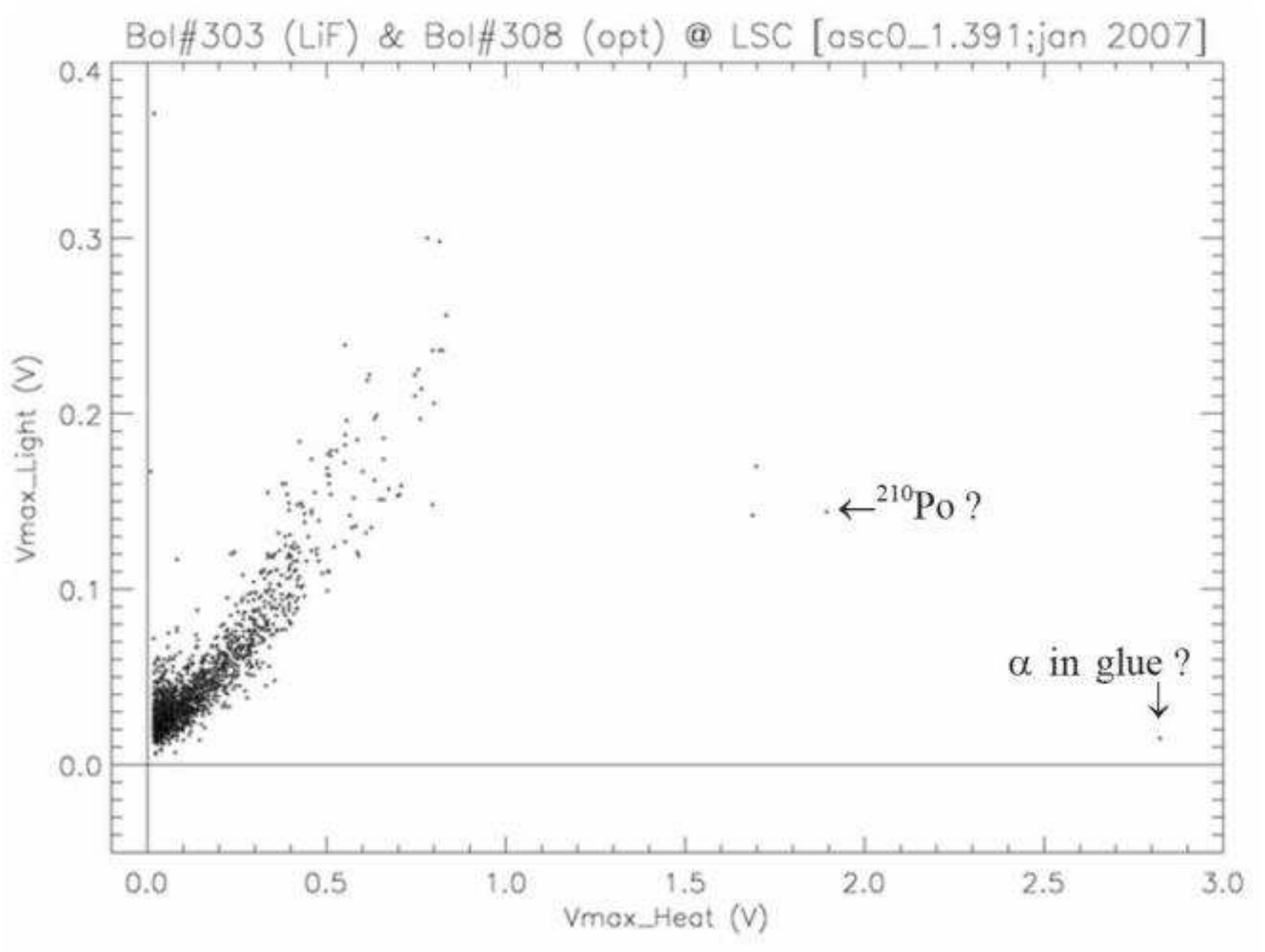

Fig. 6. Light versus heat discrimination plot during a background night at LSC (underground) as recorded by the 33 g natural LiF detector. With respect to the previous figure, the $^{241}$Am source has been removed and the strong suppression of the cosmic rays underground can be noticed. The slight $^{210}$Po alpha contamination suspected, as well as the single event detected at high energy in a non scintillating part of the detector, which might be attributed to an alpha decay in the glue, are also shown.

A flux of (2-5)$\times 10^{-6}$ n/(cm$^2$ s), in the range of published levels of neutron fluxes underground, has been derived for thermal neutrons inside the low background shielding at LSC. To increase the neutron detection efficiency (both for thermal and fast neutrons) enrichment in $^6$Li – natural abundance of $^6$Li is 7% – and/or increasing the mass of the LiF crystal are the solutions proposed for the next step of ROSEBUD in view of EURECA.

### 3. Use of radioactivity in scintillating bolometers to study the origin of the light signal dispersion. The case for a 54 g $SrF_2$ crystal

The energy dispersion in the light channel constrains the discrimination efficiency between gammas and nuclear recoils at low energy. It is therefore of the utmost importance to study its origin, but few practical tools are available for this purpose. In particular, one would wish to disentangle light yield inhomogeneities or other geometrical effects from statistical fluctuations.

Cascading alpha decays from the natural radioactive chains could provide such tools. We can illustrate this idea with data taken from a 54 g $SrF_2$ scintillating bolometer which was found to be highly contaminated in the natural radioactive chains (see Fig. 7). In alpha decays the recoiling nuclei have ranges of about 100 Å. Thus, in a cascading pair, the light emission from both decays is issued virtually from the same point in the crystal (typically cm-sized).

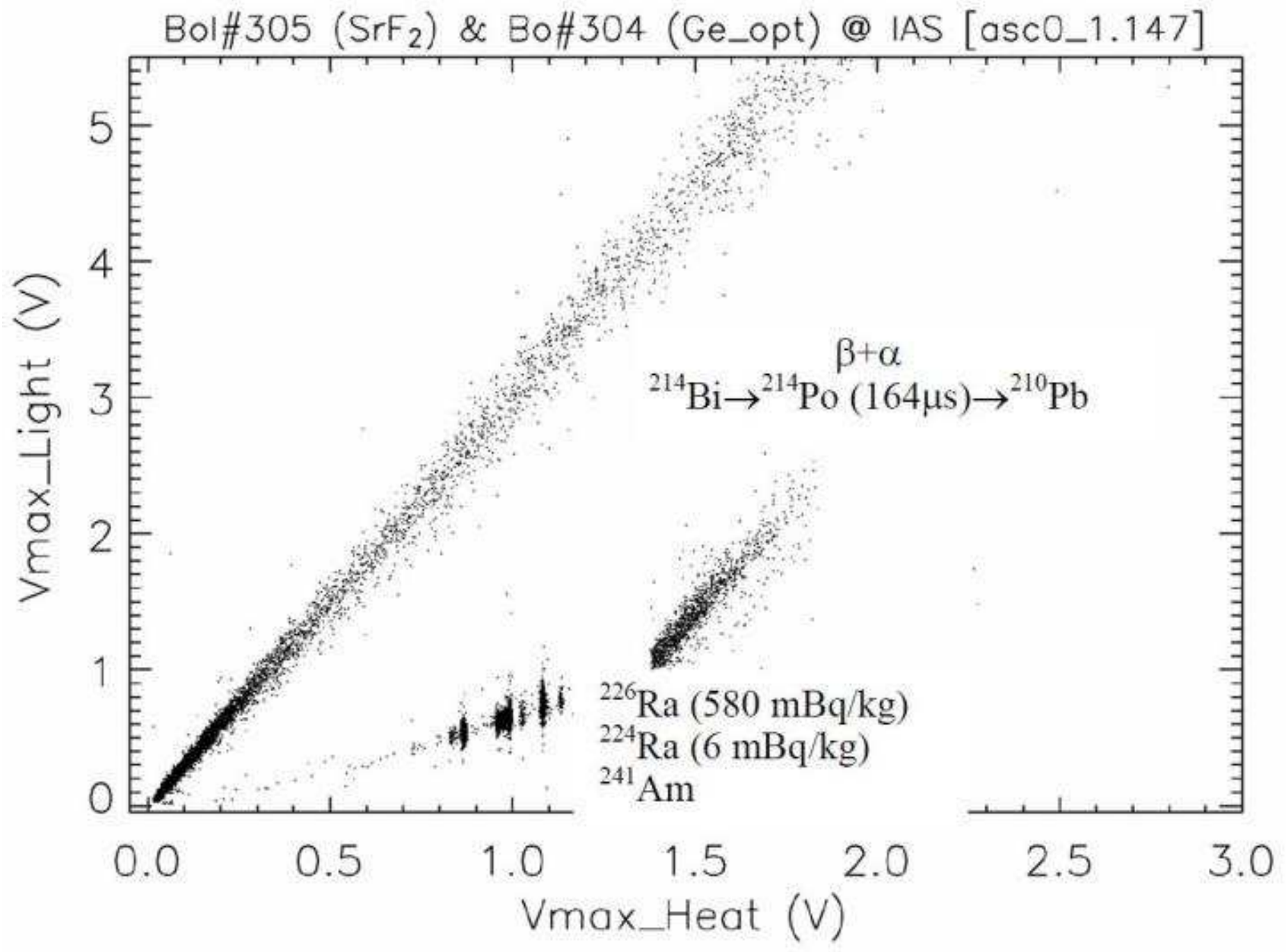

Fig. 7. Background in a 54 g $SrF_2$ bolometer at IAS which evidences the presence of a high contamination in the natural radioactive chains. A $^{241}$Am source was added for calibration purposes.

We tracked the events associated with the following decay cascades:

– $^{224}$Ra→$^{220}$Rn ($Q_\alpha$=5789 keV; $T_{½}$=3.7 d)
– $^{220}$Rn→$^{216}$Po ($Q_\alpha$=6405 keV; $T_{½}$=55 s)
– $^{216}$Po→$^{212}$Pb ($Q_\alpha$=6907 keV; $T_{½}$=150 ms)

which were easily identified, the decay constant of $^{220}$Rn being much lower than the mean rate of alpha decays seen in the detector. All decays proceed at 100% with alpha emission. Note that the last decay occurred very often in the same track as the second one, due to the 80 ms recording length chosen in the acquisition. A preliminary analysis of these data is summarized in Fig. 8.

This indicates the existence of a position dependence of the light signal that can be attributed to geometrical origin or inhomogeneities in the crystal. Alphas coming from an external $^{241}$Am facing at the detector through a collimated hole show a better energy resolution in the light channel (see Fig. 9) than those measured from the internal contaminations, supporting the above mentioned interpretation.

## 4. Conclusions

Radioactive contaminations in scintillating bolometers are very useful tools to fully characterize the detectors at a first development step. A complementary target approach, in the spirit of the future EURECA project has been initiated within the ROSEBUD project and was very rewarding. Bigger and $^6$Li enriched LiF detectors are clearly needed to monitor the neutron flux in a future cryogenic dark matter search experiment like EURECA. Commercial BGO targets suffer from $^{207}$Bi contamination at high level, but radio-pure raw material exists with a $^{207}$Bi contamination much reduced. A 91g BGO detector made from such target will be studied at LSM in 2009 within the EDELWEISS II installation.

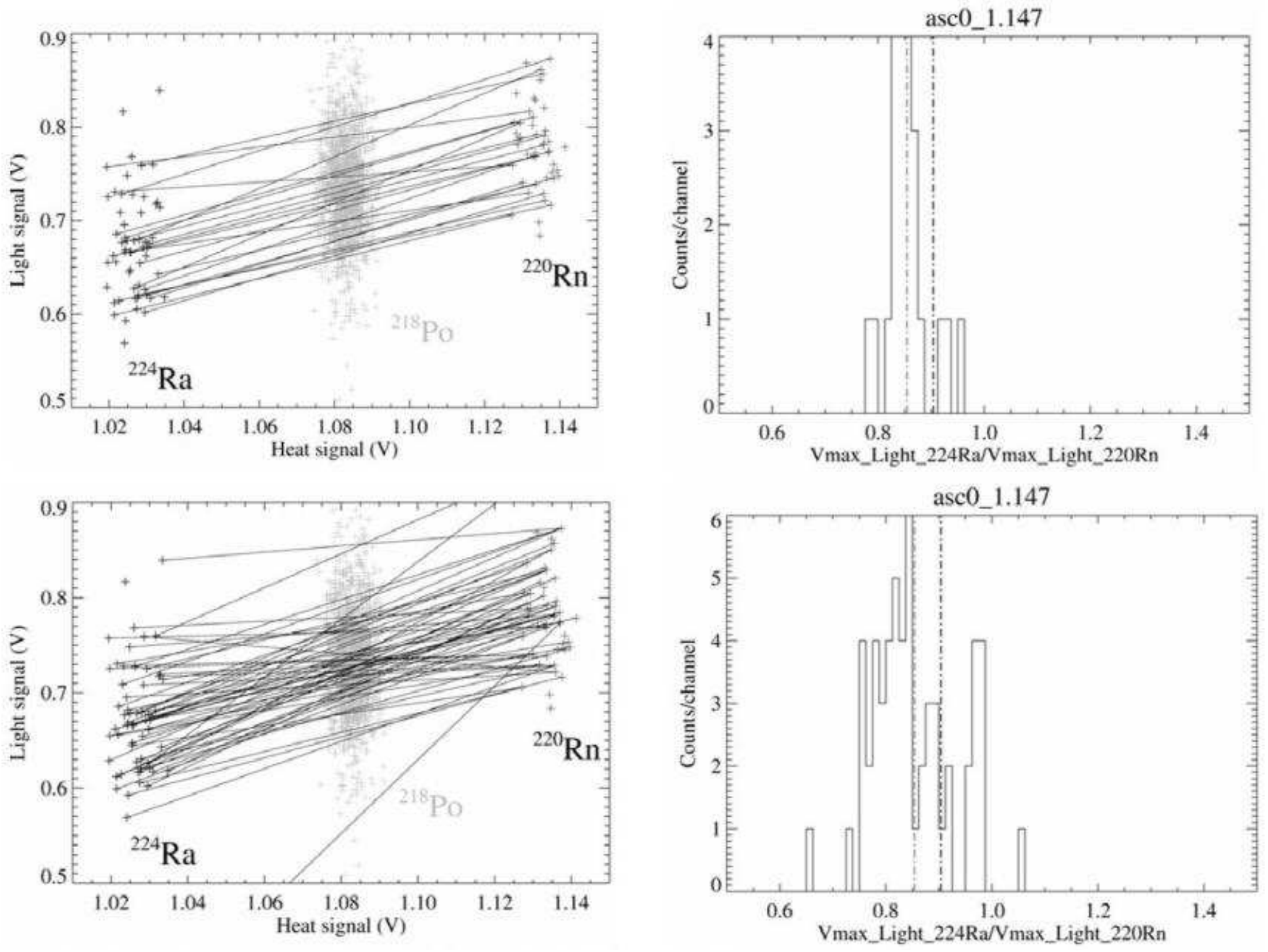

Fig. 8. Light dispersion analysis of decaying pairs in the 54 g $SrF_2$ bolometer.
Top: Associated pairs from the $^{224}Ra \rightarrow ^{220}Rn \rightarrow ^{216}Po$ are joined by lines. Most of these lines are parallel which suggests a strong correlation of the light emission with the locus of the decay (left). The distribution of the ratios of light signals issued during paired decaying events (dash-dotted line) is slightly shifted with respect to the expected ratio (~0.904), which merely reflects the increasing ionization yield of alphas with energy (right).
Bottom: Artificial, unphysical pairs are created by taking $^{220}Rn \rightarrow ^{216}Po$ decays and the following $^{224}Ra \rightarrow ^{220}Rn$ one, in a kind of time reversal (left). The resulting distribution of the light ratios is much more dispersed.

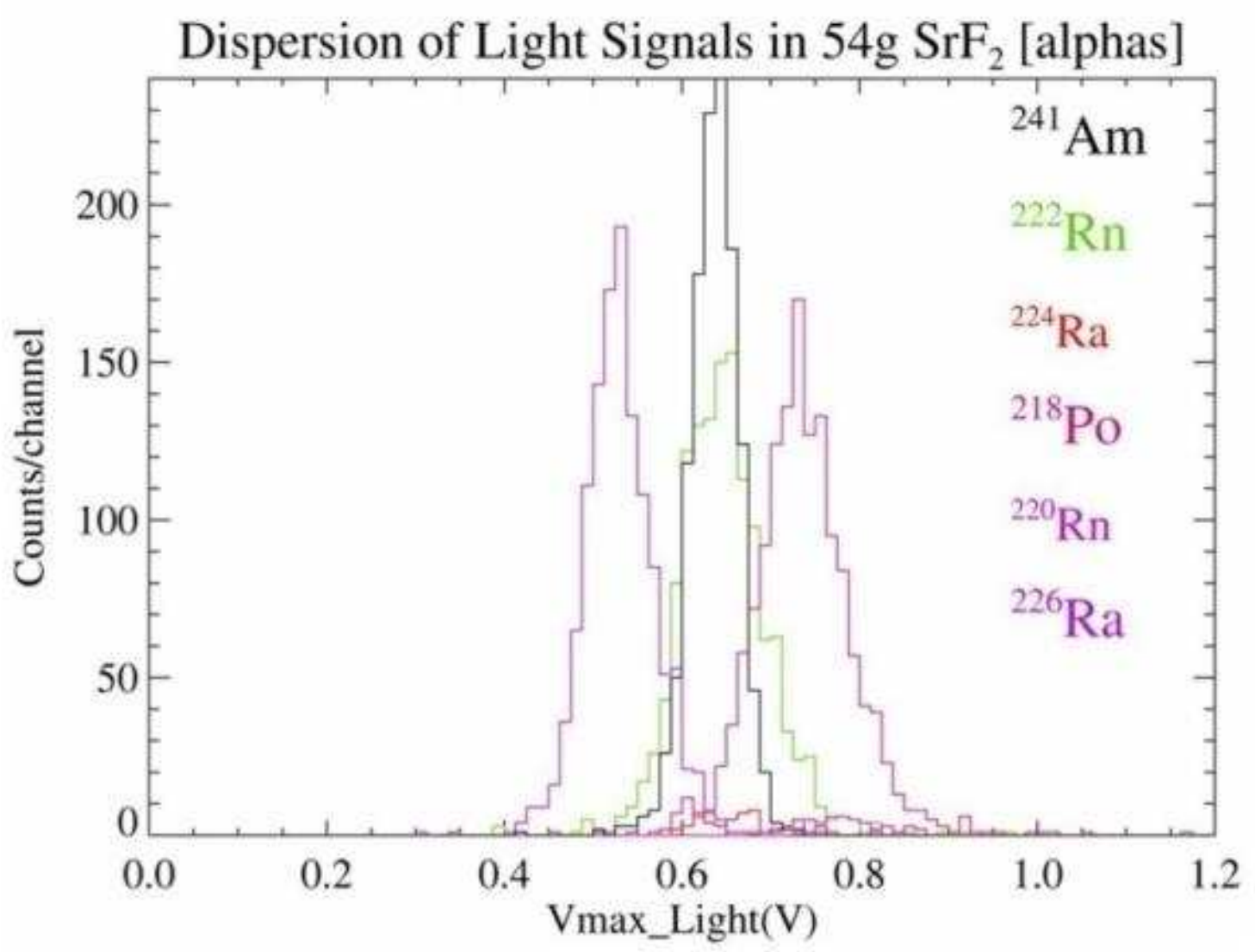

Fig. 9. Dispersion of light signals associated to alpha decays in the 54 g $SrF_2$ detector.

## Acknowledgments

This work has been partially supported by the French CNRS/INSU (MANOLIA and BOLERO projects), by the Spanish Commission for Science and Technology (MEC, Grant Nos. FPA2004-00974 and FPA2007-63777), by the Gobierno de Aragón (Group in Nuclear and Astroparticle Physics), by the EU Project ILIAS Contract No. RII3-CT-2004-506222. Y. Ortigoza was supported by a UZ/BSCH/Fundación Carolina Grant.

## References

1. L. Torres, PhD thesis, Univ. of Zaragoza (2008).
2. N. Coron et al., Proc. IDM'2008 (Identification of Dark Matter) Conference, Stockholm: PoS (idm2008) 007.
3. S. Cebrián et al., Phys. Lett. B 556 (2003) 14.
4. P. de Marcillac et al., Nature 422 (2003) 876 (together with Supplementary Information file: http://www.nature.com/nature/journal/v422/n6934/extref/nature01541-s1.pdf).
5. K. Fukuda et al., Proc. 9-th Workshop on Radiation Detectors and Their Uses, KEK Proceedings 95-7 (1995), p. 268.
6. Y. Satoh et al., Proc. 7-th Workshop on Radiation Detectors and Their Uses, KEK Proceedings 93-8 (1993), p. 186.

# R&D of tungstate and molibdate crystal scintillators to search for rare processes

**L.L. Nagornaya[1*], F.A. Danevich[2], A.M. Dubovik[1], B.V. Grinyov[1], H. Kraus[3], V.M. Kudovbenko[2], V. Mikhailik[3], S.S. Nagorny[2], D.V. Poda[2], O.G. Polischuk[2], I.A. Tupitsyna[1], Yu.Ya. Vostretsov[1]**

[1]*Institute for Scintillation Materials, 61001 Kharkiv, Ukraine*
[2]*Institute for Nuclear Research, MSP 03680 Kyiv, Ukraine*
[3]*Department of Physics, University of Oxford, Keble Road, Oxford OX1 3RH, UK*

The status of the R&D of tungstate and molybdate crystal scintillators $CdWO_4$, $ZnWO_4$, $PbWO_4$, $PbMoO_4$, $ZnMoO_4$, $MgWO_4$ is reviewed briefly. These scintillators are well suited for low count rate experiments, such as searches for double beta decay or dark matter.

## 1. Introduction

Scintillation materials, in particular tungstates and molybdates, are promising detectors for experiments to search for rare nuclear and sub-nuclear processes such as double beta decay, dark matter particles, or to investigate rare α- and β-decays. Cadmium tungstate crystal scintillators ($CdWO_4$ enriched in $^{116}Cd$ and also with natural isotopic abundance of cadmium) were successfully used in low count rate experiments to search for double beta decay processes in cadmium and tungsten [1], investigate the β-decay of $^{113}Cd$ [2], detect α activity of natural tungsten [3]. Calcium tungstate ($CaWO_4$) was further proposed as a detector for a 2β experiment with $^{48}Ca$ [4, 5]. It is also currently used by the CRESST experiment to search for dark matter particles [6, 7, 8]. $ZnWO_4$ crystal scintillators, studied for the first time as low-background detectors in [9], are operating now in the Laboratori Nazionali del Gran Sasso of INFN (Italy) as a detector to search for double beta processes in zinc and tungsten [10, 11]. Further investigations of $CdWO_4$ [12], $ZnWO_4$ and $CaMoO_4$ crystals [13, 14, 15] as scintillating bolometers for rare events experiments have been performed recently. A set of different scintillation materials is needed for the EURECA[1] cryogenic dark matter experiment, where a multi-element target is planned for the identification of a true dark matter signal. The improvement of well known scintillators ($PbWO_4$ and $PbMoO_4$), as well as the development of new materials ($ZnMoO_4$ and $MgWO_4$), having low levels of natural radioactive background and high light output, is of considerable interest for rare event search experiments.

Motivated by this, work has been started at the Institute for Scintillation Materials (ISMA, Kharkiv, Ukraine) in collaboration with the Institute for Nuclear Research (Kyiv, Ukraine) and the University of Oxford (UK), aiming to develop and optimize scintillating oxide crystals to search for dark matter and double beta decay. $CdWO_4$, $ZnWO_4$, $ZnMoO_4$, $PbWO_4$, $PbMoO_4$, crystals were grown from the melt on seeds, using the Czochralski method, in furnaces with induction heating using platinum or iridium crucibles. Raw material charges were obtained by high-temperature solid phase synthesis from metal oxides or by the co-precipitation method. Magnesium tungstate ($MgWO_4$) can not be grown using the conventional Czochralski method due to its high-temperature phase transition. A flux growth technology was developed to obtain single crystalline samples of $MgWO_4$.

* Corresponding author. *E-mail address*: nagorna@isc.kharkov.com
[1] European Underground Rare Event Calorimeter Array; www.eureca.ox.ac.uk

## 2. Results and discussion

### 2.1. $CdWO_4$ and $ZnWO_4$

Large volume radiopure $CdWO_4$ single crystals with good scintillation properties were developed at the ISMA [16] (see Fig. 1, left). Furthermore, enriched in $^{116}Cd$, $^{116}CdWO_4$ crystal scintillators were grown at the ISMA for the first time. These scintillators were successfully used in one of the most sensitive 2β decay experiment carried out in the Solotvina Underground Laboratory of the Institute for Nuclear Research (Kyiv) (see Ref. [1] and references therein).

The studies, reported in [9] and [17], have stimulated extensive research aimed to develop large-volume low-background $ZnWO_4$ crystal scintillators with good scintillation properties. First results of this R&D were published in [18]. Alongside the optimization of the process for growing large-volume $ZnWO_4$, extensive studies of the crystal quality, stoichiometry of raw material compositions, and type and concentration of dopants were performed. The scintillation properties of $ZnWO_4$ crystal scintillators are presented in Table 1. Some of the univalent dopants can substantially improve scintillation characteristics. Practically colorless $ZnWO_4$ single crystals of up to 5 cm in diameter and 10 cm length have been developed (Fig. 1, right).

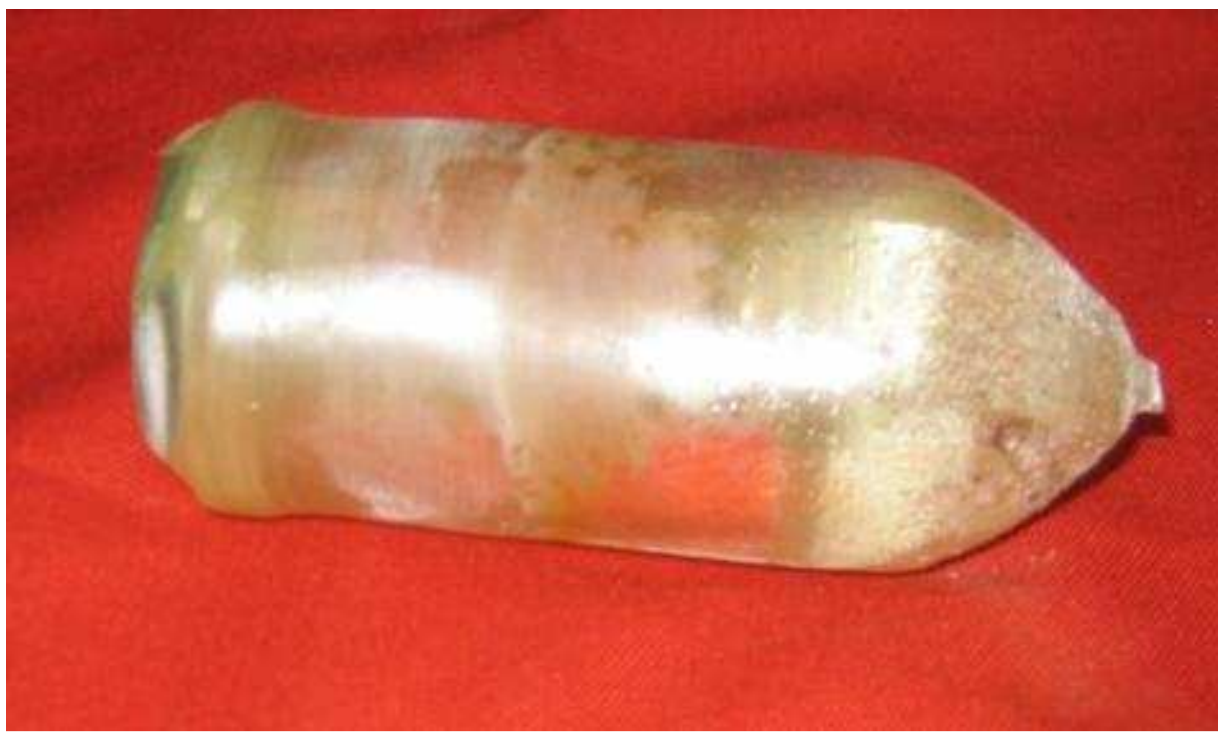

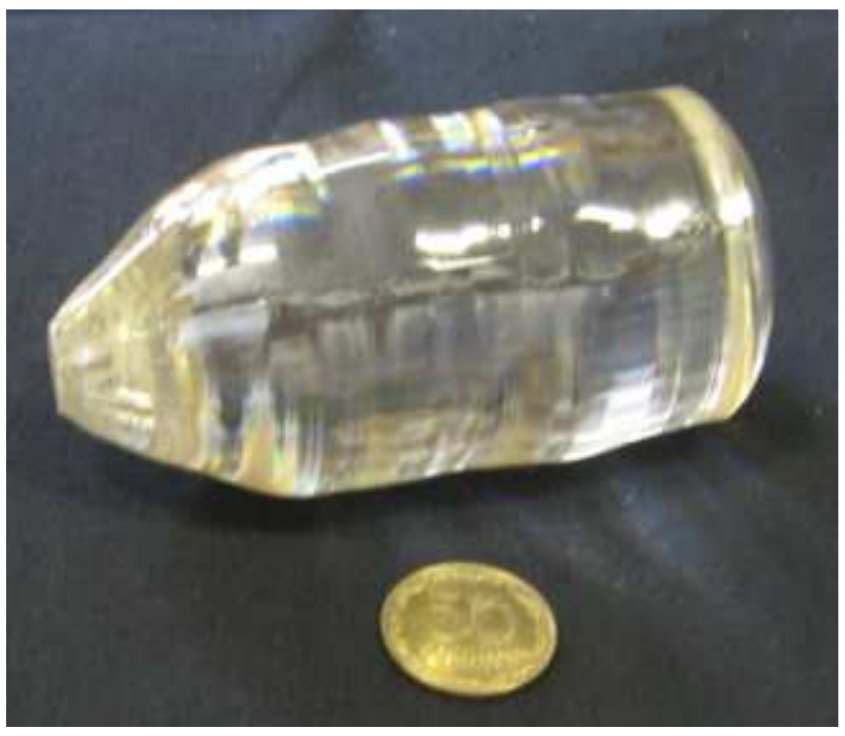

Fig. 1. $CdWO_4$ (mass 3 kg, left) and $ZnWO_4$ (≈1.3 kg, right) single crystals.

Table 1: Scintillation properties of $ZnWO_4$ crystals.

| # | Dopant | Size of samples, mm | LY [a], % $CdWO_4$ | FWHM, % at 662 keV | Afterglow, % (20 ms) |
|---|---|---|---|---|---|
| 1 | – | $10 \times 10 \times 10$ | 11 | 23 | 0.79 |
| 2 | MeF [b] | $10 \times 10 \times 10$ | 32 | 11 | 0.104 |
| 3 | $ZnF_2$, $MeO_2$ | $10 \times 10 \times 10$ | 47 | 10.2 | 0.005 |
| 4 | $MeO_2$ | $10 \times 10 \times 10$<br>$30 \times 30 \times 14$ | 47.5<br>39 | 9.3<br>11 | |
| 5 | $MeO_2$, $ZnF_2$ | $10 \times 10 \times 10$ | 50 | 8.5 | 0.002 |
| 6 | $MeO_2$ | ⬡ $40 \times 40$ | 27 | 10.7 | |
| 7 | $MeO_2$ | $\varnothing 44 \times 55$ | 15 | 13.7 | |

[a] The light output of the $ZnWO_4$ samples was measured relatively to that of a $CdWO_4$ sample of dimensions $10 \times 10 \times 10$ mm$^3$.
[b] Me is a metal.

The thermally stimulated luminescence (TSL) measured with $ZnWO_4$ samples are shown in Fig. 2. The behaviour of TSL demonstrates that there is a correlation between the afterglow and the amplitude of the peaks at $T > 233$ K. It is assumed that the traps associated with these peaks are responsible for the accumulation of charge carriers at room temperature, and this accounts for the observed slow decay process. Doping creates shallow trapping centres with activation energy so low that the trapping of carriers does not occur at room temperature. It causes a noticeable improvement of the afterglow characteristics of the $ZnWO_4$ crystals: afterglow reduces from 0.79% for the undoped crystal #1 to 0.005% for the co-doped sample #3. Taking into account these results, the process of crystal growth has been optimized and large-volume $ZnWO_4$ samples with improved scintillation properties were produced [19]. Fig. 3 shows the pulse amplitude spectra measured for the hexagonal (40 × 40 mm) $ZnWO_4$ scintillator. The energy resolution for the 662 keV γ line of $^{137}Cs$ was found to be 10.7%. It is worthwhile noting that this is the first time such a good energy resolution has been obtained for a large-volume $ZnWO_4$ (a few tens of $cm^3$) crystal scintillator.

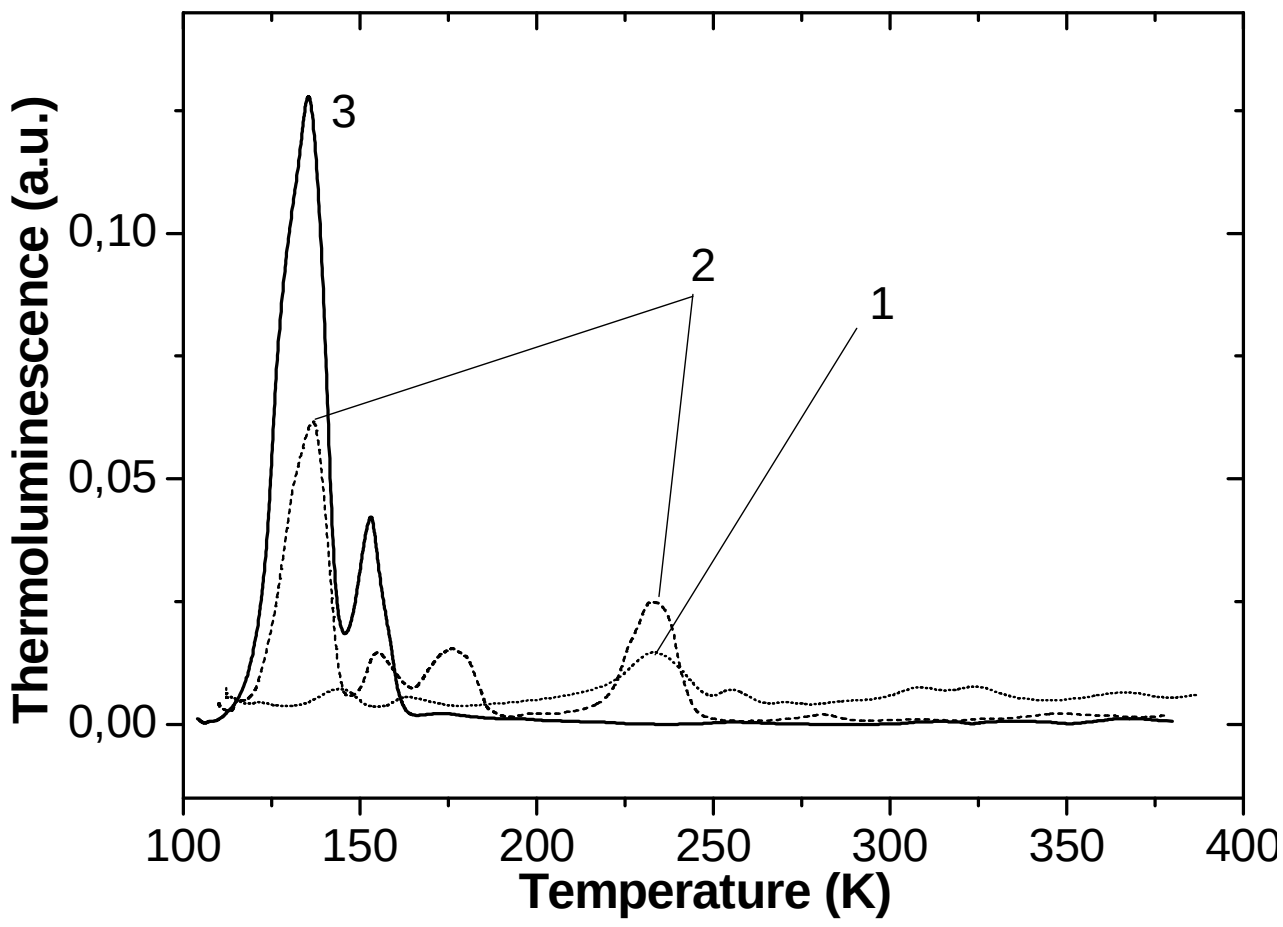

Fig. 2. Thermally stimulated luminescence of $ZnWO_4$ crystals. The numbers on the graphs correspond to the numbers in Table 1.

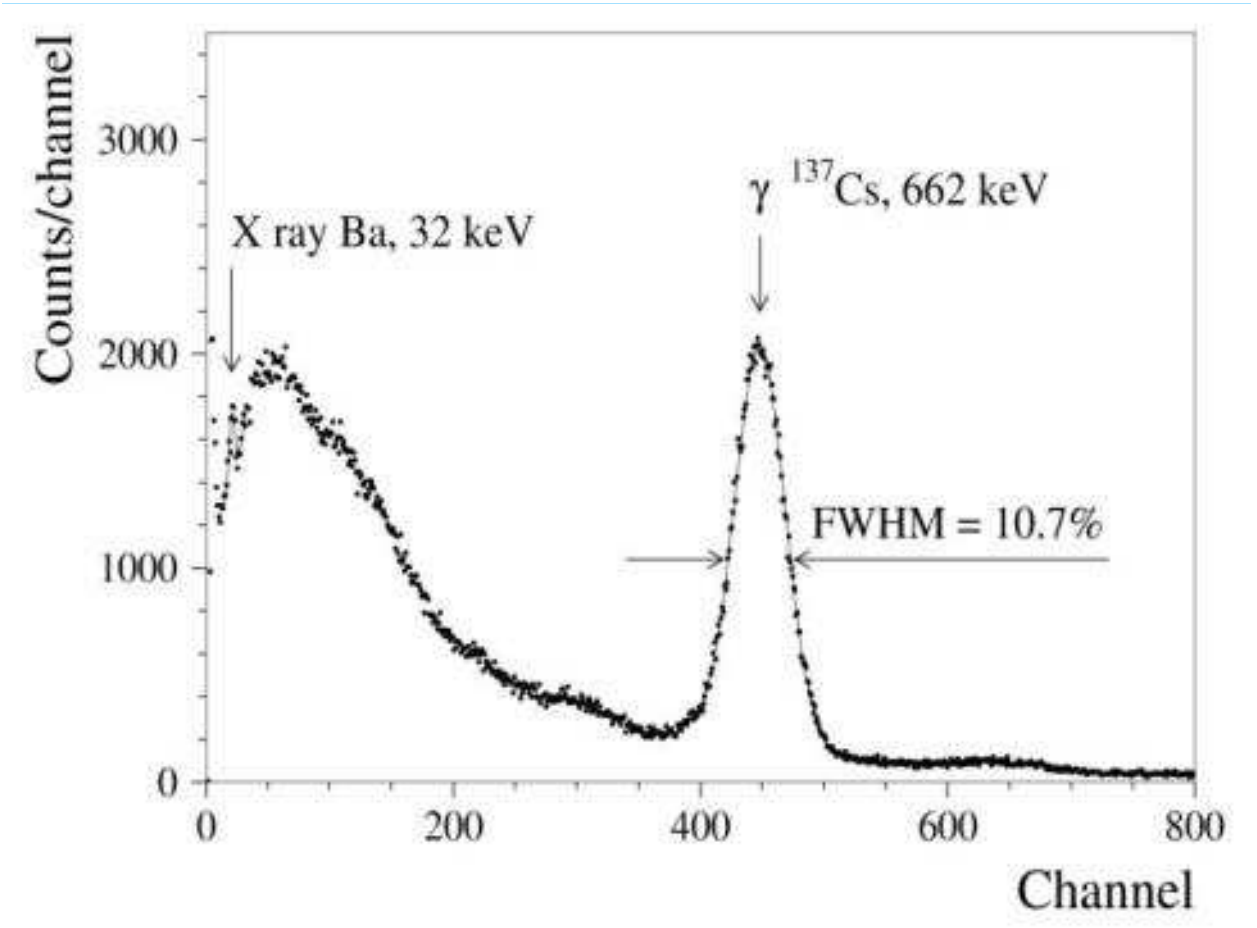

Fig. 3: Energy spectra measured for a $ZnWO_4$ detector with γ quanta of $^{137}Cs$.

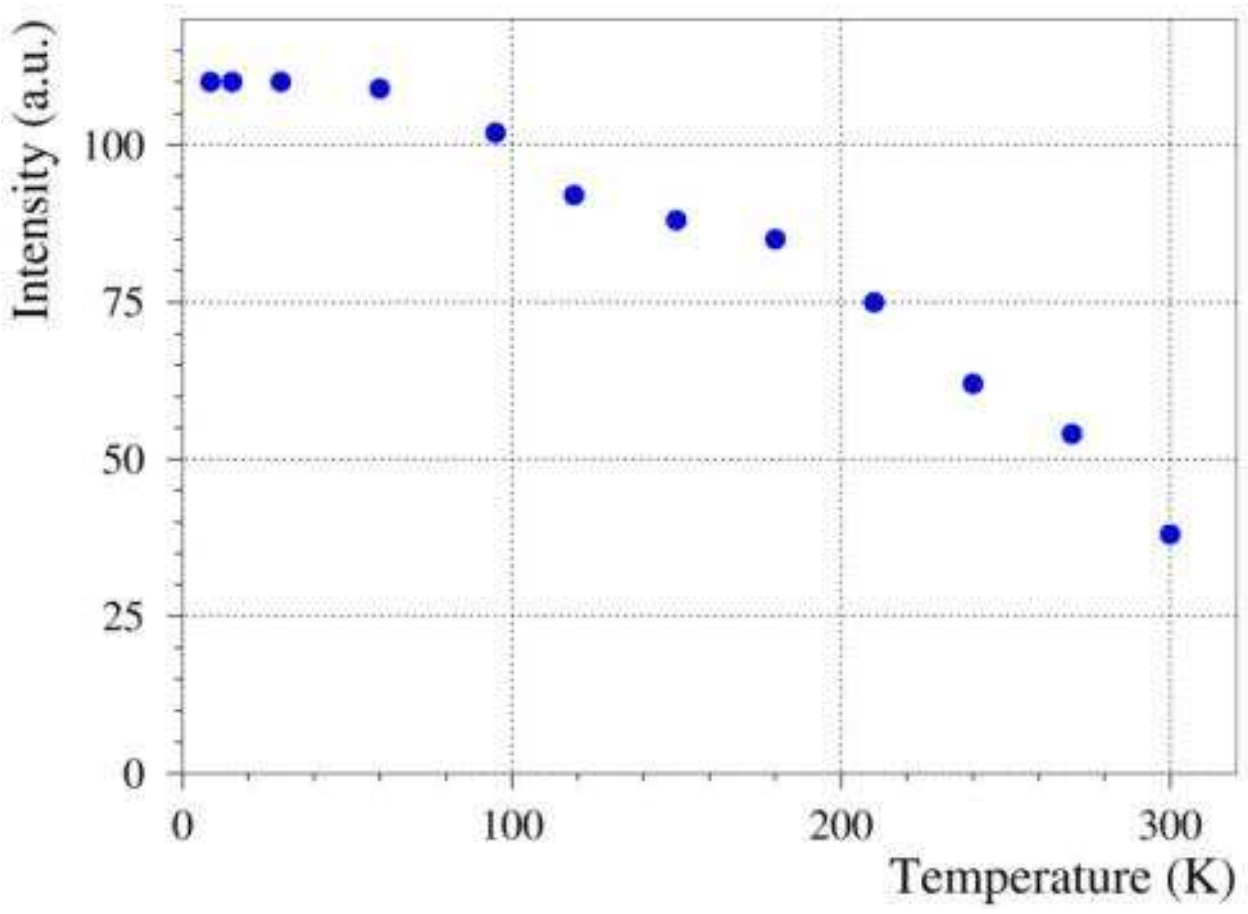

Fig. 4: Temperature dependence of the light output of the $ZnWO_4$ crystal scintillator for excitation by $^{241}Am$ α particles.

Given the strong interest in the application of $ZnWO_4$ as a cryogenic scintillation detector, we studied the light output and the decay time constants of the crystal as a function of temperature in the temperature range 7–300 K. It is shown that the light output of the crystal increases with cooling by ~60 % (Fig. 5). The relative light output of $ZnWO_4$ at 10 K is ca. 110–115% that of $CaWO_4$. This is consistent with earlier estimates reported in Ref. [20].

The temperature dependence of the decay time constants of $ZnWO_4$ is displayed in Fig. 5. The pulse shape of the $ZnWO_4$ scintillation signal can be fitted using a sum of three exponential functions with decay time constants: $\tau_1 \approx 1$ μs, $\tau_2 \approx 4$ μs and $\tau_3 \approx 25$ μs (T = 295 K), respectively. The decay time constants gradually increase with reducing temperature down to ~20 K; below this they increase significantly as temperature is lowered further.

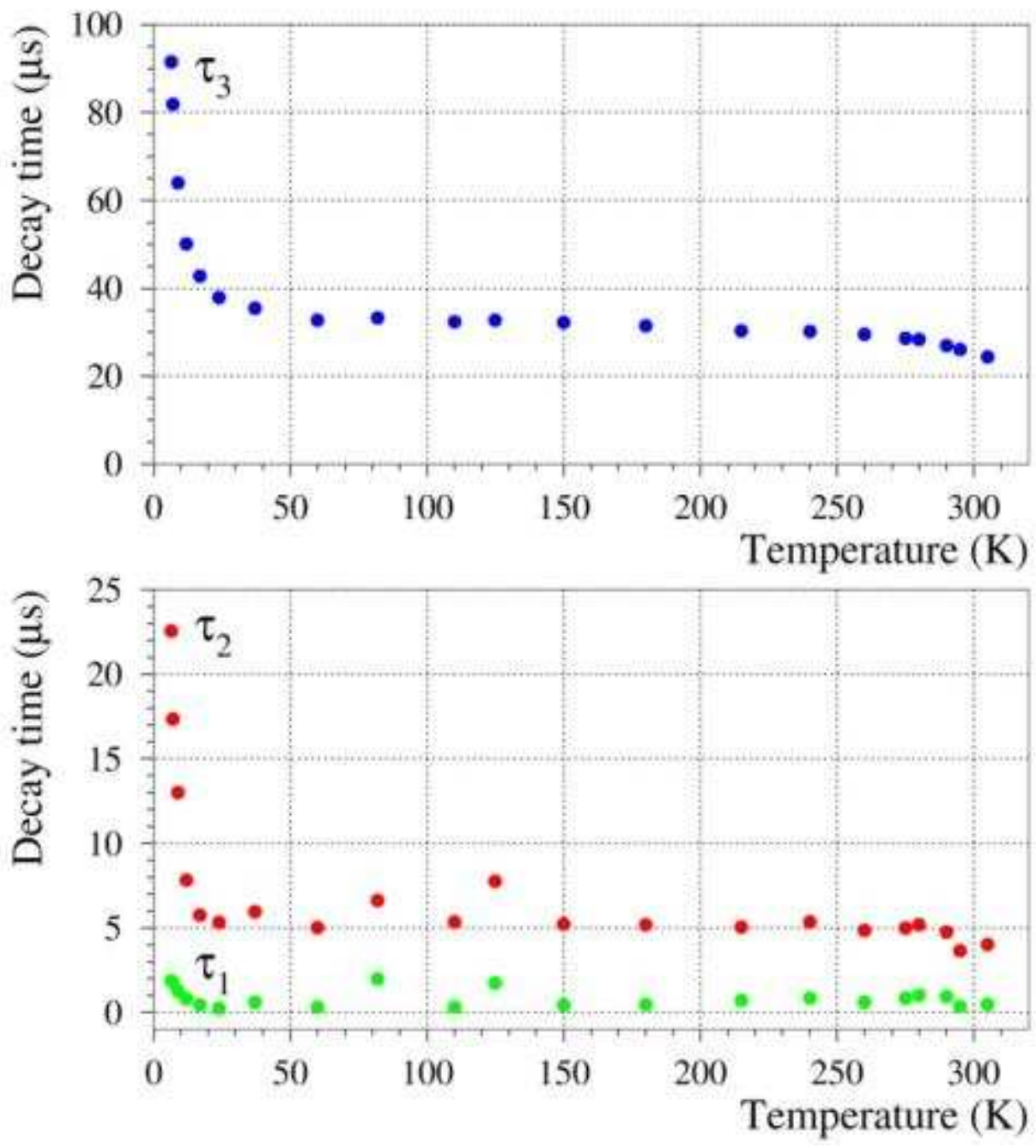

Fig. 5: Temperature dependence of the decay time constants for irradiation with $^{60}$Co γ quanta.

Measurements of the radioactive contaminations of the $ZnWO_4$ sample with dimensions 26×24×24 mm$^3$ and mass 119 g were carried out at the Solotvina Underground Laboratory (Ukraine). The results of these measurements are presented in Table 2.

Table 2: The radioactive contamination of $ZnWO_4$ crystal scintillators.

| Chain | Source | Activity (mBq/kg) | |
|---|---|---|---|
| | | $ZnWO_4$ | $ZnWO_4$ [9] |
| $^{232}$Th | $^{228}$Th | ≤ 0.1 | ≤ 0.2 |
| $^{238}$U | $^{226}$Ra | ≤ 0.16 | ≤ 0.4 |
| Total α activity | | = 2.4(3) | ≤ 20 |
| | $^{40}$K<br>$^{90}$Sr<br>$^{137}$Cs<br>$^{147}$Sm | ≤ 14<br>≤ 15<br>≤ 2.5<br>≤ 5 | ≤ 12<br>≤ 1.2<br>≤ 20<br>≤ 1.8 |

Taking into account the good scintillation properties and the low level of radioactive contamination, one can conclude that $ZnWO_4$ is one of the most promising scintillation materials for 2β decay and dark matter search experiments.

### 2.2. $PbWO_4$ and $PbMoO_4$

$PbWO_4$ crystal scintillators have been developed at the ISMA as a high-registration-efficiency detectors for high energy physics [21]. The technology for crystal production is now well established and the capability for growing large-volume high-quality $PbWO_4$ crystals exists. This is also applicable to $PbMoO_4$ which has been used for decades as optoelectronic material.

The feasibility of using lead tungstate and lead molybdate to search for rare processes has been discussed in Refs. [22, 23, 24]. It has been shown that $PbWO_4$ can be used as a 4π active shield and light guide for the $^{116}CdWO_4$ detector [23]. However, the high intrinsic radioactivity of these scintillators (~10$^2$ Bg/kg) due to $^{210}$Pb is a major limitation of these materials for their application in low-background experiments. The contribution of radioactive $^{210}$Pb can be decreased substantially by using ancient archaeological lead, where the activity of this radionuclide is typically on the level of a few mBq/kg [25]. As a next step we are going to grow $PbMoO_4$ and $PbWO_4$ crystal scintillators from ancient lead discovered in the Ukraine [26]. The activity of $^{210}Pb$ in this lead does not exceed ≈ 1 mBq/kg.

### 2.3. $ZnMoO_4$

$ZnMoO_4$ has been identified a couple of years ago as a suitable scintillation material for cryogenic rare event searches [27]. Recently, and for the first time, comparatively large $ZnMoO_4$ single crystals were grown [28]. Possible applications of this crystal include studies of the double beta-decay of $^{100}Mo$ and a search for dark matter using a multi-target detector composed of complementary $AMO_4$ scintillators (A=Ca, Zn; M=Mo, W). This motivated the development of our technological process for the preparation of the crystal, and we produced several ingots of diameter 20–30 mm and height 30–40 mm. As is seen in Fig. 6, the crystal shows a pronounced yellow colouration; its absorption spectrum exhibits an abrupt rise at ~380 nm and a band at 460 nm (see Fig. 7). This might be an indication of a high level of impurities or defects, in line with what is known of the nature of colouration of $ZnWO_4$ crystals [20]. As a next step we intend to grow $ZnMoO_4$ crystal scintillators by using high-purity, grade (5N) raw materials.

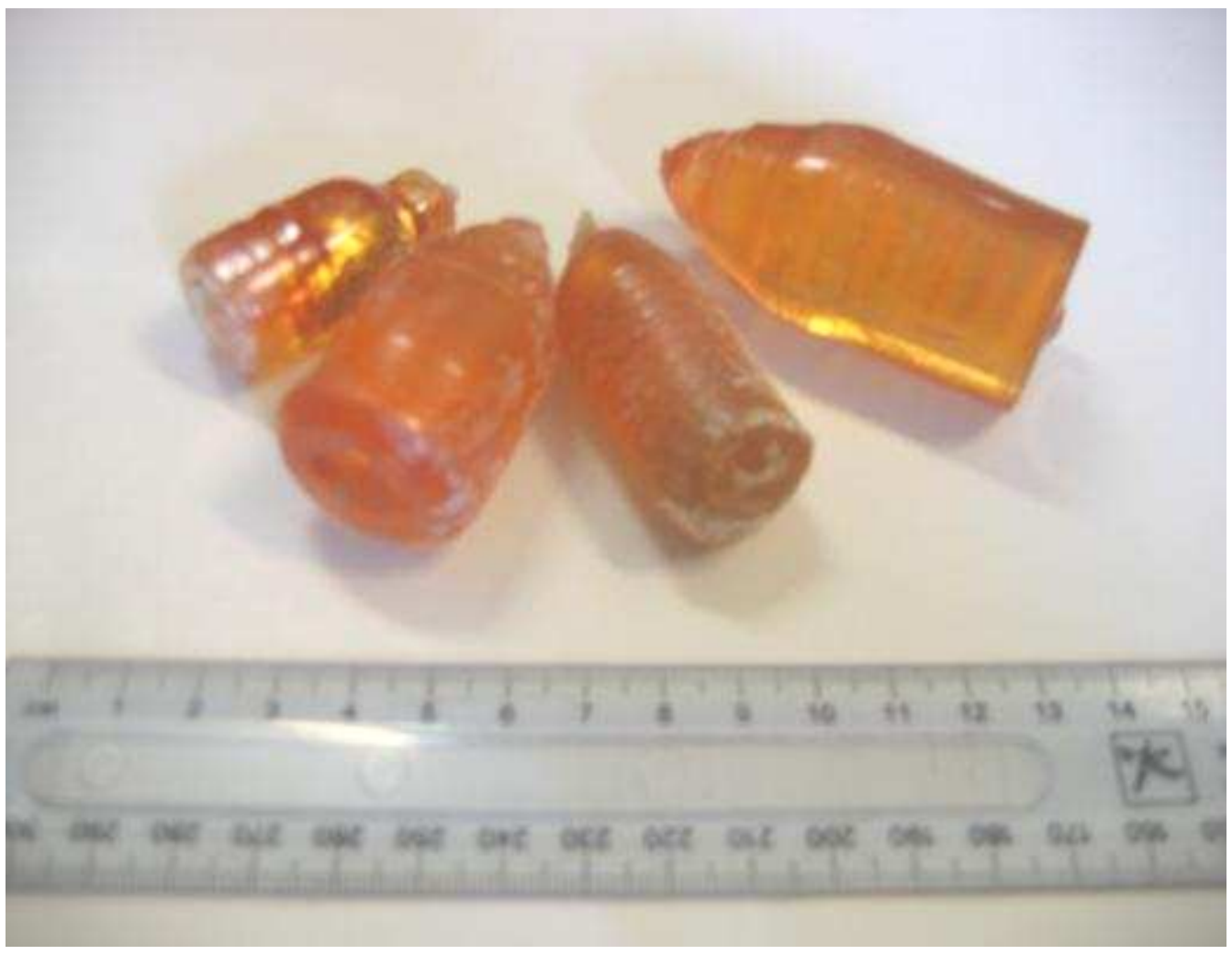

Fig. 6. $ZnMoO_4$ single crystals.

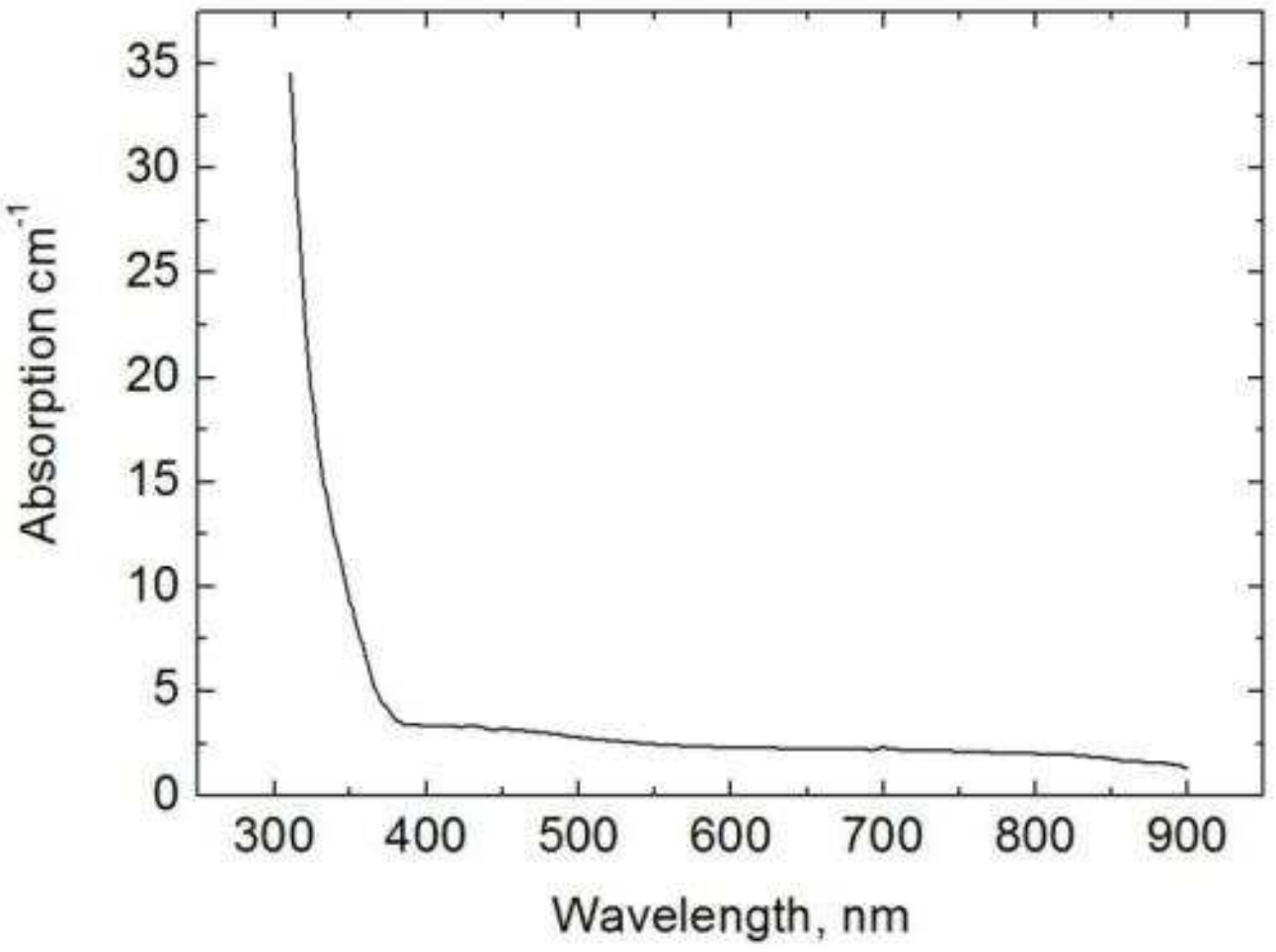

Fig. 7: Absorption spectrum of $ZnMoO_4$ crystal at room temperature.

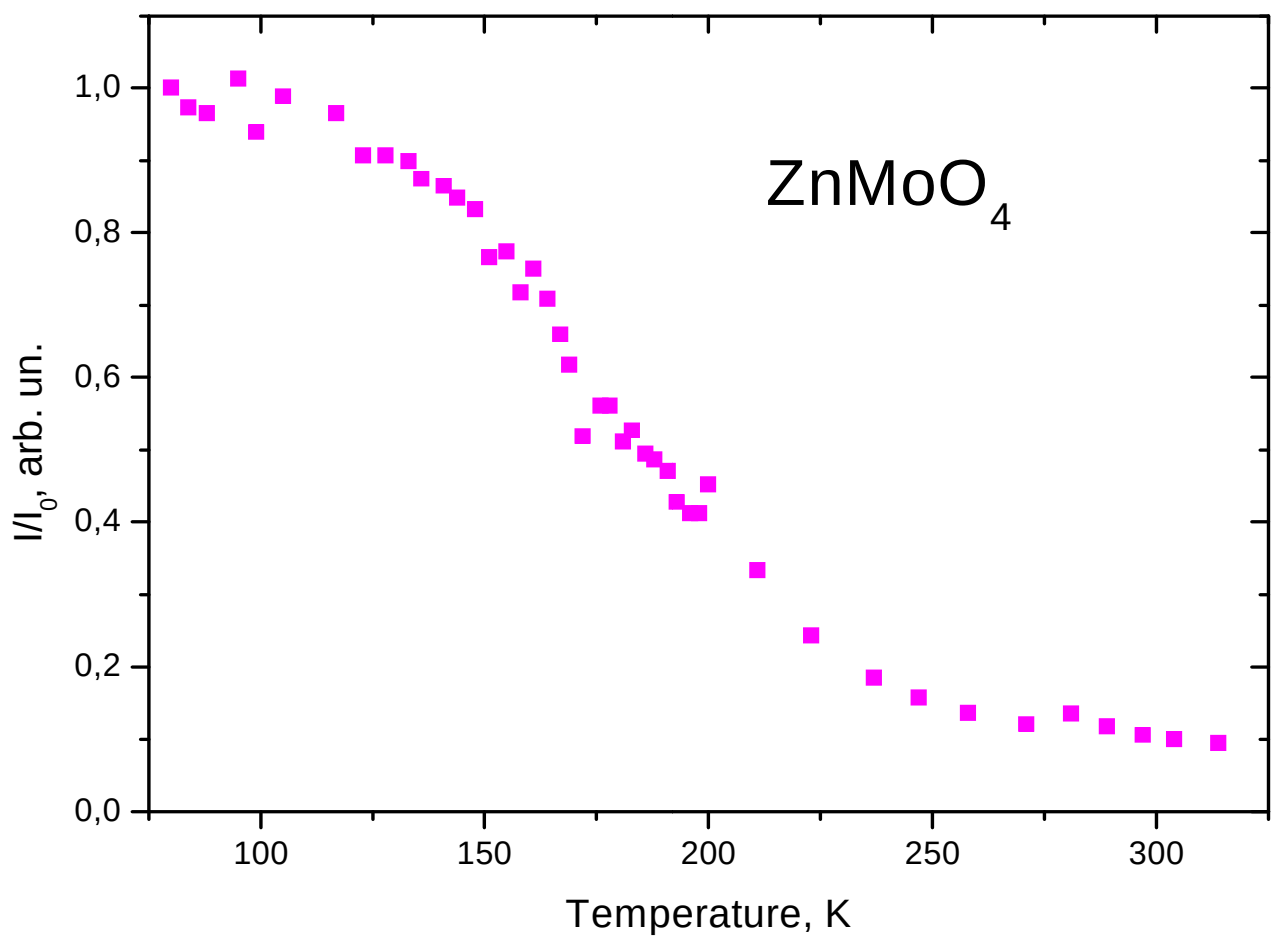

Fig. 8. Temperature dependence of X-ray luminescence intensity of $ZnMoO_4$.

The X-ray luminescence intensity of a $ZnMoO_4$ crystal was measured for various temperatures between liquid nitrogen and room temperature. The luminescence intensity increased by an order of magnitude for cooling to lower temperatures (Fig. 8).

### 2.4. $MgWO_4$

Recent studies of powder $MgWO_4$ samples demonstrated that this compound is an attractive scintillation material for cryogenic applications because of its high scintillation light output which is comparable to that of $ZnWO_4$ [29]. Due to its high-temperature phase transition, $MgWO_4$ cannot be grown by the conventional Czochralski method. Therefore, technology for flux growth of single crystalline samples of $MgWO_4$ has been developed. The crystal has intense luminescence under X-ray excitation (see Fig. 9). The broad emission band exhibits a maximum at 475 nm, agreeing well with the room temperature data obtained for a powder sample [29]. The pulse amplitude spectrum of $MgWO_4$ excited by $\gamma$ quanta of $^{241}$Am is shown in Fig. 10. An energy resolution of R = 37% for the 59.5 keV $\gamma$ line of $^{241}$Am was measured with this sample. This value is very close to the R = 35% obtained with a large $ZnWO_4$ scintillator [14]. An energy resolution of R = 15% was measured for the 662 keV $\gamma$ line of $^{137}$Cs.

The results obtained confirm the good prospect of magnesium tungstate for scintillation applications. Therefore, development of production technology for large $MgWO_4$ scintillators is now in progress.

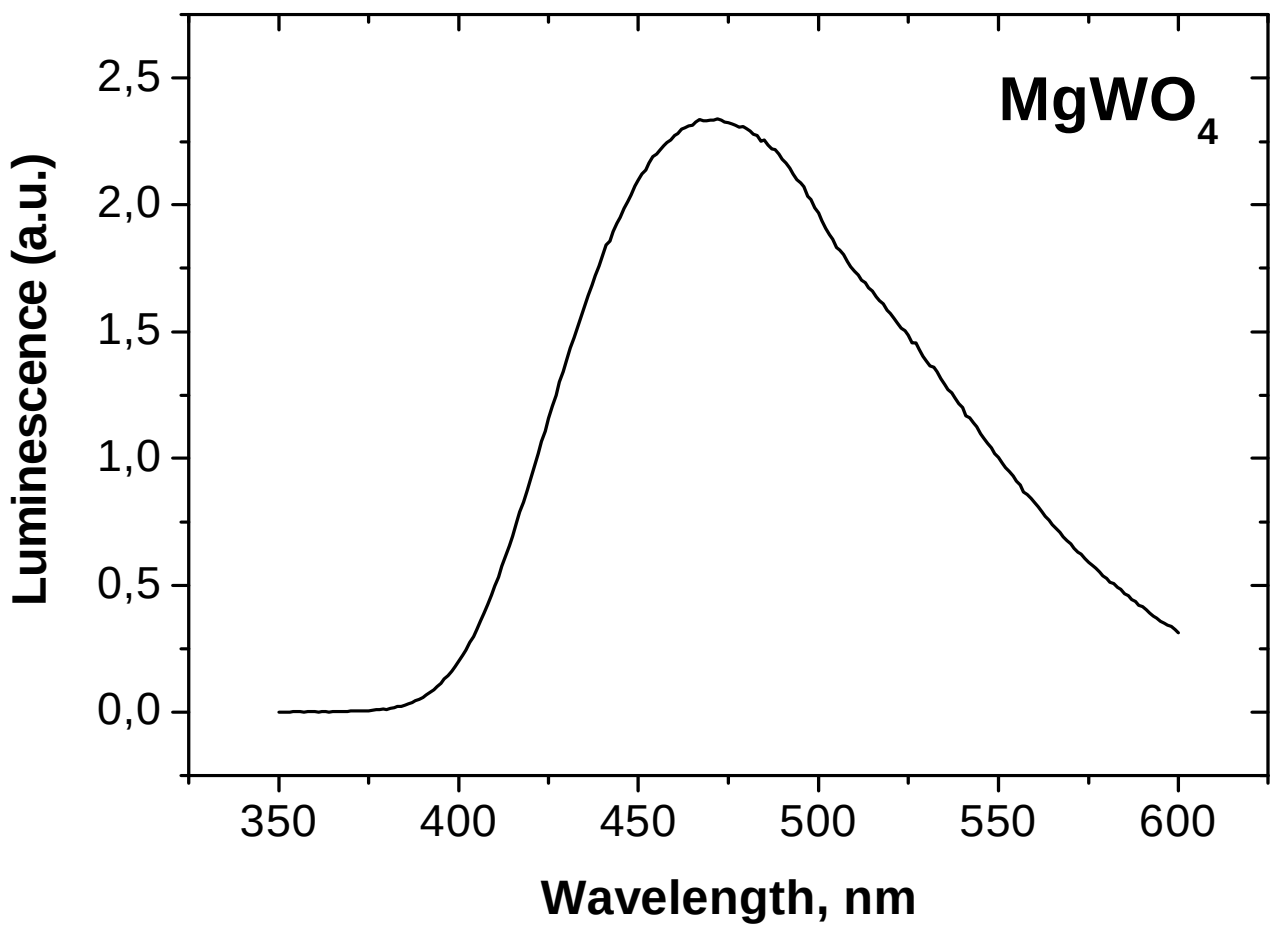

Fig. 9. X ray luminescence of $MgWO_4$ crystal.

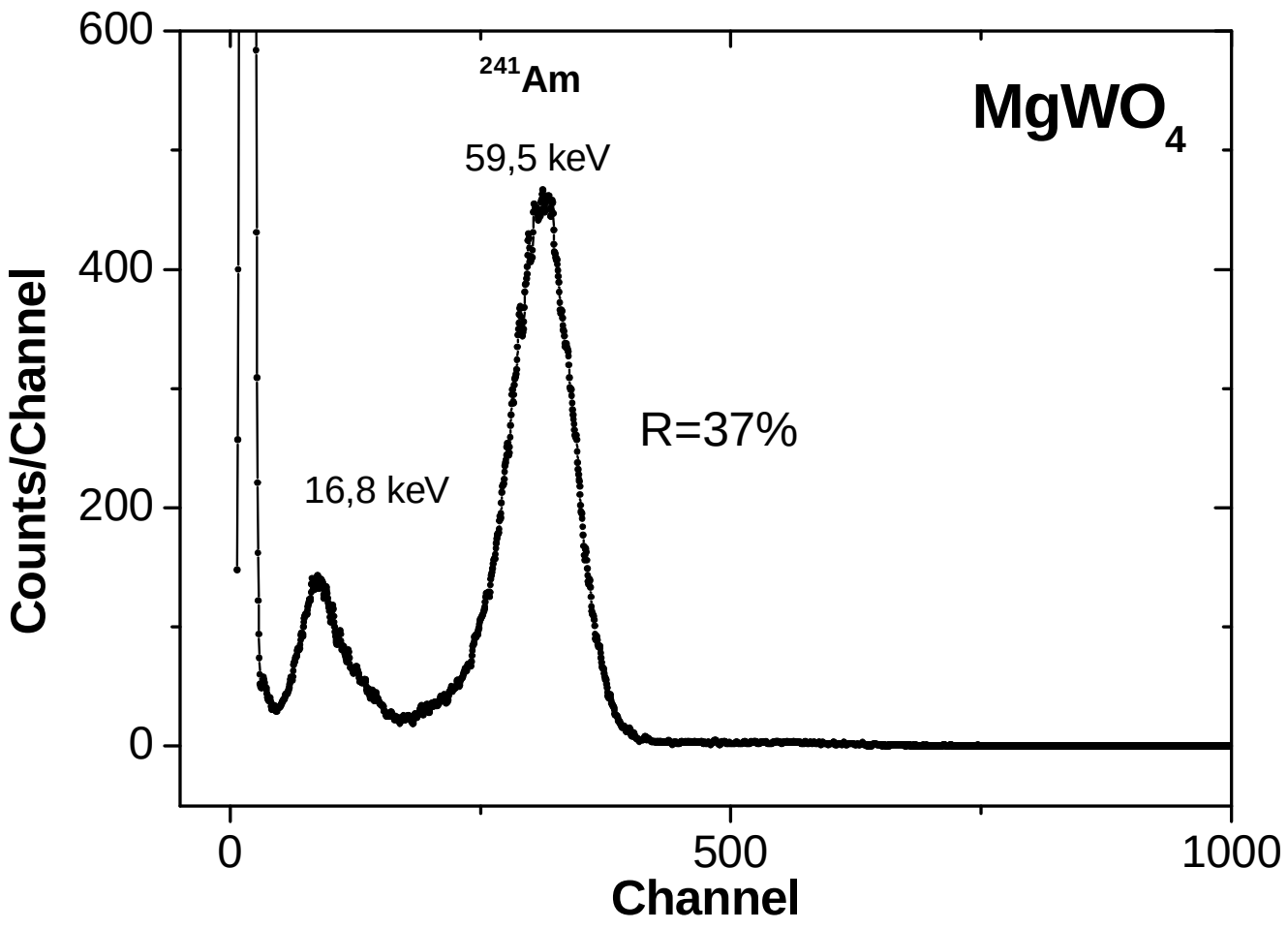

Fig. 10. Energy spectra measured by $MgWO_4$ detector with 59.5 keV $\gamma$ quanta of $^{241}$Am.

## 3. Conclusions

In this paper we reviewed the results of our efforts directed at the development of oxide scintillators for rare event search experiments. We already succeeded in producing large-volume $CdWO_4$ and $ZnWO_4$ scintillators with improved performance characteristics. Good scintillation properties at low temperatures and exceptionally low intrinsic radioactivity make zinc tungstate an excellent material for cryogenic double beta decay and dark matter experiments. Single crystal samples of $ZnMoO_4$ were produced for the first time using the Czochralski technique. We investigated the feasibility of this material for cryogenic rare event search experiments and identified ways to improve the scintillation properties of the crystal. The techniques necessary to grow $MgWO_4$ crystals of ~1 cm$^3$ volume were developed and luminescence and scintillation characteristics of this material were measured. Given the good prospects for reducing the intrinsic radioactivity of lead-based crystals we studied the temperature dependence of scintillation characteristics of $PbWO_4$ and $PbMoO_4$. We demonstrated that these crystals may also be used as cryogenic scintillators. We are planning to produce lead tungstate and molybdate with substantially lower intrinsic radioactivity using ancient lead.

## References

1. F.A. Danevich et al., Phys. Rev. C 68 (2003) 035501.
2. P. Belli et al., Phys. Rev. C 76 (2007) 064603.
3. F.A. Danevich et al., Phys. Rev. C 67 (2003) 014310.
4. Yu.G. Zdesenko et al., Nucl. Instr. Meth. A 538 (2005) 657.
5. Yu.G. Zdesenko et al., Astropart. Phys. 23 (2005) 249.
6. J. Ninković et al., Nucl. Instr. Meth. A 537 (2005) 339.
7. G. Angloher et al., Astropart. Phys. 23 (2005) 325.
8. G. Angloher et al., arXiv:0809.1829v1 [astro-ph]; submitted to Astropart. Phys.
9. F.A. Danevich et al., Nucl. Instr. Meth. A 544 (2005) 553.
10. P. Belli et al., Phys. Lett. B 658 (2008) 193.
11. P. Belli, et al., preprint ROM2F/2008/22; arXiv: 0811.2348v1 [nucl-ex]; submitted to Phys. Rev. C.
12. L. Gironi et al., $CdWO_4$ bolometers for double beta decay search, Opt. Mat., in press.
13. I. Bavykina et al., IEEE Trans. Nucl. Sci. 55 (2008) 1449.
14. H. Kraus et al., Nucl. Instr. Meth. A 600 (2009) 594.
15. I. Bavykina et al., Development of cryogenic phonon detectors based on $CaMoO_4$ and $ZnWO_4$ scintillating crystals for direct dark matter search experiments, Opt. Mat., in press.
16. S.Ph. Burachas al., Nucl. Instr. Meth. A 369 (1996) 164.
17. V.B. Mikhailik and H. Kraus, J. Phys. D: Appl. Phys. 39 (2006) 1181.
18. L.L. Nagornaya et al., IEEE Trans. Nucl. Sci. 55 (2008) 1469.
19. L.L. Nagornaya et al., Large volume $ZnWO_4$ crystal scintillators with excellent energy resolution and low background, IEEE Trans. Nucl. Sci., to be published.
20. H. Kraus et al., Phys. Lett. B 610 (2005) 37.
21. L. Nagornaya, V. Ryzhikov, Proc. of the "Crystal 2000" Int. Workshop, Chamonix, France, pp. 367-374, 1992.
22. M. Minowa et al., Nucl. Instr. Meth. A 322 (1992) 500.
23. F.A. Danevich et al., Nucl. Instr. Meth. A 556 (2006) 259.
24. Yu.G. Zdesenko et al., Instr. Exp. Technique 39 (1996) 364.
25. A. Alessandrello et al., Nucl. Instr. Meth. B 142 (1998) 163.
26. F.A. Danevich et al., Archaeological lead findings in Ukraine, AIP Conf. Proc. 897 (2007) 125; Nucl. Instr. Meth. A, in press.
27. V.B. Mikhailik et al., Nucl. Instr. Meth. A 562 (2006) 513.
28. L.I. Ivleva et al., Crystallography Reports 53 (2008) 1087.
29. V.B. Mikhailik et al., J. Phys. Cond. Matt. 20 (2008) 365219.

# Radioactive contamination of crystal scintillators

**F.A. Danevich**[*]

*Institute for Nuclear Research, MSP 03680 Kyiv, Ukraine*

Radioactive contamination of crystal scintillators, its origin and nature, experimental methods to measure, and data for several crystal scintillators are discussed.

## 1. Introduction

Experiments to search for rare processes (low-background α-, β-, γ-spectrometry, double β decay and dark matter particles search, measurements of neutrino fluxes from different sources, search for hypothetical nuclear and particle processes) require low level, the best case zero, background of a detector. Crystal scintillators are used to search for rare events both as conventional scintillation detectors and as cryogenic scintillating bolometers. Radioactive contamination of crystal scintillators plays a key role to reach low level of background. Origin and nature of radioactive contamination of crystal scintillators, experimental methods and data for several scintillation materials are briefly reviewed.

## 2. Radioactive contamination of scintillators: origin and nature

The main sources of radioactive contamination of scintillation materials are naturally occurring radionuclides of $^{232}$Th, $^{238}$U, and $^{235}$U families, and $^{40}$K. It should be stressed the secular equilibrium of U/Th chains is broken in scintillation materials as usual. Alpha active $^{147}$Sm was detected in some scintillators on a mBq/kg level. The next important group of radioactive nuclides are antropogenic $^{60}$Co, $^{90}$Sr-$^{90}$Y, $^{137}$Cs. Some scintillation crystals consist of elements having radioactive isotopes, like f.e. $^{152}$Gd in GSO, $^{113}$Cd in $CdWO_4$, $^{138}$La in $LaCl_3$ and $LaBr_3$, $^{176}$Lu in $Lu_2SiO_5$ and $LuI_3$. Cosmogenic radionuclides, i.e. created by high energy cosmic rays or/and by neutrons, were observed in some scintillation materials: $^{14}$C in liquid scintillator [1], $^{65}$Zn in $ZnWO_4$ [2], $^{152}$Eu in $CaF_2$(Eu) [3], $^{113m}$Cd in $CdWO_4$ crystal scintillators [4, 5, 6]. Origin of $^{207}$Bi in BGO is still not clear and seems to be of more complicated nature [7, 8, 9, 10].

## 3. Experimental methods

### 3.1. Low-background measurements

The highest sensitivity to measure internal contamination of crystal scintillators can be achieved in low background measurements where a scintillator is operating as a detector. Such an approach provides high efficiency of registration, especially for α and β particles. A typical low background scintillation set-up (see, for instance, [11, 12, 13, 14, 15]) consists of scintillator, light-guide to shield the scintillator from radioactivity of photomultipliers (PMT), which typically are the most contaminated details of a low background scintillation set-up, passive shield. Background of a detector can be further suppressed by using of active shield detectors surrounding a main detector, and anti-muon veto counters. Light-guides made of a scintillation material with different (relative to a main scintillation detector) scintillation decay time can serve as active anticoincidence detectors [11]. Continuous flushing of internal volume of a set-up by a radon-free gas (typically by nitrogen) allows to protect a detector from radon [12]. It is worth mentioning, if data acquisition can record amplitude, time of arrival and pulse-shape of scintillation signals, this information helps to interpret and suppress background of a scintillation detector.

[*] Corresponding author. *E-mail address:* danevich@kinr.kiev.ua

Ultra-low background HP Ge γ detectors can also be used to measure radioactive contamination of scintillation crystals (see, for instance [16, 17]. This method provides typical sensitivity at the level of mBq/kg for $^{40}$K, $^{228}$Th ($^{232}$Th), $^{226}$Ra ($^{238}$U) and $^{227}$Ac ($^{235}$U), and somewhat lower sensitivity to other parts of the U/Th chains. This method is useless to detect internal contamination by α active nuclides, and practically not sensitive to β active isotopes[1].

Long living radioactive isotopes can be also measured with the help of Inductively Coupled Plasma Mass Spectroscopy (ICP-MS). Sensitivity of this method depends on measured matrix, quality and previous using of an apparatus. For instance the sensitivity of the mass-spectrometer (Agilent Technologies model 7500a) installed in the Laboratori Nazionali del Gran Sasso of I.N.F.N. (Italy) is at the level of ~ ppb for U (which corresponds to activity 12 mBq/kg), Th (4 mBq/kg), Rb (0.9 mBq/kg of $^{87}$Rb), and Sm (0.13 mBq/kg of $^{147}$Sm), ~ ppm for K (activity of $^{40}$K: 30 mBq/kg)[2]. Unfortunately ICP-MS is practically useless to measure $^{226}$Ra and $^{228}$Th contamination (the most dangerous radionuclides for double beta decay experiments), as well as $^{210}$Pb (critical for dark matter search) due to rather low half-life of these isotopes.

### 3.2. Response of detector to γ quanta and α particles

Knowledge of a detector response to γ quanta (response function, dependence of energy resolution on energy) and α particles (energy dependence of α/β ratio[3] and energy resolution) is necessary to interpret background of the detector. Response function and dependence of energy resolution on energy of γ quanta can be measured with γ sources in wide energy interval from a few keV (5.9 keV Mn K x rays from $^{55}$Fe) up to 2.615 MeV (γ line of $^{208}$Tl). Calibration with α sources is much more complicated task because energies of commonly used α sources lie in the energy region from 5.3 to 8.8 MeV ($^{228}$Th, $^{241}$Am, $^{244}$Cm, $^{252}$Cf). To calibrate a detector at lower energies an α source with absorbers can be used (see, for instance [3, 18, 19, 20, 21, 22]). Response of scintillation detectors to α particles is non linear (see Fig. 1, where the α/β ratio measured with $CaWO_4$ crystal scintillator is presented). Alpha peaks from internal contamination of scintillators allow to extend interval of α particles energies. In addition α peaks from internal α active decays provide important test of calibration measurements with external α sources. In some scintillation crystals with anisotropic structure, α/β ratio depends on direction of α particles relatively to crystal axes. Such an effect was observed in $CdWO_4$ [18] and $ZnWO_4$ [20] crystal scintillators (see Fig. 2 where dependence of α/β ratio on direction of α irradiation relatively to crystal axes of $CdWO_4$ scintillator is presented). It leads to some worsening of energy resolution of these detectors to α particles [2, 18].

### 3.3. Time-amplitude analysis

The energy and arrival time of each event can be used to select fast decay chains from the $^{232}$Th, $^{235}$U and $^{238}$U families. The method of time-amplitude analysis is described in detail in [23, 24]. For instance, the following sequence of α decays from the $^{232}$Th family can be selected: $^{224}$Ra ($Q_\alpha$ = 5.79 MeV; $T_{1/2}$ = 3.66 d) → $^{220}$Rn ($Q_\alpha$ = 6.41 MeV; $T_{1/2}$ = 55.6 s) → $^{216}$Po ($Q_\alpha$ = 6.91 MeV; $T_{1/2}$ = 0.145 s) → $^{212}$Pb (which are in equilibrium with $^{228}$Th). As an example, the results of the time-amplitude analysis of data accumulated in the low-background experiment to search for 2β decay of $^{116}$Cd with the help of $^{116}CdWO_4$ crystal scintillators are shown in Fig. 3. The obtained α peaks (the α nature of events was confirmed by the pulse-shape analysis described below in subsection 3.4) as well as the distributions of the time intervals between events, are in a good agreement with those expected for the α decays of the $^{224}$Ra → $^{220}$Rn → $^{216}$Po → $^{212}$Pb chain [25].

---

[1] Beta activity can be detected by HP Ge detectors via registration of bremsstrahlung, however the sensitivity in this case is much lower due to low efficiency and absence of a clear signature (like peaks in γ spectra).

[2] Now the Laboratory intends to purchase a higher resolution, two orders of magnitude more sensitive device.

[3] The ''α/β ratio'' is defined as ratio of α peak position measured in the γ energy scale to the energy of α particles.

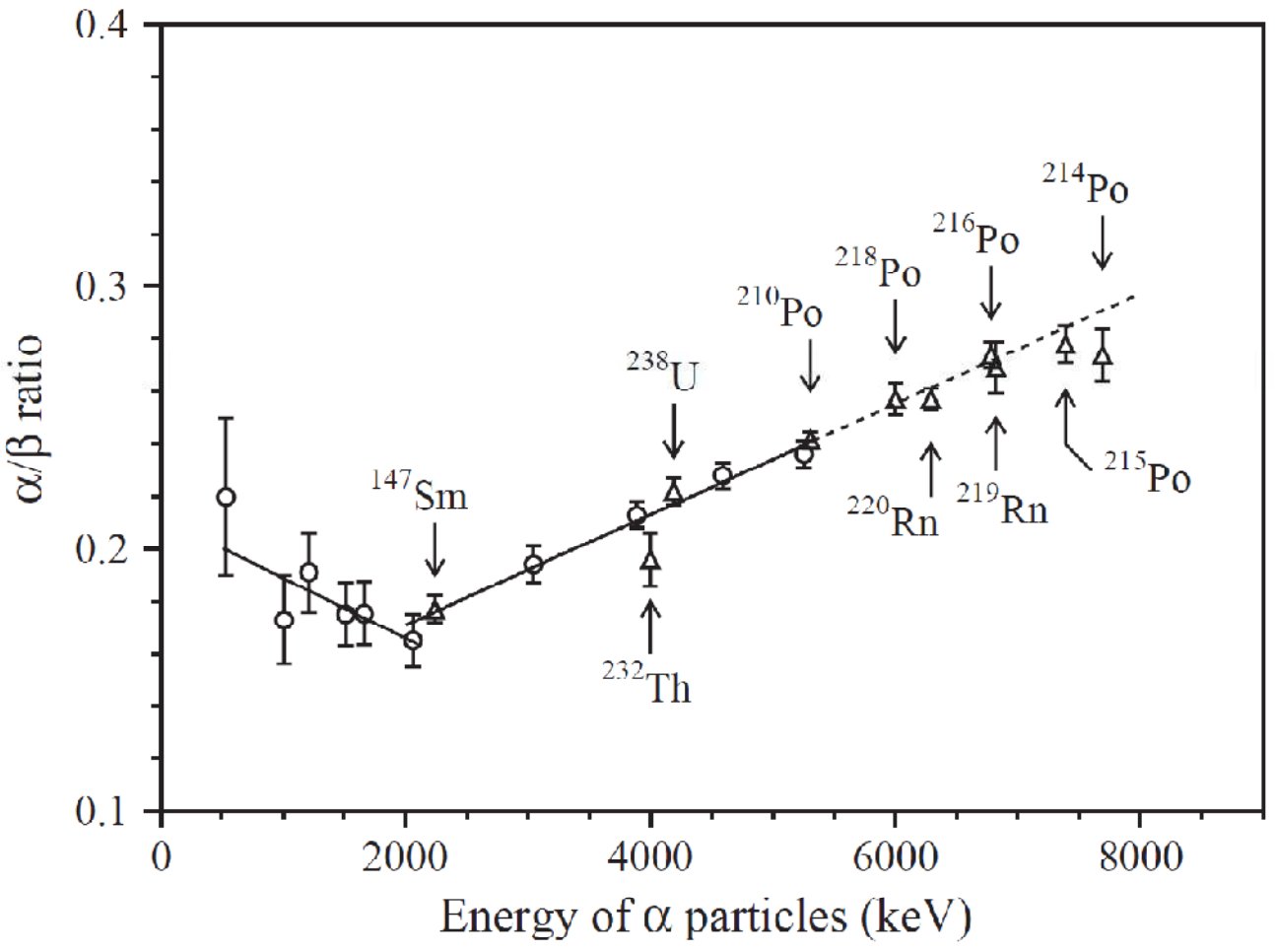

Fig. 1. Energy dependence of α/β ratio on energy measured with $CaWO_4$ crystal scintillator [20].

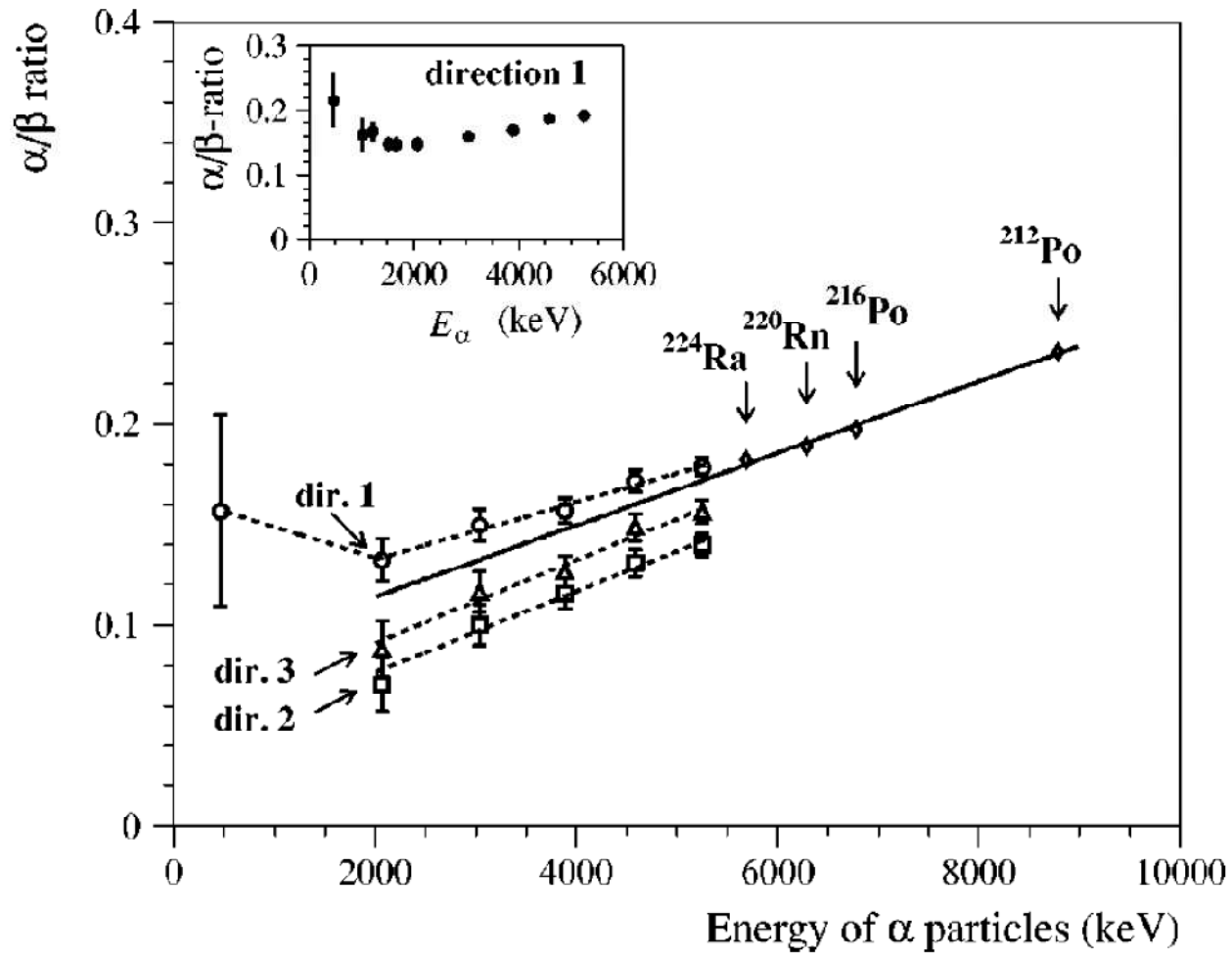

Fig. 2. Energy dependence of α/β ratio on energy measured with $^{116}CdWO_4$ crystal scintillator. α/β ratio depends on direction of irradiation relatively to crystal axes (denoted as dir. 1, dir. 2 and dir. 3). (Inset) Dependence of α/β ratio on direction measured with thin $CdWO_4$ detector to confirm increase of α/β ratio at low energies [18].

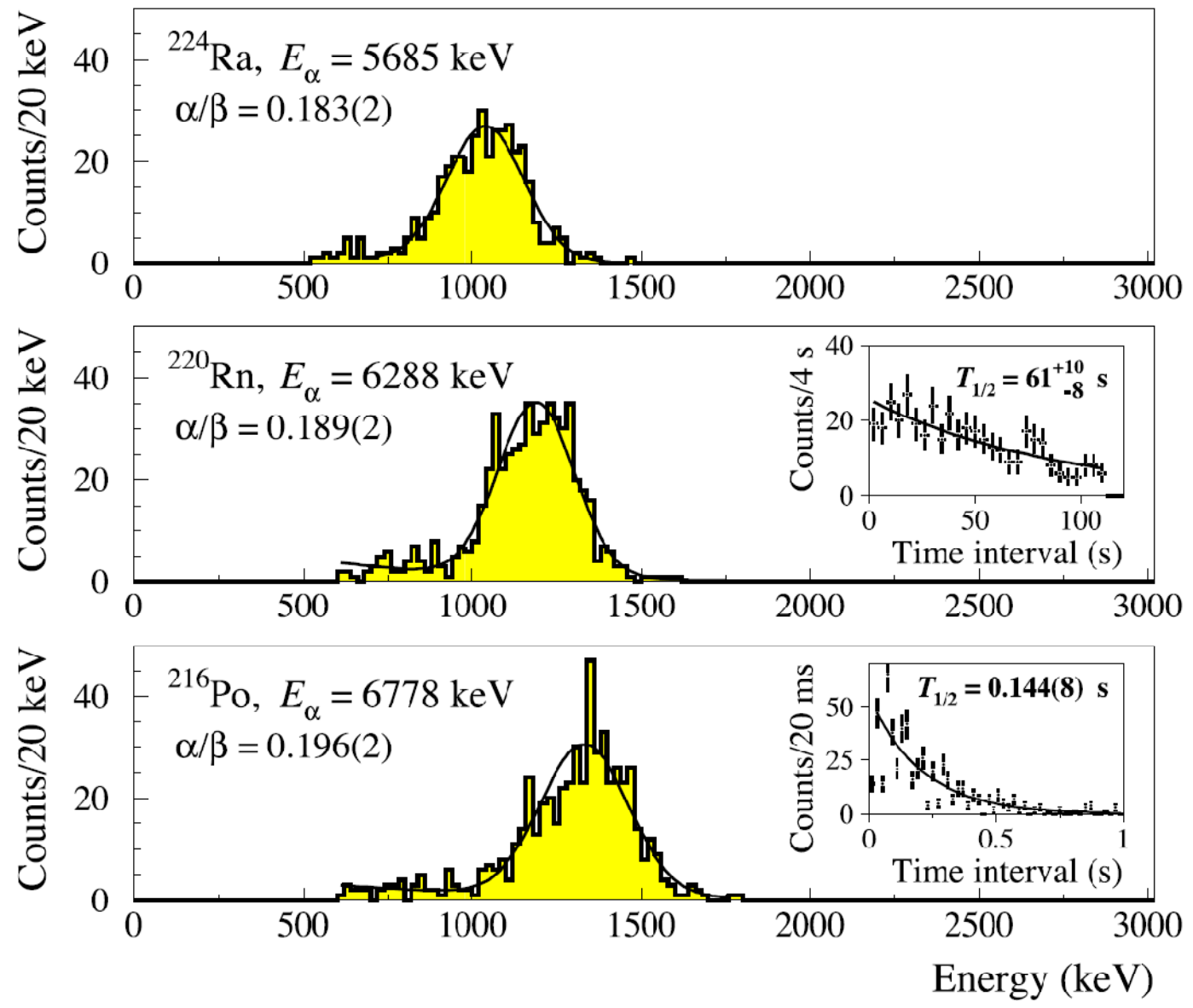

Fig. 3. The α peaks of $^{224}$Ra, $^{220}$Rn, and $^{216}$Po selected by the time-amplitude analysis from the data accumulated during 14745 h with $^{116}CdWO_4$ detector. (Insets) The time distributions between the first and second (and between second and third) events together with exponential fits are presented. Obtained half-lives of $^{220}$Rn and $^{216}$Po ($61^{+10}_{-8}$ s and 0.144±8 s, respectively) are in a good agreement with the table values [6].

Similarly the fast sequence of β and α decays: $^{214}$Bi ($Q_β$ = 3.27 MeV, $T_{1/2}$ = 19.9 m) → $^{214}$Po ($Q_α$ = 7.83 MeV, $T_{1/2}$ = 164.3 μs) → $^{210}$Pb (in equilibrium with $^{226}$Ra from $^{238}$U family) can also be selected with the help of time-amplitude analysis. In Fig. 4 one can see the energy spectra and time distributions of the sequence selected from the data accumulated in the low-background experiment to search for 2β decay of $^{160}$Gd with the help of GSO scintillator [25]. In addition the Fig. 4 illustrates a possibility to select another short chain: $^{219}$Rn ($Q_\alpha$ = 6.95 MeV; $T_{1/2}$ = 3.96 s) → $^{215}$Po ($Q_\alpha$ = 7.53 MeV; $T_{1/2}$ = 1.781 ms) → $^{211}$Pb, which is in equilibrium with $^{227}$Ac from the $^{235}$U family. In this case the events of $^{214}$Po and $^{215}$Po α decays are superimposed (see Fig. 4). Nevertheless activities of $^{226}$Ra and $^{227}$Ac can be calculated separately thanks to possibility to distinguish between broad β spectrum of $^{214}$Bi and α peak of $^{219}$Rn.

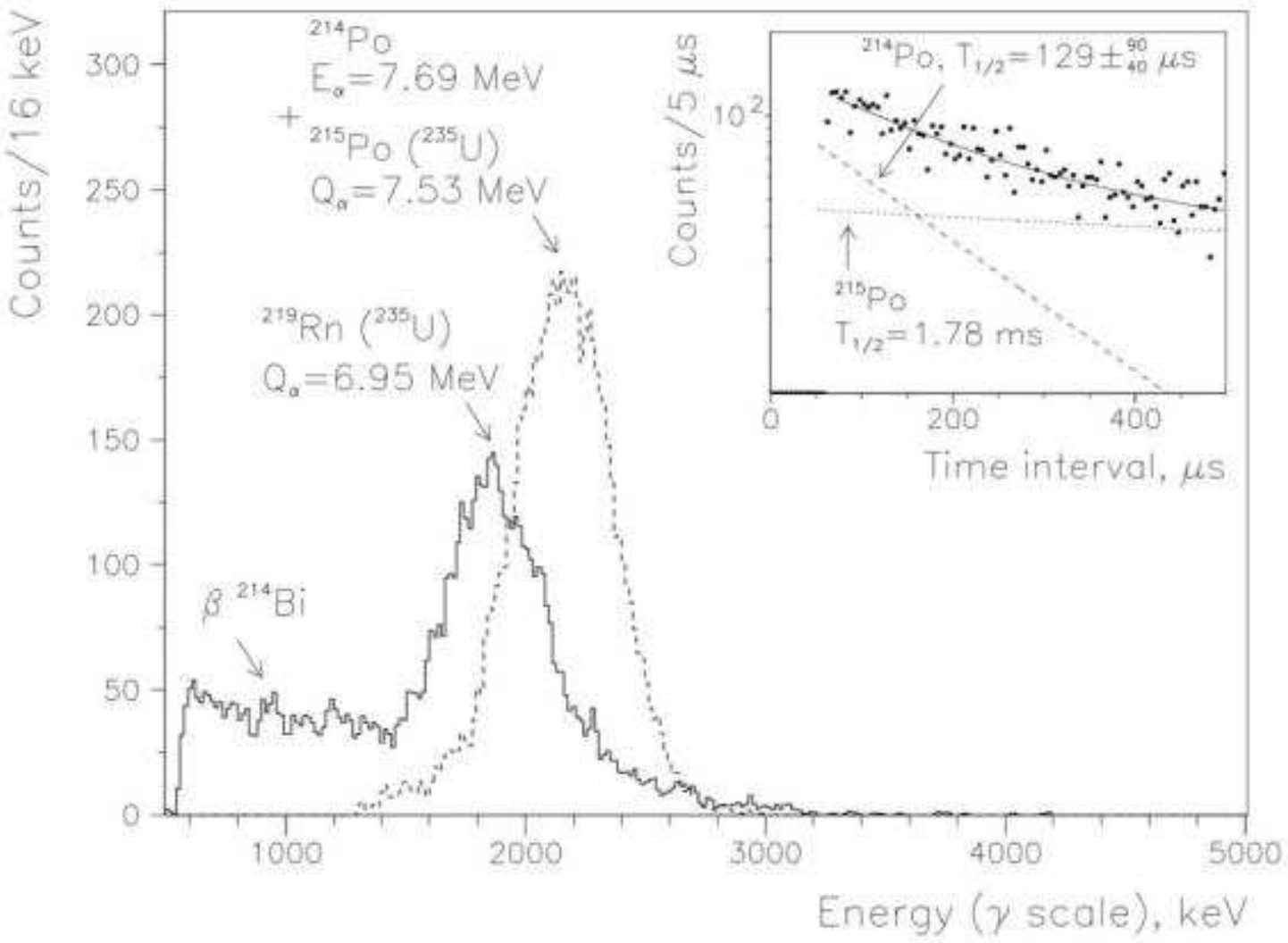

Fig. 4. The energy spectra of the sequence of β and α decays in the decay chain $^{214}$Bi → $^{214}$Po → $^{210}$Pb ($^{238}$U family) which were found by means of the time-amplitude analysis of 8609 h data accumulated with GSO scintillator. The peak with the energy in γ scale ≈ 1.8 MeV is related with the α decays of $^{219}$Rn from the chain $^{219}$Rn → $^{215}$Po → $^{211}$Pb. In the insert: the distribution of the time intervals between the first and second events together with its fit (solid line) by the sum of exponent (dashed line) with $T_{1/2}$ =129 μs (table value is $T_{1/2}$ = 164 μs) and exponent with $T_{1/2}$ = 1.78 ms corresponding to decays of $^{215}$Po from the chain $^{219}$Rn → $^{215}$Po → $^{211}$Pb (dotted line) [24].

### 3.4. Pulse-shape discrimination

Most of scintillators have slightly different decay kinetic for β particles (γ quanta) and α particles. It allows to discriminate these particles, and therefore to estimate activity of α active nuclides. Different methods can be used to realize pulse-shape discrimination. We would like to refer to the optimal filter method proposed in [26], developed in [27] for $CdWO_4$ crystal scintillators, and then successfully applied for a range of scintillators [3, 20, 21, 22, 28, 29, 30]. To realize the optimal filter method scintillation pulse shapes for α particles and γ quanta should be studied. It should be stressed pulse-shape of scintillation signals for α particles depends on energy. In some scintillators with isotropic properties pulse-shape also depends on direction of α irradiation relatively to crystal axes. As in a case with the α/β ratio such a behavior was observed in $CdWO_4$ and $ZnWO_4$ crystal scintillators [18, 22].

One can see an illustration of pulse-shape discrimination by using the optimal filter method in Fig. 5 where the scatter plot of the shape indicator (*SI*, see [27] for explanation) versus energy is shown for 171 h background data measured with the $CaWO_4$ crystal scintillator 40 × 34 × 23 mm in the low-counting experiment in the Solotvina Underground Laboratory [20]. Energy spectrum of α events selected from the data measured with the $CaWO_4$ crystal over 1734 h is presented in Fig. 6.

Another technique of background rejection can also be applied to the very fast sequence of decays from the $^{232}$Th family: $^{212}$Bi ($Q_\beta$ =2.25 MeV, $T_{1/2}$ = 60.55 m) → $^{212}$Po ($Q_\alpha$ = 8.78 MeV, $T_{1/2}$ = 0.299 μs) → $^{208}$Pb. A typical example of such an analysis is presented in Fig. 7, where the β spectrum of $^{212}$Bi, the α peak of $^{212}$Po and the distribution of the time intervals between the first and the second pulse selected from the data of low-background experiment [6] are depicted.

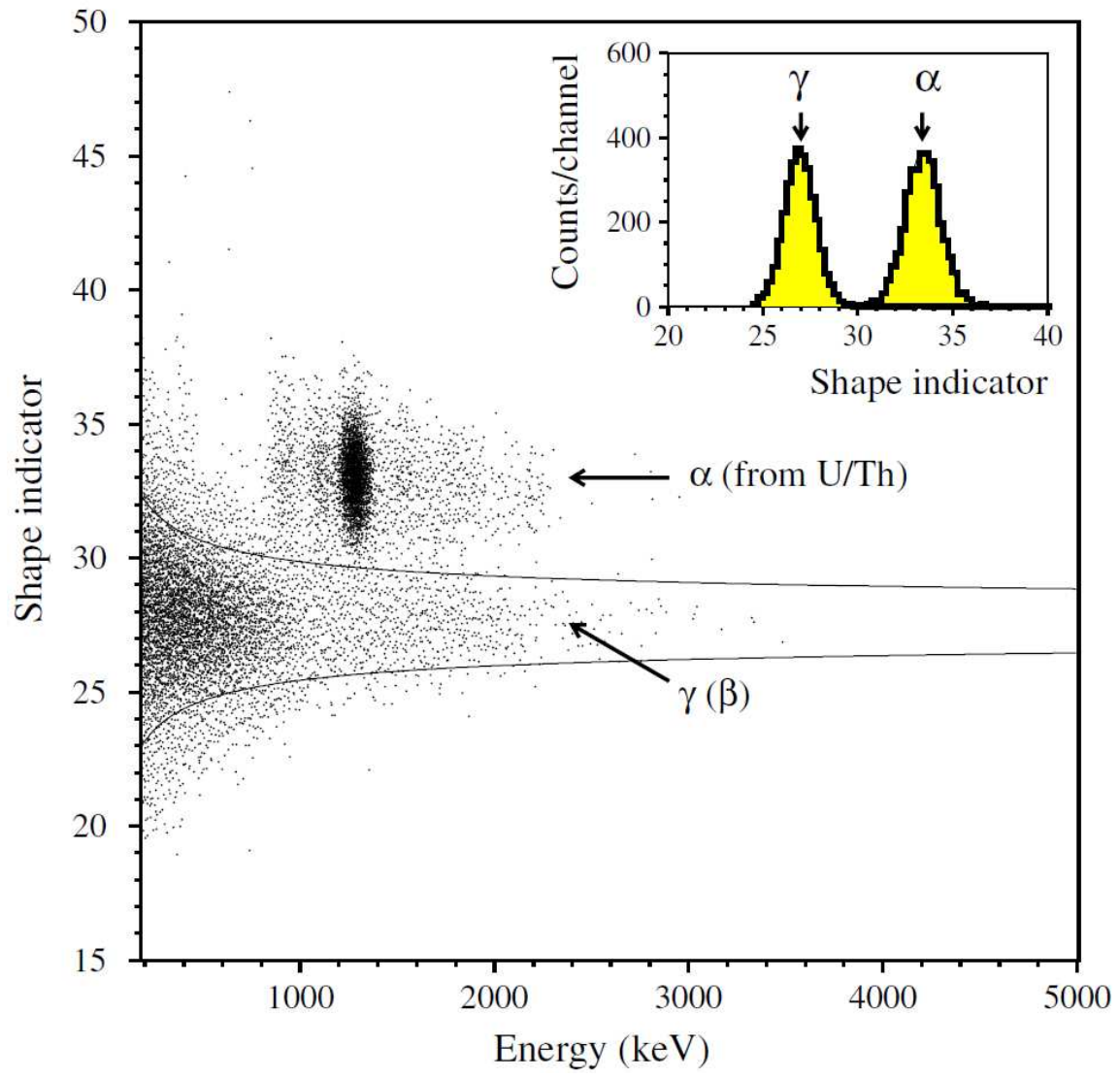

Fig. 5. Scatter plot of the shape indicator *SI* (see [27] for explanation) versus energy for 171 h background data measured with the $CaWO_4$ crystal scintillator 40 × 34 × 23 mm. Lines show ±2σ region of *SI* for γ (β) events. (Inset) The *SI* distributions measured in calibration runs with α particles ($E_α$ = 5.3 MeV which corresponds to ≈ 1.2 MeV in γ scale) and γ quanta (≈ 1.2 MeV) [19].

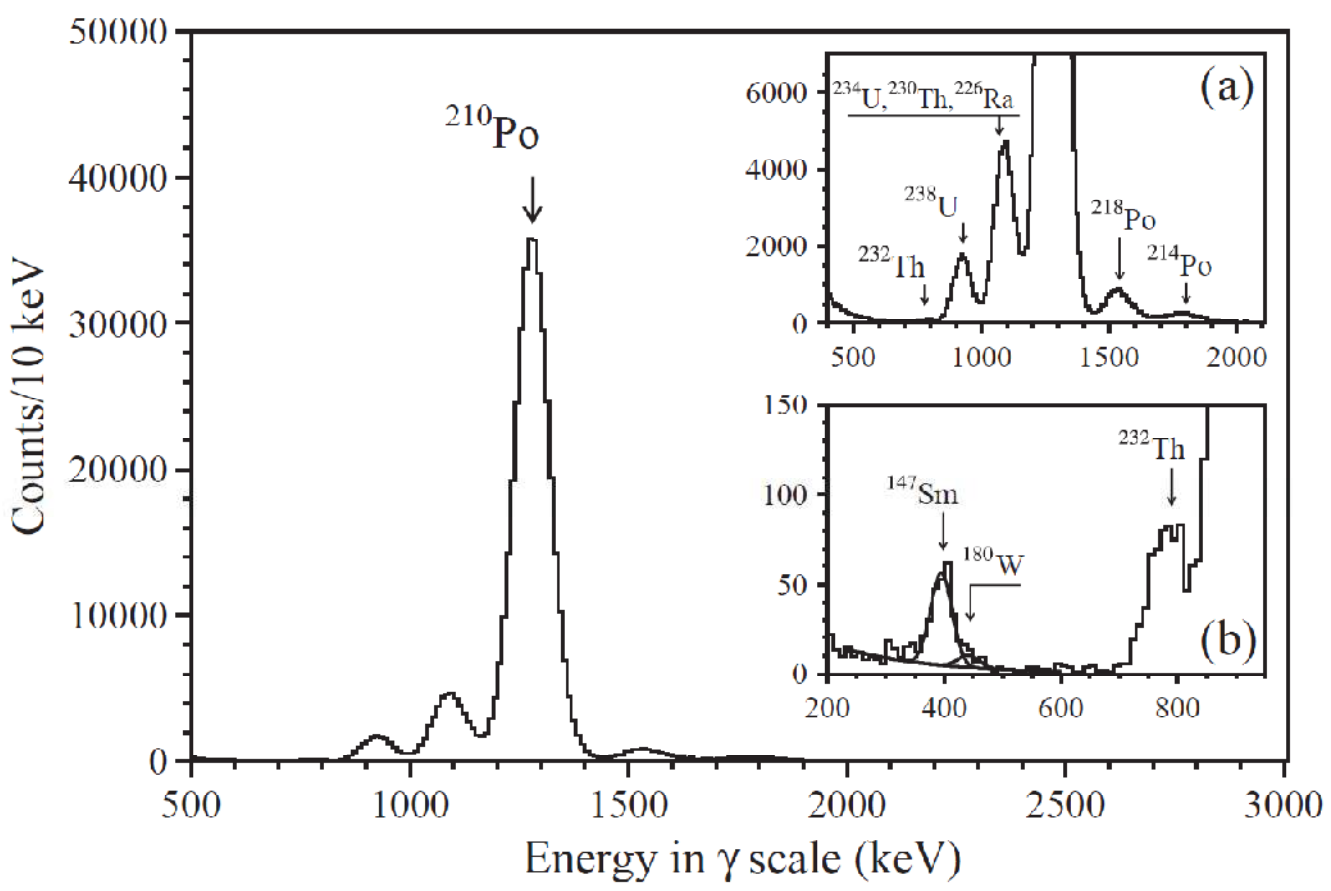

Fig. 6. Energy spectrum of α events selected by the pulse-shape analysis from background data measured over 1734 h with $CaWO_4$ detector. (Inset a) The same spectrum but scaled up. It is well reproduced by the model, which includes α decays of nuclides from $^{232}$Th and $^{238}$U families. (Inset b) Low energy part of the α spectrum [20].

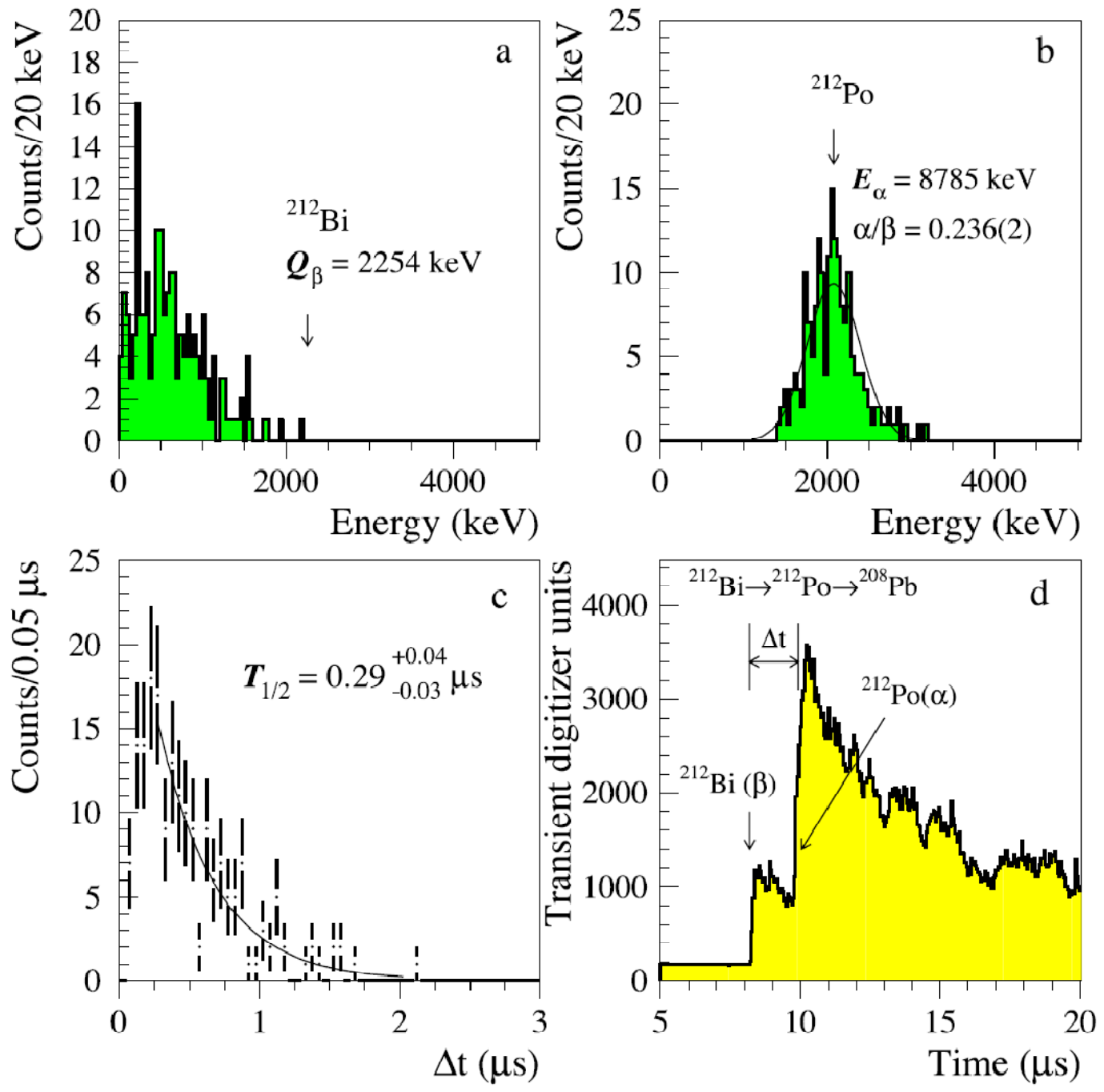

Fig. 7. The energy (a, b) and time (c) distributions for the fast sequence of β ($^{212}$Bi, $Q_β$ = 2254 keV) and α ($^{212}$Po, $E_α$ = 8785 keV, $T_{1/2}$ = 0.3 μs) decays ($^{232}$Th family) selected by the pulse-shape analysis from the background data obtained in the experiment with enriched in $^{116}$Cd cadmium tungstate crystal scintillators [6]. (d) Example of such an event in the $^{116}CdWO_4$ scintillator.

### 3.5. Energy spectra analysis

To estimate possible contamination of a scintillator, especially by β active nuclides, one can fit a low-background energy spectrum by Monte Carlo simulated models. As an example the fit of the low-background energy spectrum accumulated with GSO crystal scintillator in the experiment to search for 2β decay of $^{160}Gd$ [24] is presented in Fig. 8. The models of background were simulated with the help of the GEANT package [31, 32]. To simulate the models of background presented in Fig. 8 an event generator DECAY4 [33] was used. This generator allows to take into account a number and types of emitted particles, their energies, directions of movement and times of emission.

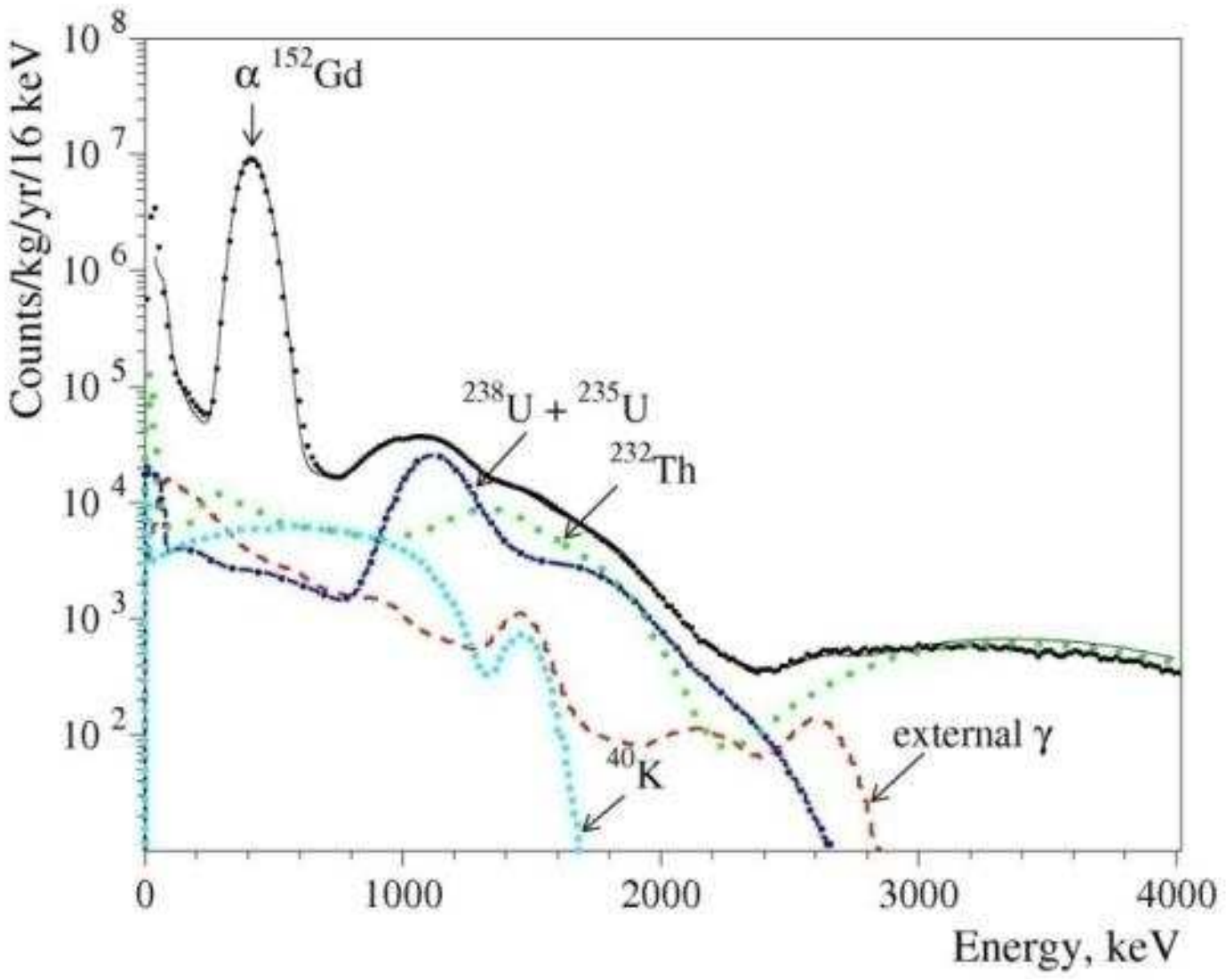

Fig. 8. The background spectrum of the GSO detector for 0.969 yr × kg of exposure (points) and the model of background (solid line) obtained by the fitting procedure in 60 – 2600 keV energy interval [24]. The most important internal ($^{40}K$, sum of $^{238}U$ and $^{235}U$, $^{232}Th$) and external (γ radiation from PMT) components of background are shown. A peak at the energy ≈ 420 keV is due to α activity of $^{152}Gd$.

### 4. Results and discussion

Radioactive contamination of crystal scintillators is presented in Table 1 where data for liquid scintillator used in the BOREXINO experiment [1] and low-background HP Ge detector [34, 35] are given for comparison. The most radiopure crystal scintillators like $ZnWO_4$, $CdWO_4$, and specially developed for low-background experiments $CaF_2$(Eu), NaI(Tl) and CsI(Tl) have rather low contamination 0.01 – 1 mBq/kg.

A level of crystal scintillators radiopurity is determined first of all by its chemical formula. For instance $CdWO_4$ and $ZnWO_4$ crystals always show low level of internal activity while scintillators containing rare earth elements (GSO, $CeF_3$) are of much higher level of radioactive trace pollution. It is due to source of rare earth mining: they usually are extracted from monazites – minerals containing a few percents of uranium and thorium. Presence of elements having radioactive isotopes in natural isotopic composition obviously determines practically unremovable[1] radioactivity of scintillators like β active $^{113}Cd$ in $CdWO_4$, α active $^{152}Gd$ in GSO, $^{138}La$ in $LaCl_3$ and $LaBr_3$, $^{176}Lu$ in $Lu_2SiO_5$ and $LuI_3$. Beta active $^{210}Pb$ is usually present in $PbWO_4$. However, this problem can be overcome by producing of lead tungstate scintillators from archaeological lead [36].

It should be mentioned an effect of concentration of radioactive pollutions in a thin (≈ mm) surface layer observed in $CdWO_4$ crystal scintillators [23]. Such an effect was not observed in GSO,

[1] We are not considering here a very expensive procedure of isotopic depletion, which can be applied to remove radioactive isotopes.

$CaWO_4$ crystal scintillators.

Table 1. Radioactive contamination of crystal scintillators (mBq/kg). Data for liquid scintillators and HP Ge detector are given for comparison.

| Scintillator | Total α activity (U + Th) | $^{228}Th$ | $^{226}Ra$ | $^{40}K$ | Particular radioactivity | References |
|---|---|---|---|---|---|---|
| $CaWO_4$ | 400 | 0.6 | 5.6 | ≤ 12 | | [20] |
| | 930 [a] | < 0.2 | 7 | | | [37] |
| $ZnWO_4$ | 0.2 | 0.002 | 0.002 | ≤ 0.4 | 0.5 ($^{65}Zn$) | [2] |
| $CdWO_4$ | 0.3 – 2 | < 0.003 – 0.039 | < 0.004 | 0.3 – 3.6 | 558 ($^{113}Cd$) | [4, 6, 18, 38, 39, 40] |
| $PbWO_4$ | $(53 – 79)\times10^3$ | ≤ 13 | ≤ 10 | | $(53 – 79)\times10^3$ ($^{210}Pb$) | [41] |
| $PbWO_4$ (from ancient lead) | | | | | ≤ 4 ($^{210}Pb$) | [36] |
| $PbMoO_4$ | | | | | $(67\text{-}192)\times10^3$ | [42] |
| $CaMoO_4$ | ≤ 10 | 0.04 | 0.13 | ≤ 3 | | [30] |
| YAG:Nd | ≤ 20 | | | | | [28] |
| BGO | | < 0.4 | < 1.2 | | $7 – 3\times10^3$ ($^{207}Bi$) | [7, 8] |
| GSO(Ce) | 40 | 2.3 | 0.3 | ≤ 14 | 1200 ($^{152}Gd$) | [24] |
| | | 100 | 1.3 | | | [43] |
| NaI(Tl) | | 0.014 | 0.045 | | | [44] |
| | 1.7 | 0.02 | 0.2 | | | [45] |
| | 0.08 | 0.009 | 0.012 | 0.6 | | [46] |
| CsI(Tl) | | 0.002 | 0.008 | | 6 ($^{134}Cs$) 14 ($^{137}Cs$) | [15] |
| $CaF_2$(Eu) | 8 | 0.13 | 1.3 | ≤ 7 | 10 ($^{152}Eu$) | [3] |
| | | 0.1 | 1.1 | | | [14] |
| $CeF_3$ | 3400 | 1100 | ≤ 60 | ≤ 330 | | [21] |
| $BaF_2$ | | 400 | 1400 | | | [47] |
| $LaCl_3$(Ce) | | ≤ 0.4 | ≤ 34 | | $21\times10^3$ ($^{138}La$) | [12] |
| $LuI_3$ | | | | | $1.7\times10^7$ ($^{176}Lu$) [b] | |
| Liquid scintillator | $10^{-6}$ | $1.2\times10^{-6}$ | $6.3\times10^{-6}$ | | 0.3 ($^{14}C$) | [1] |
| HP Ge | | $\le 2\times10^{-5}$ | $\le 2\times10^{-5}$ | | | [34, 35] |

[a] Estimated from the spectra presented in Fig. 13 of Ref. [37].
[b] Calculated value based on the half-life of $^{176}Lu$: $T_{1/2} = 3.78\times10^{10}$ y, its isotopic abundance (2.59%) [25], and chemical formula of the $LuI_3$ compound.

## 5. Conclusions

Crystal scintillators are commonly used detectors to search for rare processes in nuclear and astroparticle physics. Modern experiments to search for dark matter particles, double beta decay, rare and hypothetical processes call for extremely high level of scintillation materials radiopurity less than 0.01 mBq/kg of the total contamination, which level is not achieved for any scintillation material. Radioactive contamination of crystal scintillators varies in a wide range. The most radiopure crystal scintillators are $ZnWO_4$, NaI(Tl), CsI(Tl) whose radioactive contamination does not exceed the level of a few mBq/kg. Main sources of internal radioactivity of scintillators are daughters of U/Th families, $^{40}K$, radioactive isotopes of elements which are part of a scintillator composition. Scintillation materials containing rare earth elements have comparatively high level of U/Th contamination. Equilibrium of $^{232}$Th, $^{235}$U and $^{238}$U chains is usually broken in scintillation materials. The most sensitive method to measure internal contamination are low background measurements when a crystal scintillator to be measured operates as a detector. Time-amplitude analysis allows to detect fast sub-chains of U/Th, which are in equilibrium with $^{228}$Th (from $^{232}$Th), $^{226}$Ra ($^{238}$U), $^{227}$Ac ($^{235}$U), at the level of a few μBq/kg. Alpha active nuclides can be selected and their activity can be determined by using pulse-shape discrimination technique. Estimation of presence of β active nuclides can be realized by fit of measured energy spectra using Monte Carlo simulated models of expected background components.

### Acknowledgment

The support from the project "Kosmomikrofizyka" (Astroparticle Physics) of the National Academy of Sciences of Ukraine is gratefully acknowledged.

### References

1. G. Alimonti et al., Phys. Lett. B 422 (1998) 349.
2. P. Belli et al., preprint ROM2F/2008/22; arXiv: 0811.2348 [nucl-ex]; submitted to Phys. Rev. C.
3. P. Belli et al., Nucl. Phys. A 789 (2007) 15.
4. A.Sh. Georgadze et al., Instr. Exp. Technique 39 (1996) 191.
5. F.A. Danevich et al., Nucl. Phys. A 717 (2003) 129.
6. F.A. Danevich et al., Phys. Rev. C 68 (2003) 035501.
7. A. Balysh et al., Pribory i Tekhnika Eksperimenta 1 (1993) 118 (in Russian).
8. P. de Marcillac et al., Nature 422 (2003) 876.
9. D. Grigoriev et al., these Proceedings, p. 45.
10. N. Coron et al., these Proceedings, p. 12.
11. F.A. Danevich et al., Phys. Rev. C 62 (2000) 045501.
12. R. Bernabei et al., Nucl. Instr. Meth. A 555 (2005) 270.
13. H. Eijiri et al., Nucl. Instr. Meth. A 302 (1991) 304.
14. I. Ogawa et al., Nucl. Phys. A 721 (2003) 525.
15. H.S. Lee et al., Nucl. Instr. Meth. A 571 (2007) 644.
16. P.Belli et al., Nucl. Instr. Meth. A 572 (2007) 734.
17. O.P. Barinova et al., "Lithium molybdate crystal as a possible detector of rare nuclear events", submitted to Nucl. Instr. Meth. A.
18. F.A. Danevich et al., Phys. Rev. C 67 (2003) 014310.
19. Yu.G. Zdesenko et al., Astropart. Phys. 23 (2005) 249.
20. Yu.G. Zdesenko et al., Nucl. Instr. Meth. A 538 (2005) 657.
21. P. Belli et al., Nucl. Instr. Meth. A 498 (2003) 352.
22. F.A. Danevich et al., Nucl. Instr. Meth. A 544 (2005) 553.
23. F.A. Danevich et al., Phys. Lett. B 344 (1995) 72.
24. F.A. Danevich et al., Nucl. Phys. A 694 (2001) 375.
25. R.B. Firestone et al., Table of Isotopes, 8th ed., John Wiley & Sons, New York, 1996 and CD update, 1998.
26. E. Gatti, F. De Martini, Nuclear Electronics 2, IAEA, Vienna, 1962, p. 265.
27. T. Fazzini et al., Nucl. Instr. Meth. A 410 (1998) 213.
28. F.A. Danevich et al., Nucl. Instr. Meth. A 541 (2005) 583.
29. L. Bardelli et al., Nucl. Instrum. Meth. A 584 (2008) 129.
30. A.N. Annenkov et al., Nucl. Instr. Meth. A 584 (2008) 334.
31. CERN Program Library Long Write-Up W5013, 1994.
32. S. Agostinelli et al., Nucl. Instr. Meth. A 506 (2003) 250;
    J. Allison et al., IEEE Trans. Nucl. Sci. 53 (2006) 270.
33. O.A. Ponkratenko, V.I. Tretyak, Yu.G. Zdesenko, Phys. Atom. Nuclei 63 (2000) 1282.
34. H.V. Klapdor-Kleingrothaus et al., Nucl. Instr. Meth. A 481 (2002) 149.
35. C. Dörr, H.V. Klapdor-Kleingrothaus, Nucl. Instr. Meth. A 513 (2003) 596.

36. A. Alessandrello et al., Nucl. Instr. Meth. A 409 (1998) 451.
37. S. Cebrián, et al., Astropart. Phys. 21 (2004) 23.
38. S.Ph. Burachas etal., Nucl. Instr. Meth. A 369 (1996) 164.
39. F.A. Danevich et al., Z. Phys. A 355 (1996) 433.
40. P. Belli et al., Phys. Rev. C 76 (2007) 064603.
41. F.A. Danevich et al., Nucl. Instr. Meth. A 556 (2206) 259.
42. Yu.G. Zdesenko et al., Instr. and Exp. Technique 39 (1996) 364.
43. S.C. Wang et al., Nucl. Instr. Meth. A 479 (2002) 498.
44. J.C. Barton, J.A. Edgington, Nucl. Instr. Meth. A 443 (2000) 277.
45. J. Amaré et al., J. Phys.: Conf. Series 39 (2006) 201.
46. R. Bernabei et al., Nucl. Instr. Meth. A 592 (2008) 297.
47. R. Cerulli et al., Nucl. Instr. Meth. A 525 (2004) 535.

# Radioactive contamination of $CaWO_4$ crystal scintillators

**F.A. Danevich***, **S.S. Nagorny, A.S. Nikolaiko**

*Institute for Nuclear Research, MSP 03680 Kyiv, Ukraine*

Radioactive contamination of $CaWO_4$ crystal scintillators was measured. The samples were made from boules produced from the same raw material using recrystallization procedure, which lead to the significant difference (in the range of 0.03-1.32 Bq/kg) in $^{210}$Po contamination. Activity of $^{238}$U varies as 0.04-0.33 Bq/kg in anti-correlation with $^{210}$Po. Recrystallization improves radiopurity of $CaWO_4$ relatively to $^{210}$Po and $^{238}$U about one order of magnitude. We have not observed difference in radioactive contamination of crystal with removed ≈1 cm surface layer in comparison with the crystal made from the same boule with no surface treatment. No significant dependence in radioactive contamination of samples produced from one boule has been observed too. Recrystallization improves also scintillation properties of $CaWO_4$ crystal scintillators.

## 1. Introduction

Cryogenic phonon-scintillation detectors (see e.g. [1] and references therein) combine excellent energy resolution and low threshold with the ability to discriminate between different particles. This makes them promising tools in experimental searches for dark matter and neutrinoless double beta decay [2–4].

$CaWO_4$ crystals are now in use in CRESST experiment to search for dark matter particles [5-7]. This material is considered as one of targets in EURECA[1] [8, 9], a project of tonne-scale detector for dark matter search, where a multi-element target is planned for confirming a true dark matter signal. EURECA aims to achieve sensitivity to WIMP-nucleon scattering cross sections down to $10^{-10}$ pb, which is at least two orders of magnitude better than that of present leading experiments. Critical for success will be the suppression of background counting rate to the level less than a few events per keV per 100 kg per year in the energy interval of a few keV. To reach such extremely low background one should develop crystal scintillators with radiopurity on the level of 0.01 mBq/kg, while the radioactive contamination of present $CaWO_4$ crystals [10, 11, 12] remains to be 3 – 4 orders of magnitude higher.

The motivation of this work was to study effects of recrystallization on radioactive contamination of $CaWO_4$ crystal scintillators. Further steps to improve radiopurity of $CaWO_4$ are discussed.

## 2. Samples

Three $CaWO_4$ crystals (set 1) were produced at the Scientific Research Company "CARAT" (Lviv, Ukraine) from one of crystal boules grown by the Czochralski method from raw material in an iridium crucible with high frequency heating. Two crystals of set 2 were produced from the boule grown from the rest of melt after the first (set 1) crystallization. The crystals of set 3 were produced from the whole boule of set 1 and the rest of boules of sets 1 and 2 (Fig. 1). About 1 cm surface layer of crystal CWO3-1 of set 3 was removed. Main properties of the samples are denoted in Table 1.

---

* Corresponding author. *E-mail address:* danevich@kinr.kiev.ua

[1] European Underground Rare Event Calorimeter Array; www.eureca.ox.ac.uk

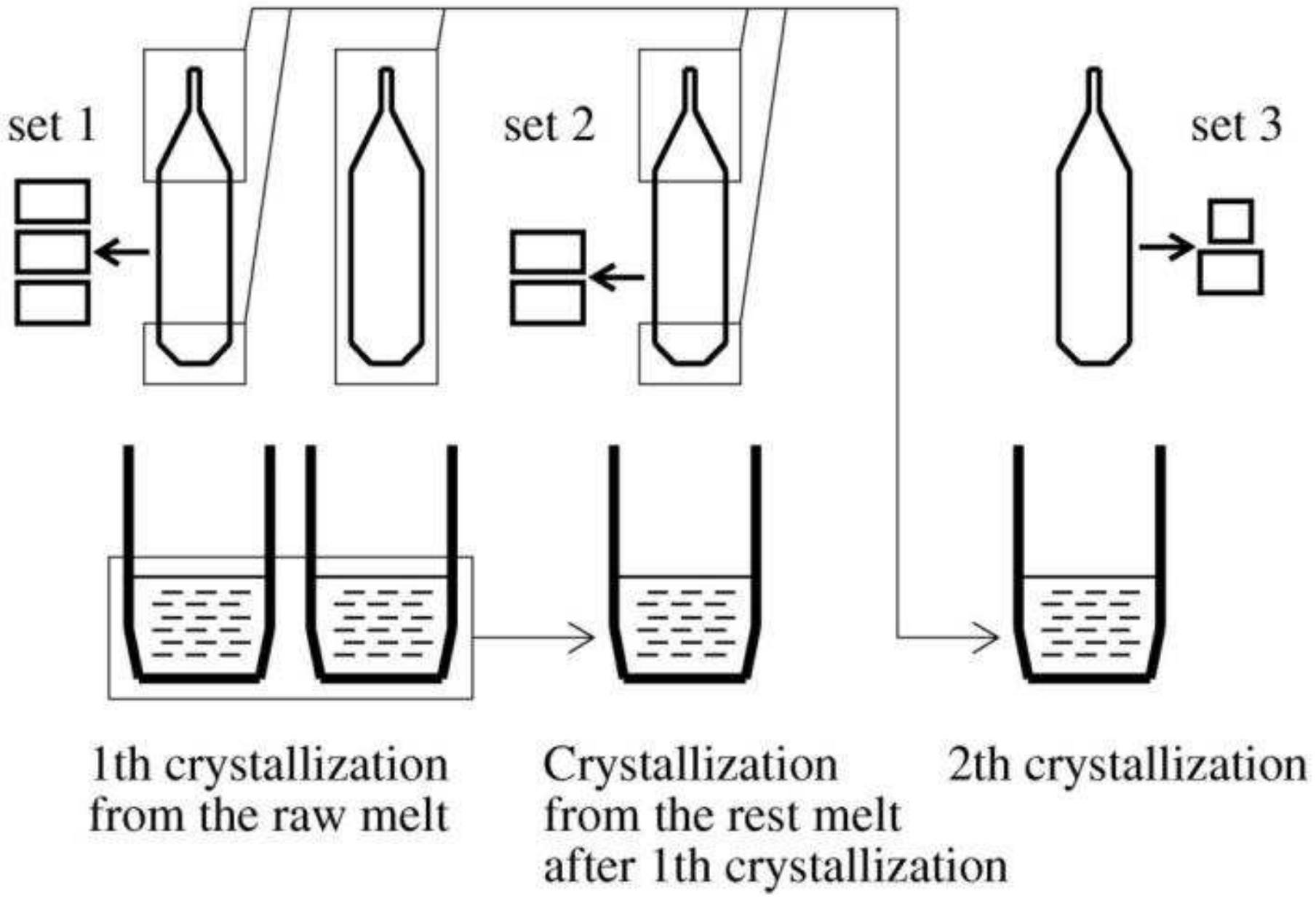

Fig. 1. Growth of $CaWO_4$ crystal scintillators. The crystals of set 2 were produced from boule grown from the rest of melt after 1th crystallization. The crystals of set 3 were grown from the whole boule and the rest of boules of sets 1 and 2 [13].

Table 1. Description of $CaWO_4$ crystal scintillators used for low-background measurements. Energy resolutions at 662 keV line of $^{137}$Cs measured with the light-guide in the low-background set-up (FWHM), $\alpha/\beta$ ratio measured with external $^{241}$Am source (external $\alpha$ source) and determined by $\alpha$ peak of $^{210}$Po presented in crystals as contamination (internal $^{210}$Po), time of low-background measurements (t) are specified.

| Set | Sample | Sizes, mm | Mass, g | FWHM, % | α/β ratio | | t, h |
|---|---|---|---|---|---|---|---|
| | | | | | External α source | Internal $^{210}$Po | |
| 1 | CWO1-1 | ∅60×42 | 740 | 11.3 | | 0.241(5) | 2.822 |
| 1 | CWO1-2 | ∅60×42 | 740 | 9.7 | | 0.242(5) | 2.583 |
| 1 | CWO1-3 | ∅60×42 | 740 | 10.0 | 0.23(1) | 0.243(5) | 2.672 |
| 2 | CWO2-1 | Prism of complicated shape | 473 | 9.8 | | 0.26(1) | 3.465 |
| 2 | CWO2-2 | 42×41×41 | 484 | 10.3 | 0.26(1) | 0.25(1) | 17.389 |
| 3 | CWO3-1 | ∅40×42 | 328 | 8.2 | | 0.233(5) | 18.644 |
| 3 | CWO3-2 | ∅60×42 | 740 | 8.3 | | 0.231(5) | 2.717 |

## 3. Measurements and results

### 3.1. Scintillation properties

Side surface of all the crystals was roughly grounded, while faces were polished. The best scintillation properties were measured for the crystals of set 3. The crystal CWO3-1 was connected optically with the help of Dow Corning Q2-3067 optical couplant to 3" photomultiplier (PMT) Philips XP2412, and covered by 3 layers of PTFE reflector tape. The measurements were carried out with 10 μs shaping time of the ORTEC 572 spectrometric amplifier. The crystal was irradiated by γ quanta of $^{137}$Cs source. The energy spectrum measured with the crystal scintillator is presented in Fig. 2. The energy resolution (full width at half of maximum, FWHM) is 6.28% for 662 keV γ quanta of $^{137}$Cs.

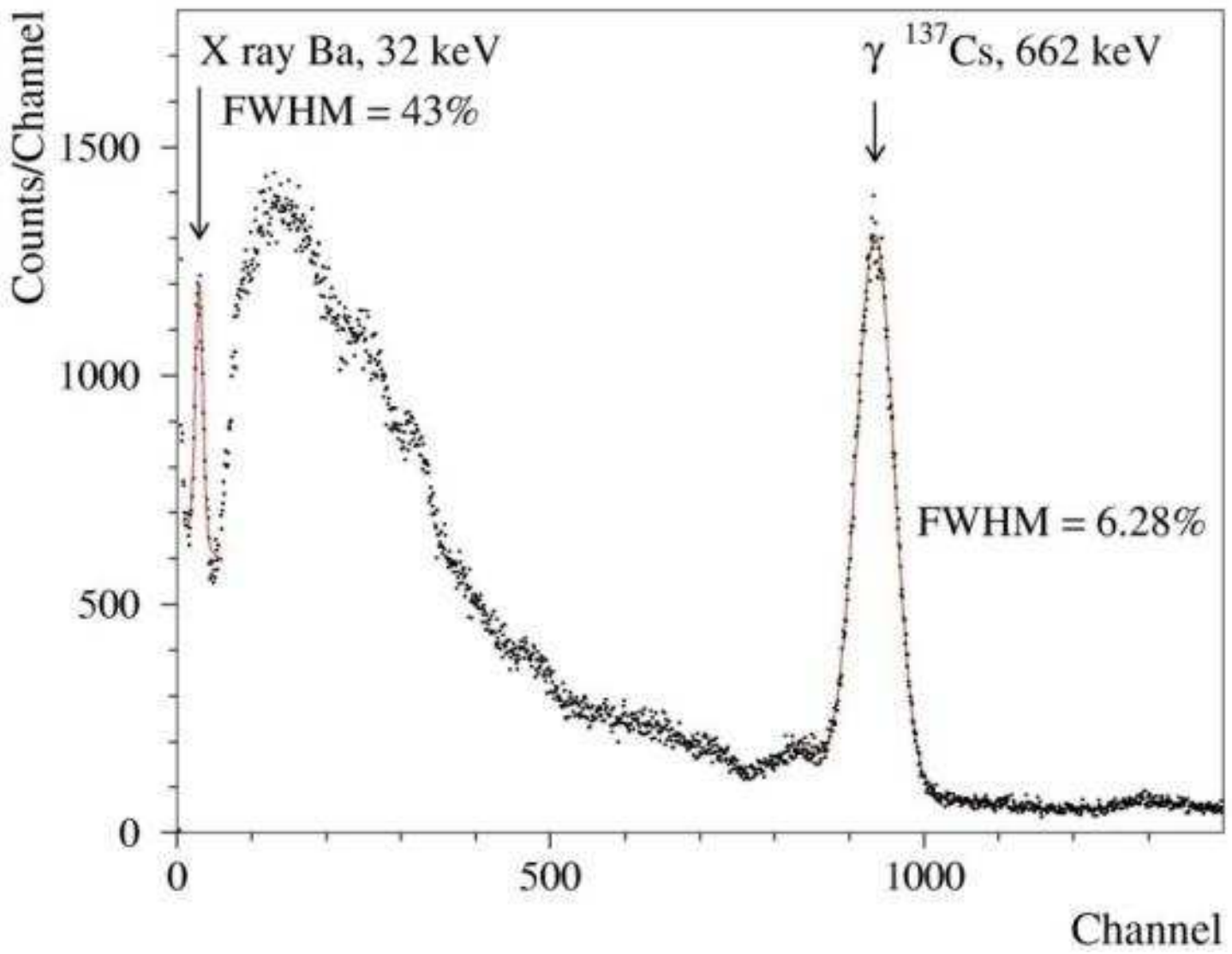

Fig. 2. Energy spectrum of $^{137}$Cs γ rays measured for a $CaWO_4$ scintillation crystal (∅42 × 40 mm, CWO3-1).

### 3.2. Low-background measurements of $CaWO_4$ crystal scintillators

*3.2.1. Low-background set-up and measurements*

Radioactive contamination of the crystals was measured in the low-background set-up installed in the Institute for Nuclear Research (Kyiv). In the set-up, a scintillation $CaWO_4$ crystal (covered by 3 layers of PTFE tape) was viewed by a 3" photomultiplier tube (XP2412) through the high purity polystyrene light-guide ∅66×120 mm. The light-guide was wrapped by aluminized mylar. The detector was surrounded by a passive shield made of high purity OFHC (oxygen free high conductivity) copper (5–12 cm) and lead (5 cm).

Energy scale and resolution of the detector were determined in calibration runs with $^{137}$Cs and $^{207}$Bi γ sources. The values of the energy resolution for 662 keV γ quanta of $^{137}$Cs are listed in Table 1. The energy calibration was checked by γ lines of 662 keV ($^{137}$Cs) and 2615 keV ($^{208}$Tl) present in the background spectra. The times of background measurements are also listed in Table 1.

The response function to α particles in wide energy interval was measured in work [11]. Above 2 MeV the α/β ratio[1] increases as $\alpha/\beta = 0.129(12) + 0.021(3) \times E_\alpha$ (where $E_\alpha$ is energy of α particles in MeV).

In the present study the α/β ratio was measured with crystals CWO1-3 and CWO2-2 using collimated α particles of a $^{241}$Am source. The dimensions of the collimator made of teflon were ∅0.75×2 mm. As it was checked by a surface-barrier detector, the energy of α particles was reduced from 5.75 to about 5.25 MeV due to passing through the collimator (2 mm of air). The values of α/β ratio measured with two crystals, as well as α/β ratios determined by analysis of α peaks from internal $^{210}$Po (see subsection 2.3.2) are presented in Table 1. One can suppose that the difference in α/β ratio is due to difference in the impurities present as trace contamination in the crystals. However we can not interpret the difference in α/β ratios as only own properties of the scintillators because it depends also on a certain characteristics of the scintillation detector: shape, sizes, surface treatment, transparency of a scintillator, etc. [14].

To discriminate α events from γ(β) background, pulse shapes of scintillation signals of $CaWO_4$ scintillators were measured with the help of a 12 bit 20 MS/s transient digitizer. Technique of pulse-shape discrimination (described in [11]) has allowed to confirm α nature of peaks in the energy spectra of the $CaWO_4$ detectors.

---

[1] The α/β ratio is defined as ratio of α peak position in the energy scale measured with γ sources to the energy of α particles. Because γ quanta interact with detector by β particles, we use more convenient term "α/β ratio".

It is important to stress that scintillation properties of the crystals of set 3 are much better (the energy resolution FWHM ≈ 8%) in comparison to set 1 (FWHM≈10–11%). One can conclude that the recrystallization procedure improves the quality of $CaWO_4$ crystal scintillators.

### *3.2.2. Radioactive contamination of $CaWO_4$ crystal scintillators*

The energy spectra of three $CaWO_4$ crystals (set 1), normalized on mass of the crystals and time of the measurements are shown in Fig. 3.

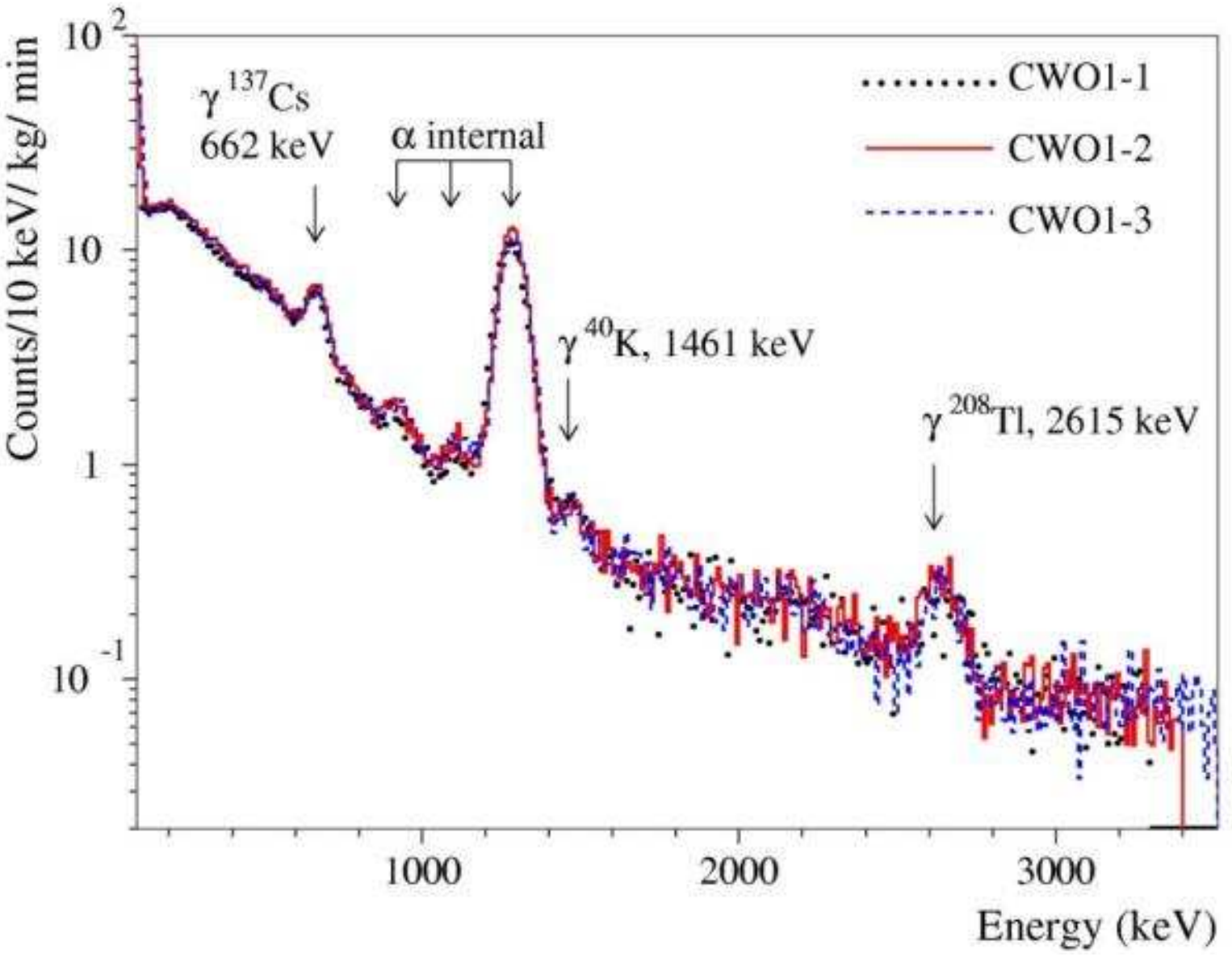

Fig. 3. Energy spectra of three $CaWO_4$ scintillators of set 1 measured in low-background set-up. The level of background of the crystals is practically indistinguishable.

There are a few peaks in the spectra caused by external γ quanta: 662 keV of $^{137}$Cs (pollution of the set-up after the Chernobyl accident), 1461 keV of $^{40}$K, and 2615 keV of $^{208}$Tl ($^{232}$Th family). The background counting rates of all three detectors are practically the same. Peaks with energies in the interval 0.85–1.3 MeV can be attributed to internal contamination by α active U/Th nuclides. The α nature of these peaks were confirmed by pulse-shape discrimination (see Fig. 4).

The peak in the energy spectra of the crystals of set 1 (see Fig. 5a, where the spectrum accumulated with the sample CWO1-2 is presented) at the energy ≈1.28 MeV can be explained by $^{210}$Po pollution (daughter of $^{238}$U, $E_\alpha$=5304 keV, $T_{1/2}$=138.376 d [15]) with the activity 1.32(2) Bq/kg. The peak near 0.9 MeV is due to $^{238}$U with the activity 0.05(1) Bq/kg, while $^{234}$U, $^{230}$Th and $^{226}$Ra contribute to the peak at 1.08 MeV with the area corresponding to activity 0.03(2) Bq/kg. Total α activity in the crystal is 1.40(3) Bq/kg. Fit of the energy spectra in the energy region 0.74–1.65 MeV by the model consisted of Gaussian functions (to describe α peaks), and exponent (to characterize background from external γ rays) is shown in Fig. 5a. Activity of $^{226}$Ra can be estimated by analysis of α peak of $^{218}$Po ($E_\alpha$=6002 keV). An α peak with the energy (in γ scale) around 1.53 MeV is expected. However, because there is no peaks in the energy spectra accumulated with the crystals of set 1 at this energy (while the peak of $^{218}$Po there is in the spectra of CWO2-2 crystal, see Fig. 4b, and Fig. 5b), we can set only limits on activity of $^{226}$Ra (see Table 2). Similarly, an α peak with the energy in γ scale near 0.78 MeV is expected for $^{232}$Th. Fit of energy spectra gives only limits on $^{232}$Th activity in all the crystals.

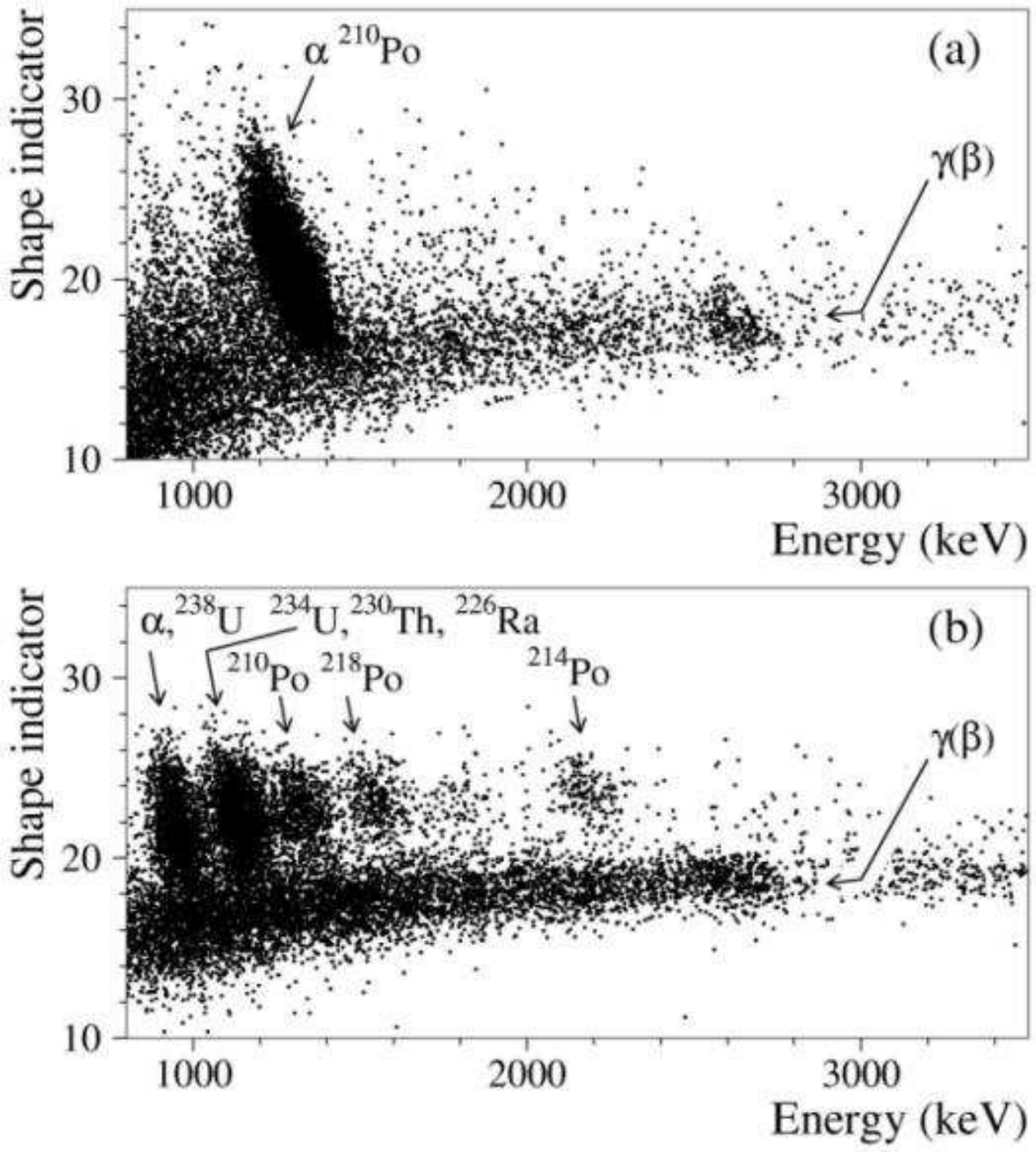

Fig. 4. Scatter plot of the shape indicator (see [11]) versus energy for background exposition with $CaWO_4$ crystal scintillators. (a) - sample CWO1-2, time of measurements 2.12 h; (b) - sample CWO2-2, measured over 1.10 h. Populations of α events are well separated from γ quanta. One can see a clear difference in the energy distributions of α events in the crystals from different sets.

Because equilibrium in thorium chain between $^{232}$Th and $^{228}$Th can be broken, we have estimated activity of $^{228}$Th by analysis of α spectra in the energy region where two α peak with energies 1.65 MeV ($^{220}$Rn, $E_{\alpha}$=6228 keV) and 1.83 MeV ($^{216}$Po, $E_{\alpha}$=6778 keV) are expected. The results of estimations of activities of α active daughters of $^{232}$Th and $^{238}$U in the crystals of set 1 are presented in Table 2.

The energy spectrum of $CaWO_4$ scintillator CWO2-2 from set 2 is shown in Fig. 5b. The activities of $^{210}$Po, $^{238}$U, sum of $^{234}$U, $^{230}$Th and $^{226}$Ra are substantially different from set 1 (see Table 2). Activity of $^{238}$U is much larger, which can be explained by accumulation of U trace impurities (activity 0.3 Bq/kg corresponds to 0.02 ppm of U, which does not contradicts to the result of ICP-MS analysis in the raw material: 0.006 ppm) in the melt, while activity of $^{210}$Po is one order of magnitude lower than that in crystals of set 1.

Finally the activities of $^{238}$U daughters in crystals of set 3 (see Fig. 5c and Table 2) differ both from set 1 and 2.

To estimate presence in the samples of β active isotopes ($^{40}$K, $^{90}$Sr–$^{90}$Y, $^{210}$Bi) the measured background spectra of the $CaWO_4$ detectors were simulated with the GEANT4 package [16, 17]. Initial kinematics of particles emitted in β decays of nuclei and subsequent nuclear de-excitation process was generated with the DECAY0 event generator [18]. There are no clear peculiarities in the spectra which could be referred to the internal trace contamination by the β active nuclides. Therefore only limits on activities of $^{40}$K, $^{90}$Sr – $^{90}$Y, and $^{210}$Bi ($^{210}$Pb) can be set on the basis of the experimental data. With this aim the background spectra were fitted in the energy interval 0.1–3.0 MeV by the model, which includes also Monte Carlo simulated distributions to describe external γ background from materials of the set-up. Limits on activities of β active radionuclides in the $CaWO_4$ crystals are presented in Table 2.

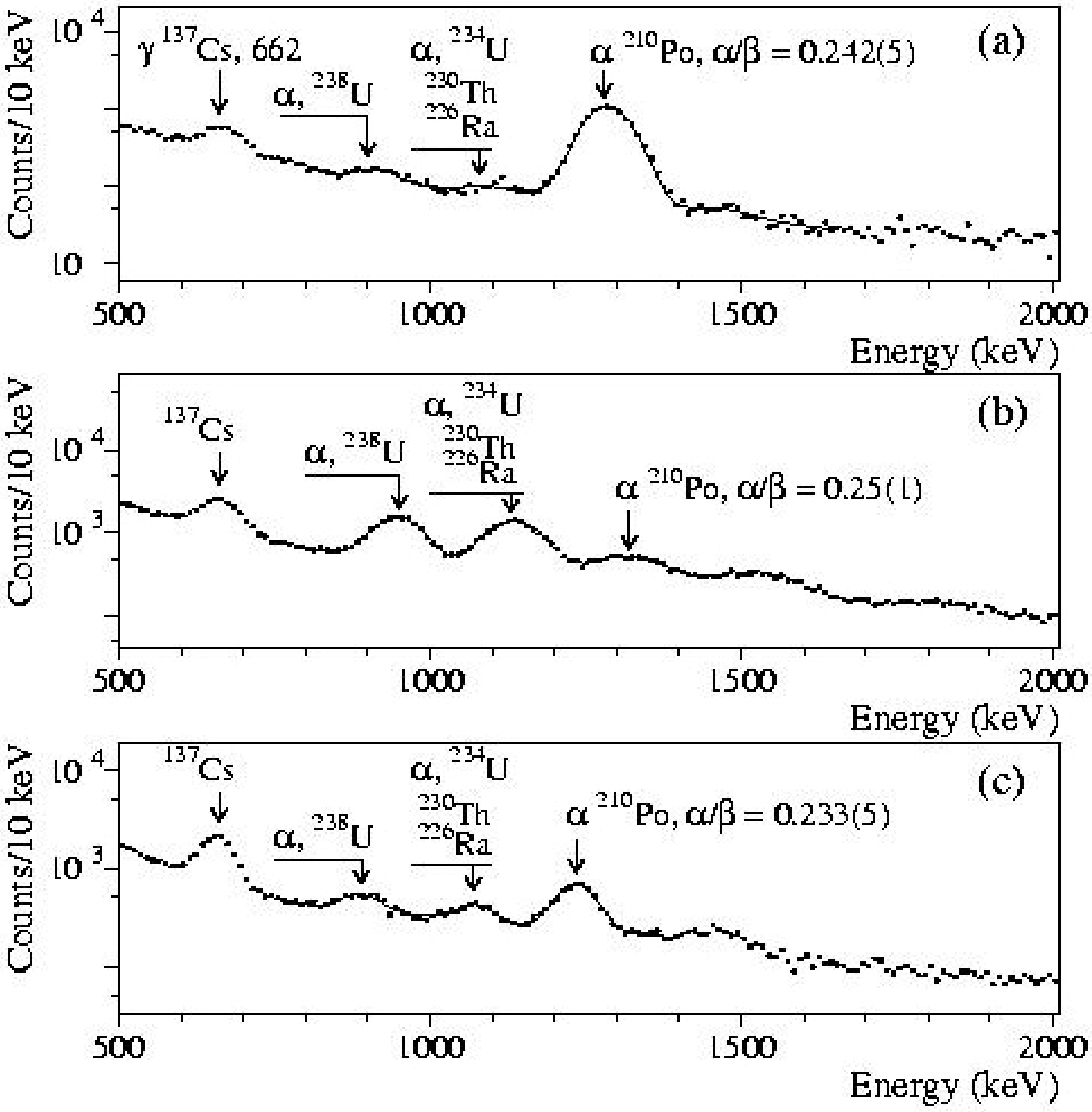

Fig. 5. Part of the energy spectra of $CaWO_4$ crystal scintillators measured in low-background set-up (a) - CWO1-2, 2.583 h; (b) - CWO2-2, 17.389 h; (c) - CWO3-2, 18.644 h. Fits of alpha peaks are shown by solid lines.

Table 2: Radioactive contamination of $CaWO_4$ crystals. The data for $CaWO_4$ from [11] and $CaMoO_4$ from [19] are given for comparison.

| Nuclide | Activity (mBq/kg) in sample | | | | | | | | |
|---|---|---|---|---|---|---|---|---|---|
| | 1-1 | 1-2 | 1-3 | 2-1 | 2-2 | 3-1 | 3-2 | [11] | [19] |
| $^{232}$Th | ≤ 9 | ≤ 14 | ≤ 22 | ≤ 48 | ≤ 25 | ≤ 30 | ≤ 34 | 0.7(1) | ≤1.5 |
| $^{228}$Th | ≤ 6 | ≤ 14 | ≤ 11 | ≤ 10 | ≤ 12 | ≤ 8 | ≤ 13 | 0.6(2) | 0.04(2) |
| $^{238}$U | 38(14) | 54(12) | 67(14) | 330(17) | 268(20) | 51(8) | 48(11) | 14(5) | ≤1.5 |
| $^{226}$Ra | ≤ 10 | ≤ 11 | ≤ 37 | 72(16) | 107(11) | ≤ 21 | ≤ 15 | 5.6(5) | 0.13(4) |
| $^{210}$Pb | ≤ 3800 | ≤ 4100 | ≤ 4000 | ≤ 4300 | ≤ 4000 | ≤ 2300 | ≤ 3400 | ≤430 | ≤17 |
| $^{210}$Po | 1244(17) | 1316(17) | 1243(16) | 26(9) | 55(6) | 151(8) | 176(10) | 291(50) | ≤8 |
| Total α activity U/Th | 1310(30) | 1400(30) | 1340(30) | 670(30) | 580(30) | 240(20) | 260(20) | 400 | |
| $^{40}$K | ≤ 370 | ≤ 400 | ≤ 390 | ≤ 440 | ≤ 980 | ≤ 210 | ≤ 330 | <12 | ≤3 |
| $^{90}$Sr–$^{90}$Y | ≤ 440 | ≤ 480 | ≤ 480 | ≤ 500 | ≤ 1140 | ≤ 260 | ≤ 380 | <70 | ≤23 |

The summary of the measured radioactive contamination of the $CaWO_4$ crystals is given in Table 2 where also data obtained in [11] for $CaWO_4$ and in [19] for $CaMoO_4$ are given for comparison.

## 4. Discussion

Behaviour of activities of U daughters in the crystals of sets 1, 2 and 3 is presented in Fig. 6 where the effect of anti-correlation between Po and U is clearly seen. One can suppose that the decrease of $^{210}$Po activity in the crystals of set 3 in comparison with set 1 is also due to evaporation of Polonium from melt.

An equilibrium of $^{238}U$ chain in the crystal is broken: activity of $^{238}U$ and $^{226}Ra$ in the crystals are substantially lower than that of $^{210}Po$. We assume that radioactive contamination of $CaWO_4$ crystal scintillators is coming mainly from $CaCO_3$ used to prepare raw compound for crystal growing. Such a conclusion we made taking into account extremely low radioactive contamination of $CdWO_4$ [20] and $ZnWO_4$ [21] crystals produced from practically the same tungsten oxide. Direct measurements of $CaCO_3$ and $WO_3$ compounds in the Solotvina Underground Laboratory with the help of low-background large volume $CdWO_4$ crystal scintillator have shown the same result: radioactive contamination of $CaCO_3$ is approximately 15 times higher than that of $WO_3$ [22].

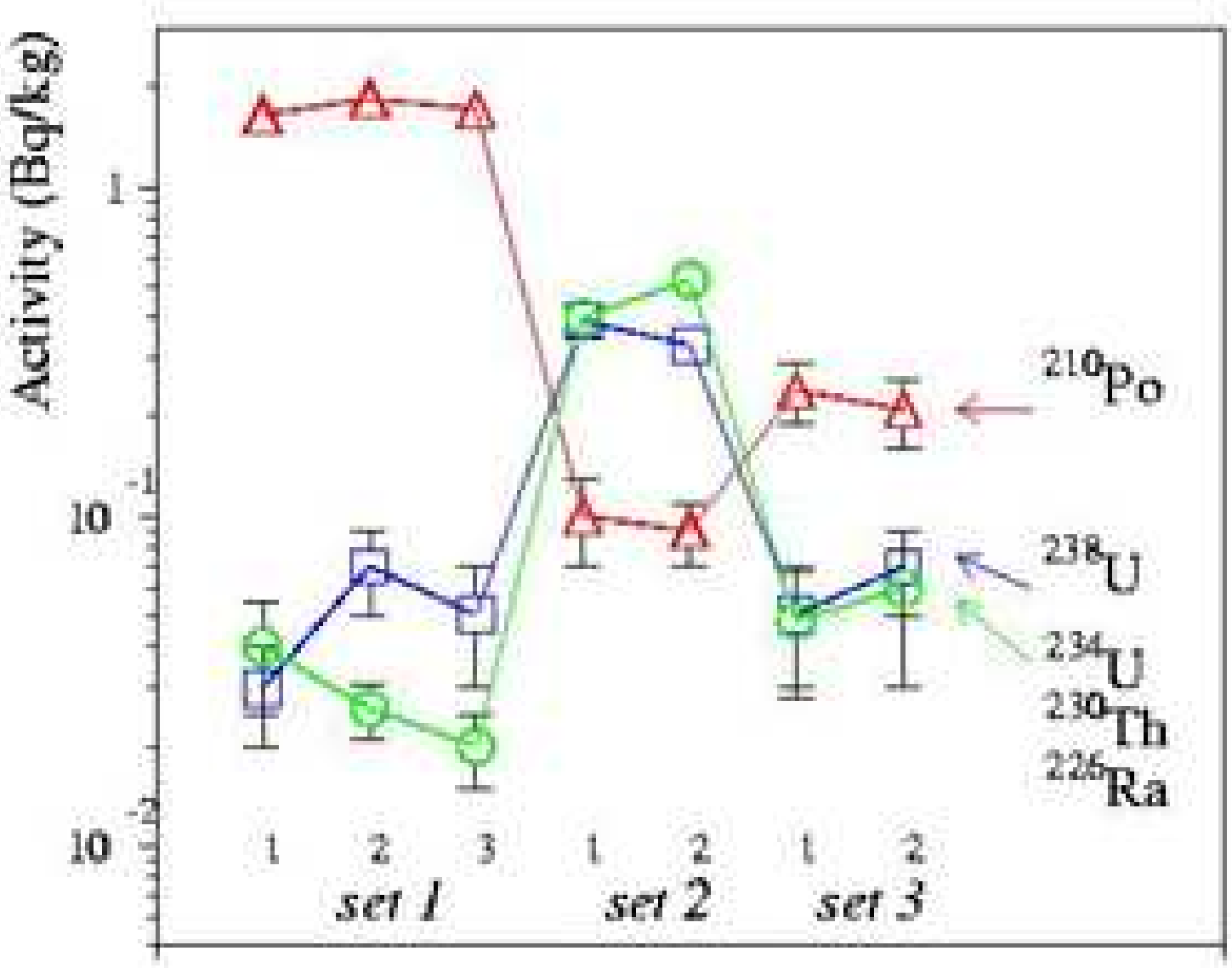

Fig. 6. Dependence of radioactive contamination of $CaWO_4$ crystals on growing conditions.

We do not observe a significant difference in contamination of samples from different parts of crystal boules.

It should be stressed we have not observed difference in radioactive contamination of the crystals of set 3 (CWO3-2 with removed ≈ 1 cm surface layer, and CWO3-1 with no machining of side surface). Such an effect was observed in $CdWO_4$ crystal scintillators [23].

A further improvement of radiopurity of $CaWO_4$ by a factor of ~ $10^4$ is necessary to satisfy requirements of the EURECA project. As a first step we are going to realize the following program:

1. Deep purification of raw materials is supposed to be the most important issue that needs addressing. Metal purification by vacuum distillation, zone melting, and filtering are very promising approaches, while further study is necessary for the purification of Ca in order to achieve the required low levels.
2. Two to five step recrystallization, involving inspection and assessment of the produced scintillators after each step.
3. Screening at all stages through ultra-low background γ, α, β spectroscopy is needed in the production of compounds for crystal growing (choice of raw materials, quality control of purified elements and compounds).

All work should be done using highly pure reagents, lab-ware and water. All chemistry should be done in a clean room, and, as far as possible, in nitrogen atmosphere. Careful protection from radon has to be foreseen at all stages of crystal production and storage.

It is important to stress that scintillation properties of the crystals of set 3 are much better (the energy resolution measured in low-background set-up with light-guide is FWHM≈8%) in comparison to set 1 (FWHM≈10–11%). One can conclude that the recrystallization procedure with the main aim to reach higher level of radiopurity will improve the quality of $CaWO_4$ crystal scintillators too.

## 5. Conclusions

Scintillation properties and radioactive contamination of seven samples of $CaWO_4$ crystal scintillators were measured in low-background set-up. The samples were made from three boules produced from the same raw material using recrystallization procedure, which lead to the significant difference in $^{210}$Po (in the range 0.03–1.32 Bq/kg), and $^{238}$U (0.04–0.33 Bq/kg) contaminations. Recrystallization improves radiopurity of $CaWO_4$ relatively to $^{210}$Po and $^{238}$U by one order of magnitude. The equilibrium of $^{238}$U chain is broken in the crystals. We have found that recrystallization procedure improves scintillation quality of $CaWO_4$ crystal scintillators too.

We have not observed difference in radioactive contamination of the crystals with removed ≈1 cm surface layer (in comparison to sample with no surface treatment), as well as in radioactive contamination of the crystals produced from one boule.

A program consisting in deep raw materials purification, a few step recrystallization with careful screening at all stages by ultra-low background α, β and γ spectrometry is proposed to obtain radiopure $CaWO_4$ crystal scintillators requested by the EURECA dark matter experiment.

## Acknowledgments

The support from the project “Kosmomikrofizyka” (Astroparticle Physics) of the National Academy of Sciences of Ukraine is gratefully acknowledged. The authors would like to thank V.V. Kobychev for Monte Carlo simulation with the help of GEANT4 package.

## References

1. V.B. Mikhailik, H. Kraus, J. Phys. D: Appl. Phys. 39 (2006) 1181.
2. A. Alessandrello et al., Phys. Lett. B 420 (1998) 109.
3. P. Meunier et al., Appl. Phys. Lett. 75 (1999) 1335.
4. S. Cebrián et al., Astropart. Phys. 21 (2004) 23.
5. J. Ninković et al., Nucl. Instr. Meth. A 537 (2005) 339.
6. G. Angloher et al., Astropart. Phys. 23 (2005) 325.
7. J. Jochum, Prog. Part. Nucl. Phys. 57 (2006) 357.
8. H. Kraus et al., J. Phys.: Conf. Series 39 (2006) 139.
9. H. Kraus et al., Nucl. Phys. B (Proc. Suppl.) 173 (2007) 168.
10. Yu.G. Zdesenko et al., Astropart. Phys. 23 (2005) 249.
11. Yu.G. Zdesenko et al., Nucl. Instrum. Meth. A 538 (2005) 657.
12. C. Cozzini et al., Phys. Rev. C 70 (2004) 064606.
13. I.M. Solskii, G. Stryganyuk, "Effect of impurity segregation on the properties of single-crystal scintillators", presented at Workshop on Radiopure Scintillators for EURECA (RPSCINT’2008), 9 - 10 September 2008, Institute for Nuclear Research, Kyiv, Ukraine.
14. L. Bardelli et al., Nucl. Instrum. Meth. A 584 (2008) 129.
15. R.B. Firestone et al., *Table of Isotopes*, 8th ed., John Wiley & Sons, New York, 1996 and CD update, 1998.
16. S. Agostinelli et al., Nucl. Instr. Meth. A 506 (2003) 250.
17. J. Allison et al., IEEE Trans. Nucl. Sci. 53 (2006) 270.
18. O.A. Ponkratenko, V.I. Tretyak, Yu.G. Zdesenko, Phys. At. Nucl. 63 (2000) 1282; V.I. Tretyak, to be published.
19. A.N. Annenkov et al., Nucl. Instrum. Meth. A 584 (2008) 334.
20. F.A. Danevich et al., Phys. Rev. C 68 (2003) 035501.
21. P. Belli et al., Preprint ROM2F/2008/22, submitted to Phys. Rev. C.
22. F.A. Danevich et al., LPD KINR technical report 1/2007 (unpublished).
23. F.A. Danevich et al., Phys. Lett. B 344 (1995) 72.

# Incidental radioactive background in BGO crystals

**D. Grigoriev[b*], G. Kuznetcov[a], I. Novoselov[a], P. Schotanus[c], B. Shavinski[a], S. Shepelev[a], V. Shlegel[a], Ya. Vasiliev[a]**

[a] *Nikolaev Institute of Inorganic Chemistry, SB RAS, 630090 Novosibirsk, Russia*
[b] *Budker Institute of Nuclear Physics, SB RAS, 630090 Novosibirsk, Russia*
[c] *SCIONIX Holland BV, 3980 CC Bunnik, The Netherlands*

In this paper some cases of unusual internal radioactive background in BGO crystals are described. Routinely produced at the Nikolaev Institute of Inorganic Chemistry during nearly quarter of century BGO crystals have low radioactive background, caused by $^{207}Bi$ contamination. But a few batches of BGO crystals incidentally have higher internal radioactive background with activity up to 10 Bk/kg related with other contaminations. One type of the background is pure alpha radioactivity. It is caused by $^{210}Po$ contamination and has technogenic origin. The other background is identified as gammas coming from short living $^{214}Bi$ and $^{214}Pb$ isotopes. It indicates the pollution of crystals by long living isotopes, most likely $^{226}Ra$. Unusually high rate suggests a high probability of technogenic origin of this background too.

## 1. Introduction

BGO ($Bi_4Ge_3O_{12}$) is one of the most commonly used scintillating crystals. Its components have not long living radioactive isotopes. Recently discovered instability of $^{209}Bi$ itself causes a negligible background rate because of very long decay time ($T_{1/2}=2\times10^{19}$ years) [1]. The main source of the internal radioactive background of the BGO crystals is contamination by $^{207}Bi$ with typical activity 1–3 Bk/kg [1, 2, 3, 4]. It corresponds to $^{207}Bi$ concentration of order of $10^{-15}$ g/g. A study of the dependence of $^{207}Bi$ contamination on the source of bismuth shows that using bismuth from lead ore mines creates an order of magnitude higher gamma background compared to lead free mines [2]. Therefore the author suggests the hypothesis of $^{207}Bi$ production via interaction of the cosmic rays protons with lead $^{206}Pb+p\rightarrow^{207}Bi$. A measurement of $^{207}Bi$ background on earth surface and in deep salt mine shows the same rate [3]. It eliminates the hypothesis of gamma pollution from direct activation of $^{209}Bi$ in a crystal by cosmic muons via reaction $^{209}Bi(\mu^-,2n)^{207m}Pb$ ($T_{1/2}=0.8$ s). The upper limits on the contamination by $^{238}U$ and $^{232}Th$ are set as low as $10^{-11}$ g/g [3]. It corresponds to 2–3 orders of magnitude lower radioactive background compared to $^{207}Bi$ activity.

BGO crystals are grown at the Nikolaev Institute of Inorganic Chemistry (NIIC) from beginning of 1980-s [5, 6]. Contamination by $^{207}Bi$ is usually low. However some crystal batches have a substantial intrinsic radioactive background from other sources. In this paper, the types of backgrounds have been identified as gamma emitting isotopes for some crystals and pure alpha radioactivity for others.

## 2. Gamma radioactive background

The main source of gammas in $^{207}Bi$ decay is cascade transition $^{207m}Pb\rightarrow{}^{207}Pb$. It gives single peaks at 570, 1063 keV and sum peak at 1633 keV in BGO background spectrum.

The high background rates about 10 Hz/kg with lines similar to $^{207}Bi$ were observed in one batch of the BGO crystals in 1999. The gamma energies from $^{214}Bi$ decays (609, 1120 and 1764 keV) are close to ones from $^{207}Bi$ (Fig. 1). The lower energy lines from $^{214}Pb$ decays in gamma spectra are hidden under Compton tails and other low energy background. So it is difficult to discriminate those isotopes from intrinsic low rate background spectra of BGO crystals. The gamma spectrum from intrinsic background radiation of the BGO crystal has been measured with the low background HPGe detector. The lines 350 keV from $^{214}$Pb and 609 keV from $^{214}$Bi are clearly seen. No associated with $^{207}$Bi gamma lines have been observed in the spectrum.

[*] Corresponding author. *E-mail address:* D.N.Grigoriev@inp.nsk.su

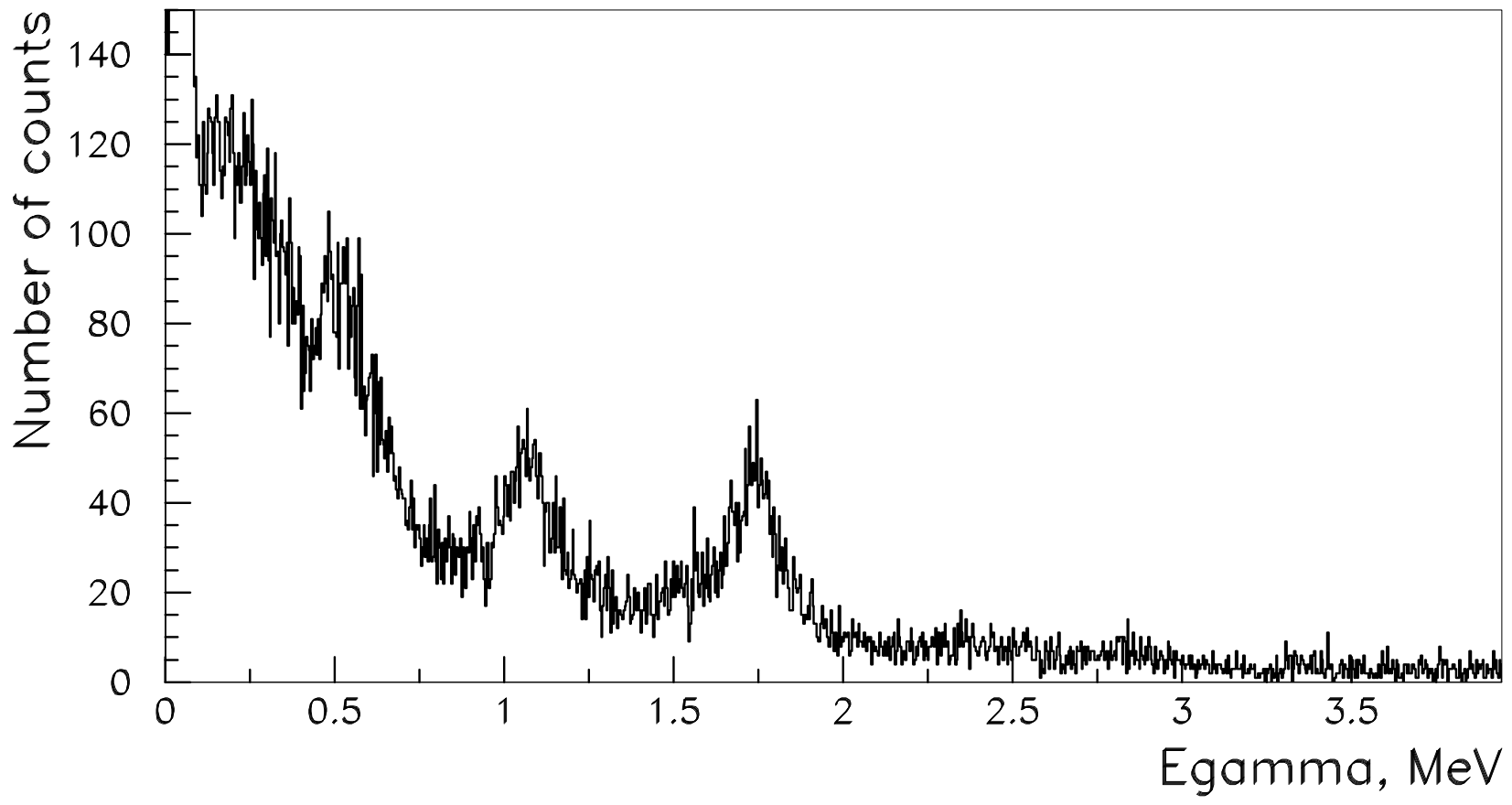

Fig. 1 Sample of intrinsic gamma background in BGO crystal.

Both $^{214}$Pb and $^{214}$Bi isotopes are far from stable isotopes and have short decay time ($T_{1/2}$= 27 and 20 minutes respectively) to be produced by cosmic ray interactions. Besides they are part of the natural radioactive background coming from $^{238}$U decay chain. Therefore a presence of $^{214}$Pb and $^{214}$Bi isotopes indicates a pollution of the crystals by long living isotopes from $^{238}$U decay chain. To explain observed gamma background rate the $^{238}$U concentration should be of order of $10^{-6}$–$10^{-7}$ g/g. It is unlikely to be true. Such level of contamination will affect optical properties of a crystal. More reasonable hypothesis is $^{226}$Ra contamination at level of a few units of $10^{-13}$ g/g. The alphas from radium decay chain give energy deposition in BGO equivalent to slightly more than 1 MeV gammas because of quenching factor of about 5. Therefore alpha lines merge with the 1120 keV gamma line.

We suggest the following hypothesis. During raw material purification and crystallization processes bismuth is cleaned from uranium and thorium but not from radium. An indirect support of this hypothesis comes from BGO crystal alpha background spectrum in Ref. [1]. There are no $^{238}$U and $^{235}$U lines and only trace amount of $^{232}$Th in this spectrum. But alpha lines from radium isotopes and their decay products are clearly seen. Anyway the pollution by natural radioactive isotopes in this crystal is orders of magnitude higher than usually. A source of pollution is untraceable but its large level suggests high probability of its technogenic origin.

### 3. Pure alpha radioactive background

A background counting rate of about 10 Hz/kg was observed in some crystals. The spectrum shows a clear peak with energy deposition equivalent to about 1 MeV gammas (Fig. 2). This peak has no associated Compton part of spectrum. It indicates not a gamma origin of the peak but alpha.

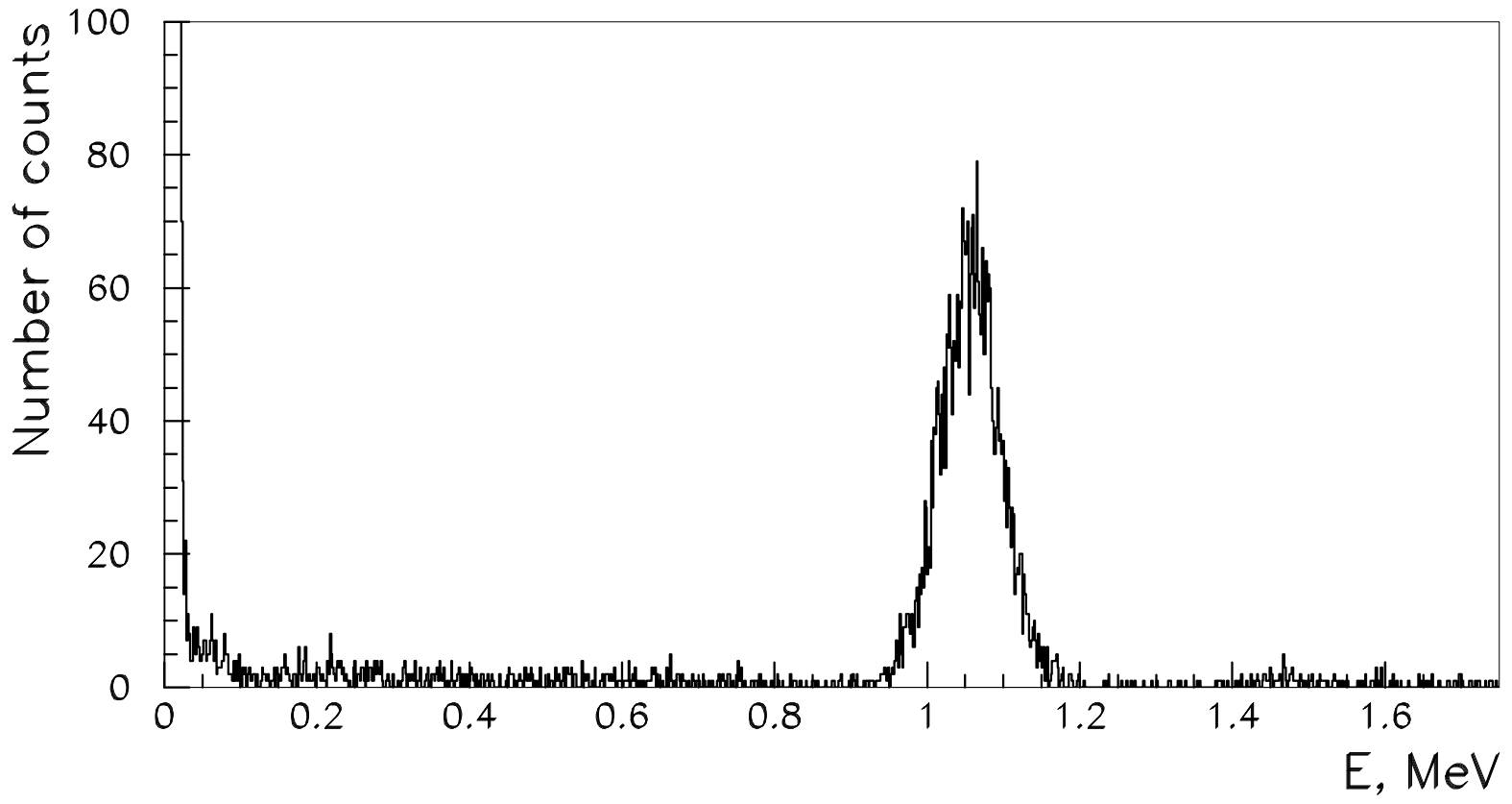

Fig. 2. Intrinsic BGO background spectrum.

To check the alpha hypothesis a background free crystal has been coupled to a PMT and the suspected crystal has been placed on top of it as radioactive source (Fig. 3). No additional rate was observed. Measurement with a low background HPGe detector shows no additional gamma background too. It demonstrates that radiation particles do not leave the crystal. That supports the alpha origin of the background.

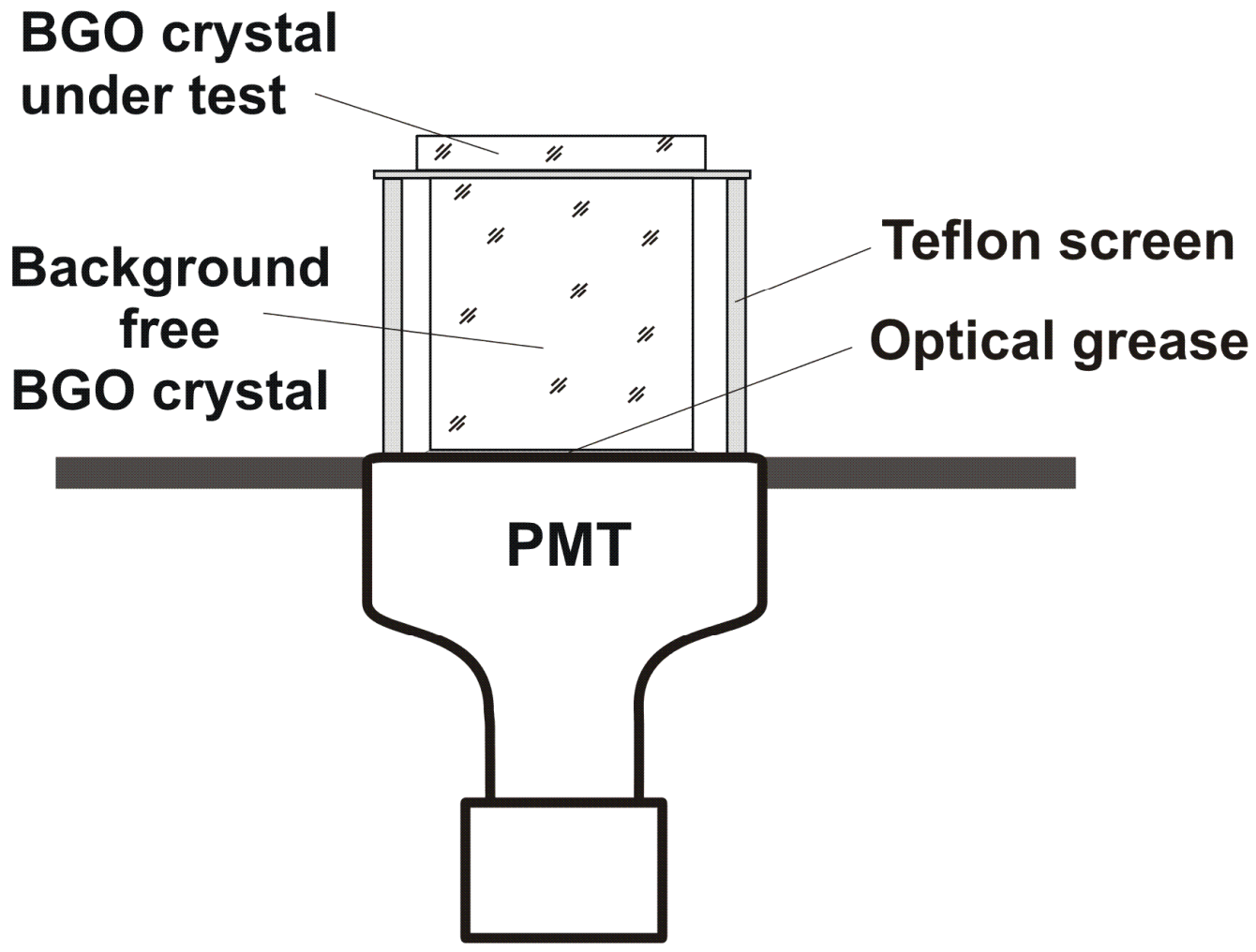

Fig. 3. Scheme of a background set-up with 2 BGO crystals.

The typical energy of alpha particles is several MeV. But response to alpha particles in heavy scintillating crystals is much lower compared to gammas because of quenching effect. This effect is nonlinear. So the peak position cannot be used to identify source of alpha particles by measurement of an energy deposition in BGO crystal. On another hand the absence of gamma rays requires that radioactive series must be short, preferably consisting of one nuclide. The good candidate is $^{210}$Po isotope. The measured with alpha spectrometer energy of emitted alphas is 5.3 MeV in accordance with $^{210}$Po radiation. The observed background rate corresponds to $^{210}$Po pollution lower than $10^{-16}$ g/g.

The absence of gamma rays shows rather neutron activation origin of $^{210}$Po isotope due to reaction chain $^{209}$Bi+n → $^{210}$Bi ($\beta^-$, 5 days) → $^{210}$Po than natural radioactivity. The $^{214}$Bi and $^{214}$Pb gamma lines must be present in case of uranium or radium origin of the polonium. Pure Bi metal for Bi oxide production is refined at NIIC from commercial Bi, having 99.9% purity. Residuals from electrolyzes (one stage of refining) are enriched by possible radioactive impurities. The gamma spectrum of selected residuals was measured with Ge-Li detector. Many associated with $^{152}$Eu lines are seen (Fig. 4). This isotope has half life of 13 years and is absent in natural isotope composition.

But $^{151}Eu$ has large cross section of neutron activation (σ=9200 b). So the tiny amount of $^{151}Eu$ can produce substantial radioactive background after neutron irradiation. The presence of the radioactive europium isotope confirms hypothesis of technogenic origin of the $^{210}Po$ contamination.

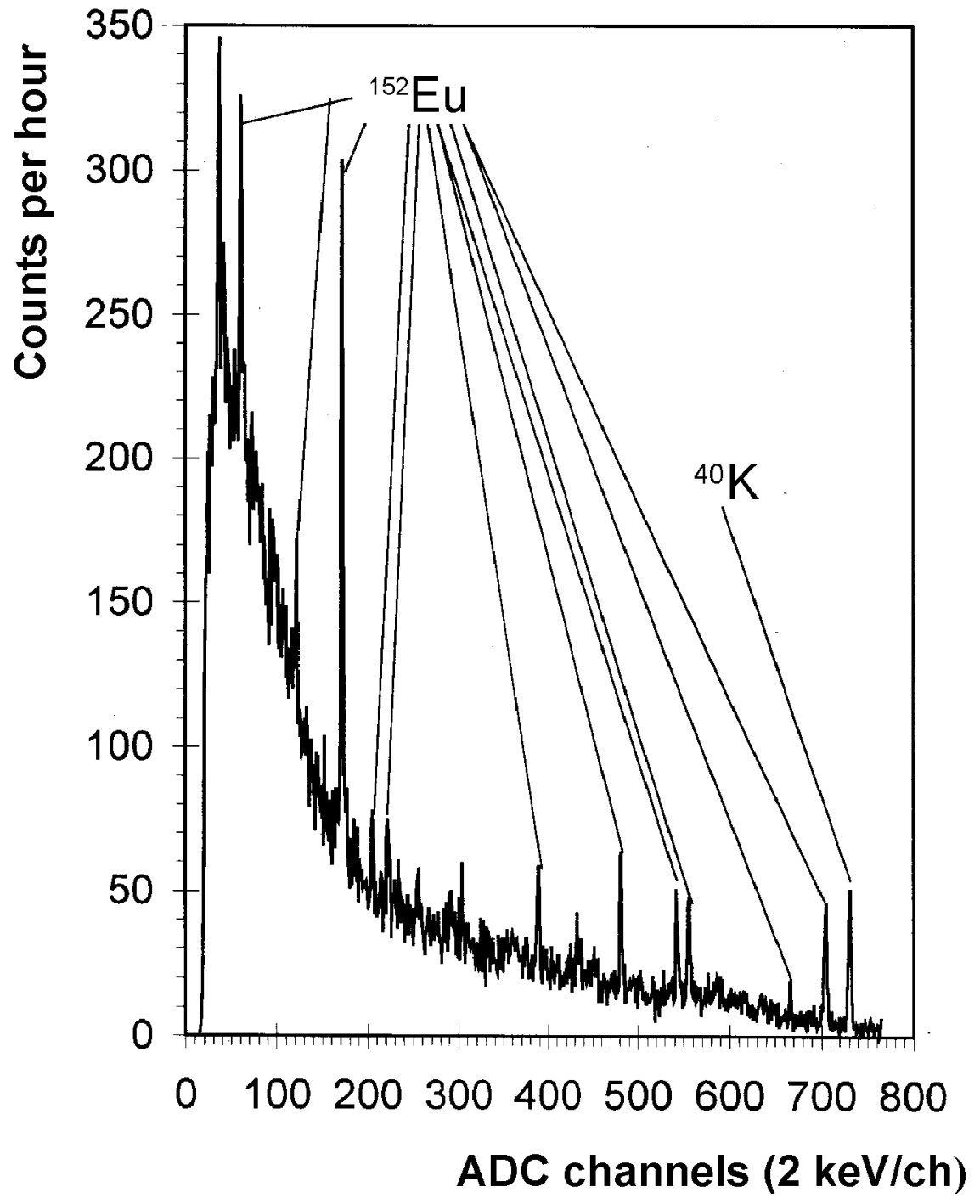

Fig. 4. Gamma spectrum of Bi electrolyzes residuals measured with Ge-Li detector.

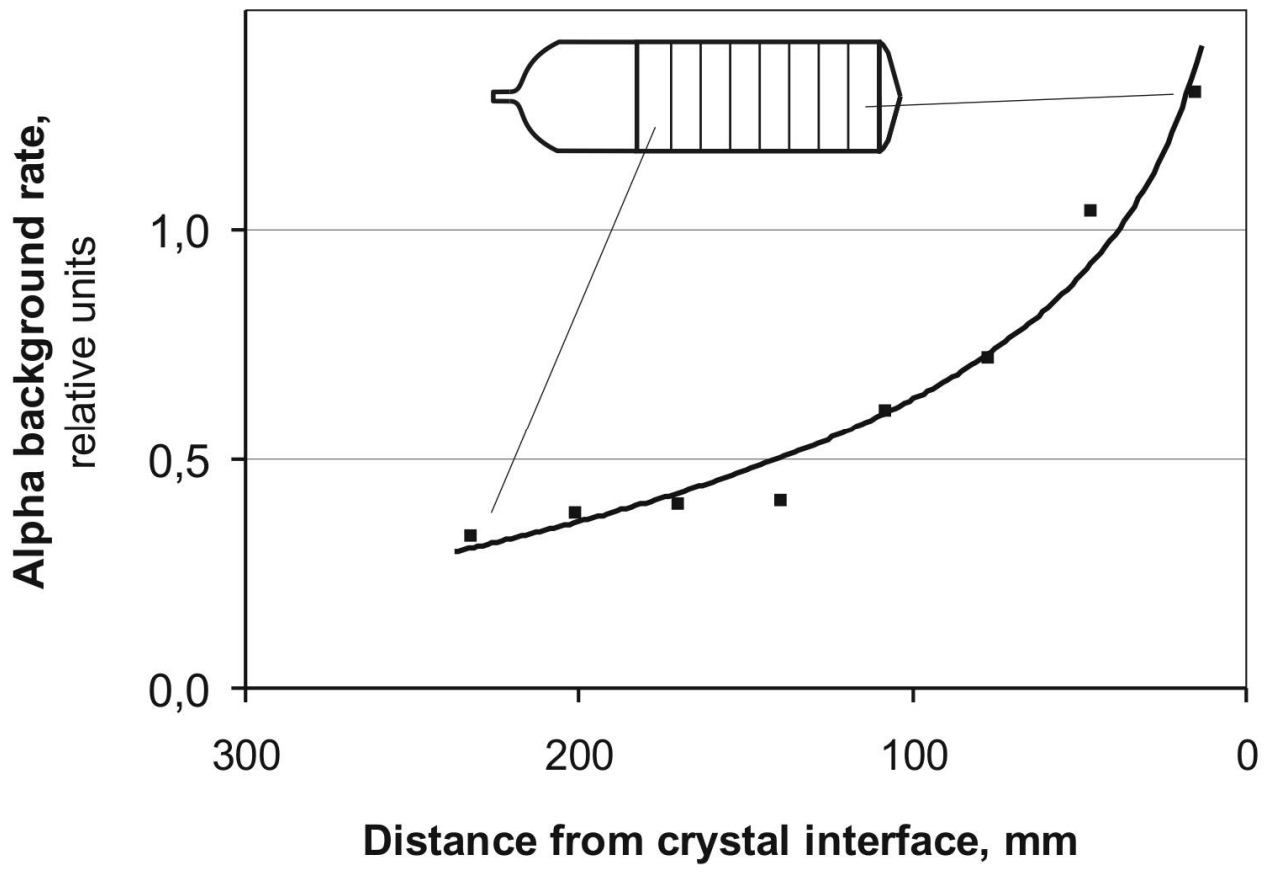

Fig. 5. Alpha background rate vs distance from crystal interface.

The measurements of alpha background rate along the crystal show an increasing rate from beginning to the end of crystal (Fig. 5). It means enhancement of the $^{210}Po$ concentration in melt and decreasing in grown crystals. Therefore BGO crystals could be purified from $^{210}Po$ contamination by recrystallization. To check this, high alpha background BGO crystals were selected and used as charge to grow a new BGO crystal. The background activity of re-grown crystal is about 3.5 times less than that of initial crystals. So recrystallization of BGO crystals decreases alpha background.

So far we know there is no literature data on this type of BGO radioactivity. First time it was seen in 2003. However, it happens more and more frequently later and can become a problem for low background experiments. A probable reason is following. The main source of bismuth is lead production industry. Production of lead decreases now because of ecology reasons [7]. Therefore more and more bismuth on market comes from recycling and not from primary production. The bismuth is widely used at nuclear power stations. Observed radioactivity is small from point of view of safety rules. So polluted by $^{210}Po$ bismuth can come from utilization of used Bi at nuclear power stations.

## 4. Conclusion

The BGO crystals grown in the Nikolaev Institute of Inorganic Chemistry have a low intrinsic radiation background caused by $^{207}Bi$ contamination. However a small fraction of the crystals have a substantially higher background counting rate. Two types of background were found. The first one is a pure alpha background caused by $^{210}Po$ isotope contamination. This contamination comes most probably from neutron irradiation. So it has technogenic origin. The second one is gamma background coming from short living $^{214}Pb$ and $^{214}Bi$ isotopes. They are part of natural radioactive background and indicate the presence of long living isotopes, most likely $^{226}Ra$. Moreover, the unusually large concentration of those isotopes suggests a high probability of technogenic origin. The precise sources of radioactive contamination are untraceable. One possible explanation is the recycling of bismuth used at nuclear power stations. Primary bismuth from lead free ore mines should be used to grow BGO crystals for low background applications.

## References

1. P. de Marcillac et al., Nature 422 (2003) 876. See also Supplementary Information.
2. T.A. Lewis, Nucl. Instrum. Meth. A 264 (1988) 534.
3. A.Ya. Balysh et al., Instrum. Exp. Tech. 1 (1993) 118.
4. F.A. Danevich, Proc. 8th Int. Conf. on Inorganic Scint. and their Use in Sci. and Industrial Applications (SCINT'2005), Alushta, Crimea, Ukraine, September 19-23, 2005, p. 403.
5. Ya.V. Vasiliev et al., Nucl. Instrum. Meth. A 379 (1996) 533.
6. Yu.A. Borovlev et al., J. Crystal Growth 229 (2001) 305.
7. A.V. Naumov, Russ. J. Non-Ferrous Metals 48 (2007) 10.

# Radiopurity of $ZnWO_4$ crystal scintillators

**D.V. Poda***

*Institute for Nuclear Research, MSP 03680 Kyiv, Ukraine*

Recently $ZnWO_4$ was proposed as a perspective material for the low counting experiments to search for dark matter and double beta decay. Such experiments demand high radiopurity of $ZnWO_4$ crystal scintillators. Radiopurity of a $ZnWO_4$ scintillator (produced in the Institute for Scintillation Materials, Kharkiv, Ukraine) was measured in the Solotvina Underground Laboratory at a depth of ≈1000 m w.e. The radioactive contaminations of the $ZnWO_4$ sample (26 × 24 × 24 mm) don't exceed 0.1–10 mBq/kg (depending on radionuclide). Taking into account good scintillation properties at low temperatures, $ZnWO_4$ crystal scintillator is one of the best candidates for a cryogenic double beta decay and dark matter experiments.

## 1. Introduction

Zinc tungstate ($ZnWO_4$) is very promising scintillator material. It has been proposed as a detector to search for double beta decay [1]. The first low-background measurement with the small $ZnWO_4$ sample (mass of 4.5 g) was performed in the Solotvina Underground Laboratory (Ukraine) in order to study its radioactive contamination, and to search for double beta decay of zinc and tungsten isotopes [2]. In addition, possibilities to apply $ZnWO_4$ crystals for dark matter search were discussed in this work. Luminescence of $ZnWO_4$ at low temperature was studied in Ref. [3]. Authors consider $ZnWO_4$ crystal scintillator as a good material for cryogenic WIMP dark matter experiments. The results of Refs. [2] and [3] gave reasons for extensive R&D in the Institute for Scintillation Materials (ISMA, Kharkiv, Ukraine) in order to produce high quality large-volume $ZnWO_4$ crystal scintillators [4, 5]. As a result of these studies, the large volume $ZnWO_4$ crystals with the improved scintillation properties were developed [4, 5]. Two large volume $ZnWO_4$ samples (with masses of 0.12 kg and 0.7 kg) grown in the ISMA were used in the low background experiment at the underground Gran Sasso National Laboratories of the INFN (Italy) to search for double beta processes in Zn and W isotopes [6, 7]. The new improved half-life limits on the 2β processes in $^{64,70}$Zn, $^{180,186}$W have been set, in particular, the half-life limits on double electron capture and electron capture with positron emission in $^{64}$Zn were established as: $T_{1/2}^{2\nu 2K} \geq 6.2 \times 10^{18}$ yr, $T_{1/2}^{0\nu 2\varepsilon} \geq 1.1 \times 10^{20}$ yr, $T_{1/2}^{2\nu\varepsilon\beta^+} \geq 7.0 \times 10^{20}$ yr, and $T_{1/2}^{0\nu\varepsilon\beta^+} \geq 4.3 \times 10^{20}$ yr (all the limits are at 90% C.L.) [6, 7]. Further, investigations of $ZnWO_4$ crystals as scintillating bolometers for dark matter search experiments have been performed recently [8, 9, 10]. All these studies are very important to choice the $ZnWO_4$ crystals as a promising target for cryogenic dark matter search experiment, in particular for EURECA[1] [11, 12], where a multi-element target with the total mass up to 1 t is planned for confirming a dark matter signal.

In the present work, radiopurity of $ZnWO_4$ crystal scintillators for the next generation double beta decay and cryogenic dark matter experiments is discussed. The results were previously published in [4, 6, 7].

* Corresponding author. *E-mail address*: poda@kinr.kiev.ua

[1] European Underground Rare Event Calorimeter Array; www.eureca.ox.ac.uk

## 2. Measurements and results

Radioactive contamination of a clear, slightly pink coloured $ZnWO_4$ crystal scintillator ($26 \times 24 \times 24$ mm, mass of 119 g), produced in the ISMA from monocrystal grown by the Czochralski method was investigated in the Solotvina Underground Laboratory at a depth of ≈1000 m w.e. The scintillator was viewed by a special low radioactive 5" PMT (EMI D724KFLB) through a high pure quartz light-guide 10 cm in diameter and 33 cm long. The crystal and the light-guide were wrapped by the PTFE reflection tape. The detector was surrounded by a passive shield made of teflon (3–5 cm), plexiglass (6–13 cm), high purity copper (3–6 cm) and lead (15 cm). For each event in the detector, the amplitude of a signal and its arrival time were recorded. In addition, scintillation pulses of the $ZnWO_4$ scintillator were digitized with a 20 MHz sampling frequency. The energy spectrum measured with the $ZnWO_4$ crystal scintillator during 44.7 h in the low background set-up is presented in Fig. 1.

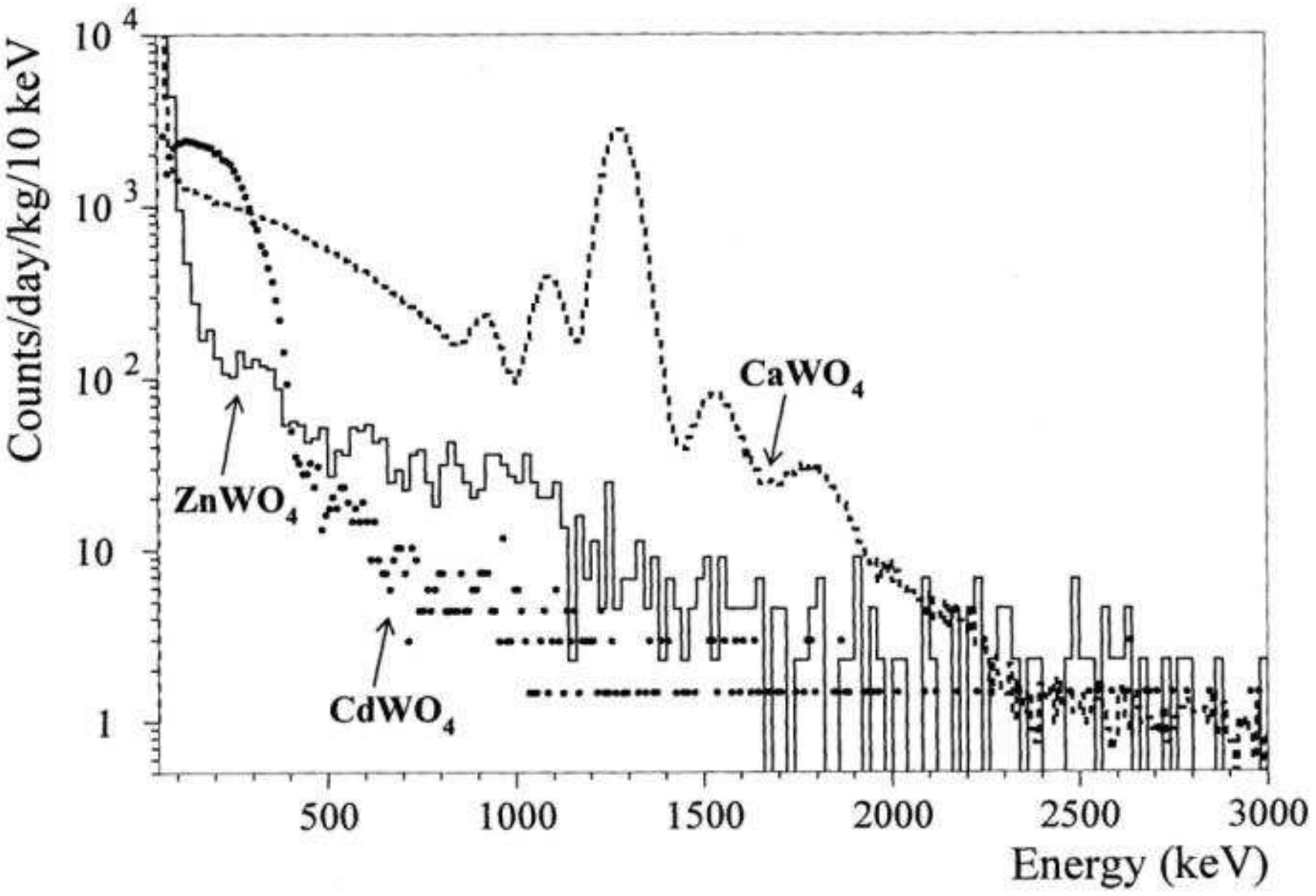

Fig. 1. Energy spectra of $ZnWO_4$ (119 g, 44.7 h), $CaWO_4$ (189 g, 1734 h), and $CdWO_4$ (448 g, 37 h) scintillation crystals measured in the low background set-up. The $CaWO_4$ crystal is considerably polluted by radionuclides from U/Th chains. Beta decay of $^{113}$Cd dominates in the low energy part of the $CdWO_4$ spectrum. The background of the $ZnWO_4$ detector is caused mainly by external $\gamma$ rays.

The spectra of widely used calcium tungstate ($CaWO_4$) and cadmium tungstate ($CdWO_4$) scintillators measured in similar conditions are given for comparison (the spectra are normalized to the measurement time and the detector mass). The background of the $ZnWO_4$ detector is substantially lower than that of $CaWO_4$ and is comparable with that of $CdWO_4$ above ≈ 0.5 MeV. Note, that below 0.5 MeV the counting rate of the $ZnWO_4$ detector is one order of magnitude lower than that of $CdWO_4$. Obviously, it is due to presence in the $CdWO_4$ crystals of the $\beta$ active $^{113}$Cd isotope (natural abundance 12.22%; $Q_\beta$ = 320 keV; $T_{1/2} = 8.04 \times 10^{15}$ yr [13]).

There are no peculiarities in the spectrum measured with $ZnWO_4$ detector (see Fig. 1) which can be interpreted as certain radioactivity. Therefore, only limits on contaminations of the crystal by nuclides from U/Th families as well as by $^{40}$K, $^{90}$Sr-$^{90}$Y, $^{137}$Cs, and $^{147}$Sm were set on the basis of the experimental data. With this aim the background spectrum was fitted in different energy intervals by simple model composed of an exponential function (to describe external $\gamma$ rays) and background components searched for. The latter were simulated with the GEANT4 package [14]. The initial kinematics of the particles emitted in the decay of nuclei was given by an event generator DECAY0 [15]. The activities of $^{40}$K, $^{232}$Th, and $^{238}$U inside the PMT were taken from Ref. [16]. Because equilibrium of U/Th families in crystals is expected to be broken, different parts of the families ($^{208}$Tl, $^{210}$Bi, $^{214}$Bi, $^{234m}$Pa) were considered separately. The pulse-shape discrimination between $\gamma/\beta$ events and $\alpha$ particles [2, 17] was used to estimate the total $\alpha$ activity of U/Th, while the fast chains $^{214}$Bi $\rightarrow$ $^{214}$Po$\rightarrow$ $^{210}$Pb (it gives activity of $^{226}$Ra from $^{238}$U family) and $^{220}$Rn $\rightarrow$ $^{216}$Po$\rightarrow$ $^{212}$Pb ($^{228}$Th from $^{232}$Th) were selected with the help of the time-amplitude analysis [17, 18].

The summary of the measured radioactive contamination of the $ZnWO_4$ scintillator (or limits on their activities) is given in Table 1 in comparison with the results presented in [2], and with the present-day results of radiopurity of $CaWO_4$, and $CdWO_4$ crystal scintillators.

Table 1. The radioactive contamination of $ZnWO_4$, $CaWO_4$, and $CdWO_4$ crystal scintillators.

| Chain | Source | Activity (mBq/kg) | | | |
|---|---|---|---|---|---|
| | | $ZnWO_4$ Present study | $ZnWO_4$ [2, 19] | $CaWO_4$ [19, 20] | $CdWO_4$ [13, 17, 19] |
| $^{232}$Th | $^{228}$Th | ≤ 0.1 | ≤ 0.2 | = 0.6(2) | ≤ 0.004 – = 0.039(2) |
| $^{238}$U | $^{226}$Ra | ≤ 0.16 | ≤ 0.4 | = 5.6(5) | ≤ (0.004 – 0.04) |
| Total $\alpha$ activity | | = 2.4(3) | ≤ 20 | = 20 – 400 | = 0.26(4) – 2.3(3) |
| | $^{40}$K | ≤ 14 | ≤ 12 | ≤ 12 | = 0.3(1) |
| | $^{90}$Sr | ≤ 15 | ≤ 1.2 | ≤ 70 | ≤ 0.2 |
| | $^{137}$Cs | ≤ 2.5 | ≤ 20 | ≤ 0.8 | ≤ 0.3 – = 0.43(6) |
| | $^{147}$Sm | ≤ 5 | ≤ 1.8 | = 0.49(4) | ≤ (0.01 – 0.04) |

The limits on activities of $^{40}$K and $^{147}$Sm obtained in the present study are worse than the limits from [2]. It is due to slightly higher background of the $26 \times 24 \times 24$ mm detector near the energy of 0.3 MeV (where alpha peak of $^{147}$Sm is expected), and in the energy region of 0.5 – 0.9 MeV where the maximum of $^{40}$K spectrum is expected.

## 3. Conclusions

The radioactive contamination of large volume $ZnWO_4$ crystal scintillator ($26 \times 24 \times 24$ mm, mass of 119 g) has been measured in the low background set-up at the Solotvina Underground Laboratory [4]. Only upper limits at the level of ~0.1 mBq/kg could be set for the contamination by nuclides from the U/Th families, while radioactive contamination by $^{40}$K, $^{90}$Sr-$^{90}$Y, $^{137}$Cs, and $^{147}$Sm were found to be less than a few mBq/kg. Alpha activity at the level of 2.4 mBq/kg (daughters of U/Th) was detected in this scintillator [4].

In addition, we wish to refer to the recent measurements of the intrinsic radioactivity, which has been performed in the mentioned double beta decay experiment with large volume (0.1–0.7 kg) $ZnWO_4$ scintillators at the underground Gran Sasso National Laboratories of the INFN [7]. Pulse-shape discrimination technique and time-amplitude analysis were used to determine very low activity at the level of $\mu$Bq/kg ($^{228}$Th and $^{226}$Ra), and total $\alpha$ activity (U/Th) 0.2–0.4 mBq/kg [7].

Thus, summarizing the results of measurements of the radioactive contaminations of the $ZnWO_4$ samples we can conclude that $ZnWO_4$ crystal scintillators are extremely radiopure detectors.

## Acknowledgments

These studies were supported in part by the project "Kosmomikrofizyka" (Astroparticle Physics) of the National Academy of Sciences of Ukraine. Author is grateful to his colleagues from the Lepton Physics Department of the Institute for Nuclear Research (Kyiv, Ukraine) and the DAMA group from the National Institute of Nuclear Physics (Instituto Nazionale di Fisica Nucleare, Italy) for common work in the measurements of radioactive contamination of $ZnWO_4$ and for useful discussions. The author is thankful to the group from Department of Technology of Monocrystals's Growth of the Institute for Scintillation Materials (Kharkiv, Ukraine) for the development and growth of the radiopure large volume $ZnWO_4$ samples with high scintillation properties.

## References

1. F.A. Danevich et al., Prib. Tekh. Eksp. 5 (1989) 80 [Instrum. Exp. Tech. 32 (1989) 1059].
2. F.A. Danevich et al., Nucl. Instr. Meth. A 544 (2005) 553.
3. H. Kraus et al., Phys. Lett. B 610 (2005) 37.
4. L.L. Nagornaya et al., IEEE Trans. Nucl. Sci. 55 (2008) 1469.
5. L.L. Nagornaya et al., IEEE Trans. Nucl. Sci., to be published.
6. P. Belli et al., Phys. Lett. B 658 (2008) 193.
7. P. Belli et al., Preprint ROM2F/2008/22; arXiv:0811.2348v1 [nucl-ex], submitted to Phys. Rev. C.
8. I. Bavykina et al., IEEE Trans. Nucl. Sci. 55 (2008) 1449.
9. H. Kraus et al., Nucl. Instr. Meth. A, in press.
10. I. Bavykina et al., Opt. Mat., in press.
11. H. Kraus et al., J. Phys. Conf. Series 39 (2006) 139.
12. H. Kraus et al., Nucl. Phys. B (Proc. Suppl.) 173 (2007) 168.
13. P. Belli et al., Phys. Rev. C 76 (2007) 064603.
14. S. Agostinelli et al., Nucl. Instrum. Meth. A 506 (2003) 250;
    J. Allison et al., IEEE Trans. Nucl. Sci. 53 (2006) 270.
15. O.A. Ponkratenko, V.I. Tretyak, Yu.G. Zdesenko, Phys. At. Nucl. 63 (2000) 1282;
    V.I. Tretyak (to be published).
16. F.A. Danevich et al., Nucl. Phys. A 643 (1998) 317.
17. F.A. Danevich et al., Phys. Rev. C 68 (2003) 035501.
18. F.A. Danevich et al., Nucl. Phys. A 694 (2001) 375.
19. F.A. Danevich et al., AIP Conf. Proc. 785 (2005) 87.
20. Yu.G. Zdesenko et al., Nucl. Instr. Meth. A 538 (2005) 657.

# Production of high-purity metals

**G.P. Kovtun[a*], A.P. Shcherban'[a], D.A. Solopikhin[a], V.G. Glebovsky[b]**

*[a] National Science Center "Kharkov Institute of Physics and Technology" (NSC KIPT), 1, Academicheskaya str., Kharkov 61108, Ukraine*
*[b] Institute of Solid State Physics, Russian Academy of Sciences, Chernogolovka, Russia*

The interest in high-purity metals, first and foremost, is associated with their applications in atomic energy production, microelectronics, space engineering, medicine and fundamental science researches.

There are several reasons which restrain the deep purification of metals. Factors of the first group are associated with peculiarities of the behavior of impurities in metal being purified, their mutual interactions and interactions with the base material. The second, most substantial group, is related with infusion of impurities from structural materials in the processes of purification. The analysis indicates that, while the impurity comes in from the contact devices, the topmost impurity concentration does not depend on its content in the original substance.

One of the most radical ways of enhancing the efficiency of metal purification is to employ consecutively a number of refinement techniques with different mechanisms of impurity separation. In this case, a more effective separation of different classes of impurities is to be expected in comparison when only one technique, although very effective, is used many times.

The present paper brings forward the results of our R&D on obtaining some of high-purity metals (Cr, Ga, Cd, Zn, Te) by heating and distillation.

The physical-chemical foundation of the distillation technique processes is based on metal separation during evaporation (condensation) owing to the difference in pressure of saturated vapor [1, 2]. This separation is characterized by the separation coefficient α.

In the course of the molecular in-vacuum evaporation, when the evaporated components do not return to the melt, with the molar concentrations of the components being proportional to the molar evaporation rates, the separation degree is determined by the expression:

$$\alpha_i = \frac{P_A^0}{P_B^0}\frac{\gamma_A}{\gamma_B}\sqrt{\frac{M_B}{M_A}}, \qquad (1)$$

where $P_A^0, P_B^0, \gamma_A, \gamma_B, M_A, M_B$ – the vapor elasticity, molar fraction, activity coefficient and molar weight of the base material A and impurity B.

In accordance with the main postulates of the metal distillation theory [1-3] and considering the low impurity content in the metal refined, simplified equations were derived in order to assess the content variation of metallic impurities in the initial and condensed metal upon the in-vacuum distillation:

$$\frac{X_1}{X_0} = \left(\frac{G_1}{G_0}\right)^{\alpha_I - 1}, \qquad (3)$$

$$\frac{X_k}{X_0} = \frac{1-(1-G_k/G_0)^{\alpha_I}}{G_k/G_0}, \qquad (4)$$

[*] Corresponding author. *E-mail address:* gkovtun@kipt.kharkov.ua

where $X_0$ and $X_1$ – the initial and final content of the impurity in the original metal; while $X_k$ stands for content in the condensate; $G_0$, $G_1$, $G_k$ – the initial metal masses (in the beginning and in the end of the process) and that of the condensate, respectively, $\alpha_i$ – the separation coefficient of ideal binary alloy.

As a rule, most of the metals that are subjected to refinement via the in-vacuum distillation techniques possess high-volatile and low-volatile groups of impurities. Efficiency of the metal purification is considerably enhanced, if one sequentially removes various groups of impurities with different volatility. An example of such an approach would be deep purification of chromium, gallium, zinc, cadmium, tellurium via combination of the heating and distillation in vacuum [4-8].

Fig. 1 shows the chromium distillation device. The chromium evaporation was made with the aid of electron beam heating from the mesh crucible (1) that was made of refractory metal wiring (Ta, Nb, Mo). At first, the crucible containing chromium was placed off the condensation zone (Position I) in order to remove the highly volatile impurities at T ~ 1270 K. Afterwards, the crucible was placed in the condensation zone (Position II), where at T ~ 1620 K chromium was abstracted to 70…80% with subsequent condensation onto the gradient temperature column (5).

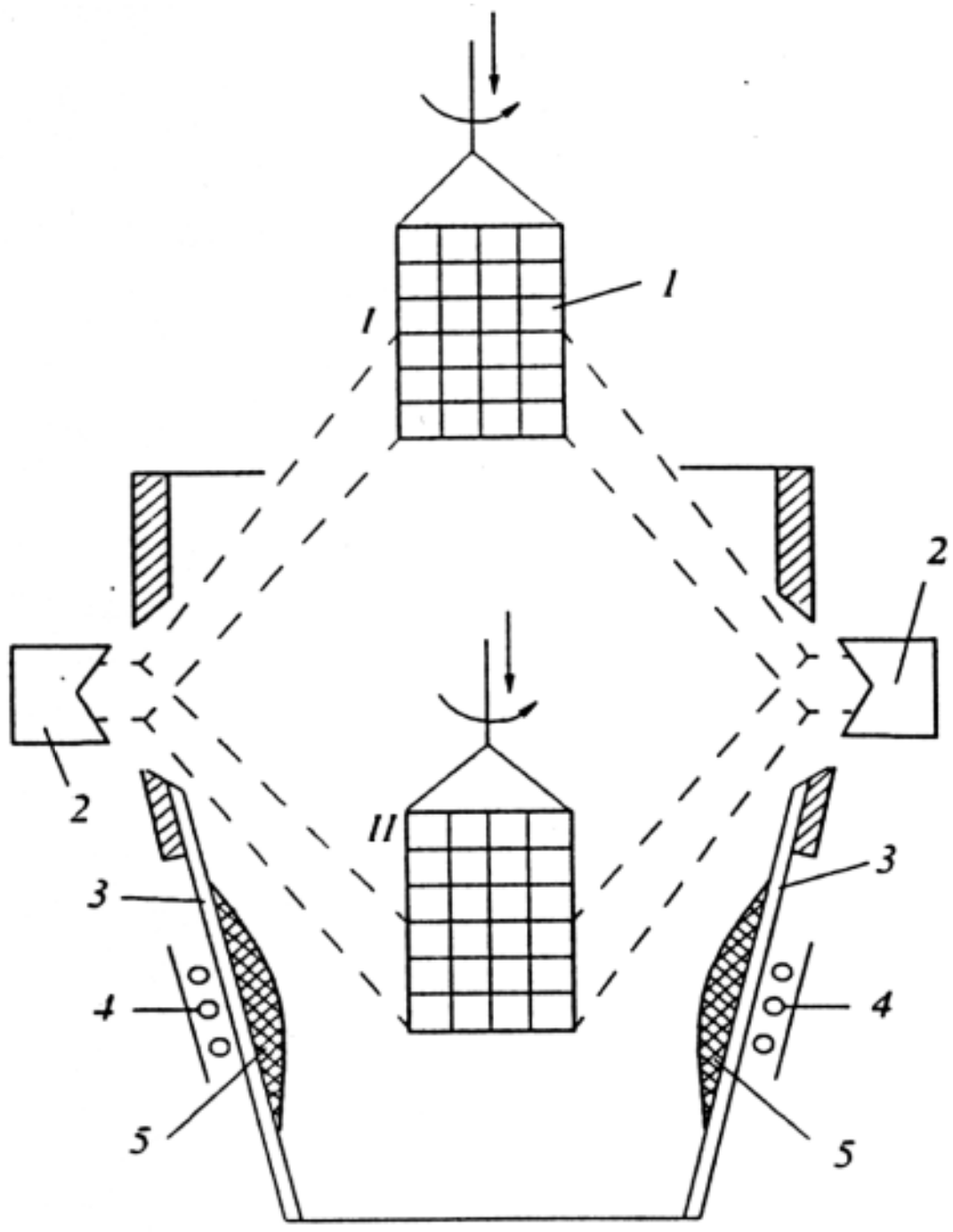

Fig. 1. Scheme of device for chromium refinement via heating (I) and in-vacuum sublimation (II): 1 – mesh crucible with chromium; 2 – electron beam gun; 3 – column; 4 –column heaters; 5 – chromium condensate.

The content of main impurities of chromium and its residual resistance value $R_{res} = R_{293K}/R_{4,2K}$ are given in Table 1.

A study on variation of gaseous impurities vs. condensation temperature indicated that their content in the abstracted chromium decreased with increasing condensation temperature.

Even more effective was the use of heating with subsequent distillation in the case of gallium refinement. Fig. 2 shows the scheme of gallium device for refinement involving heating and in-vacuum distillation [6].

At the first stage of the refinement (Fig. 2) (the condenser 7 closed with the valve 5), gallium had, via heating at 1320 K, highly volatile impurities removed from it (Cd, Zn, Mg, Pb, gases, etc.). In order to remove the low-volatile impurities (Al, Cu, Cr, Fe, Ni, Si, etc.) the second stage involved (the upper branch pipe 4 closed with the valve 5) abstraction of gallium into the condenser 7 at the temperature 1520 K, the abstracted share being 80%.

Table 1. Impurity content and $R_{res}$ of various chromium kinds.

| Material | Impurity content, $\times 10^{-3}$ wt.% | | | | | | | | | $R_{res}$ |
|---|---|---|---|---|---|---|---|---|---|---|
| | O | N | H | C | Fe | Al | Si | Cu | Ni | |
| Initial | 7 | 4 | 0.7 | 2 | 8 | 3 | 5 | 3 | 5 | ~4 |
| Condensate | 2 | 0.6 | 0.4 | 1 | 5 | 1 | 1 | 0.5 | < 1 | 90 |

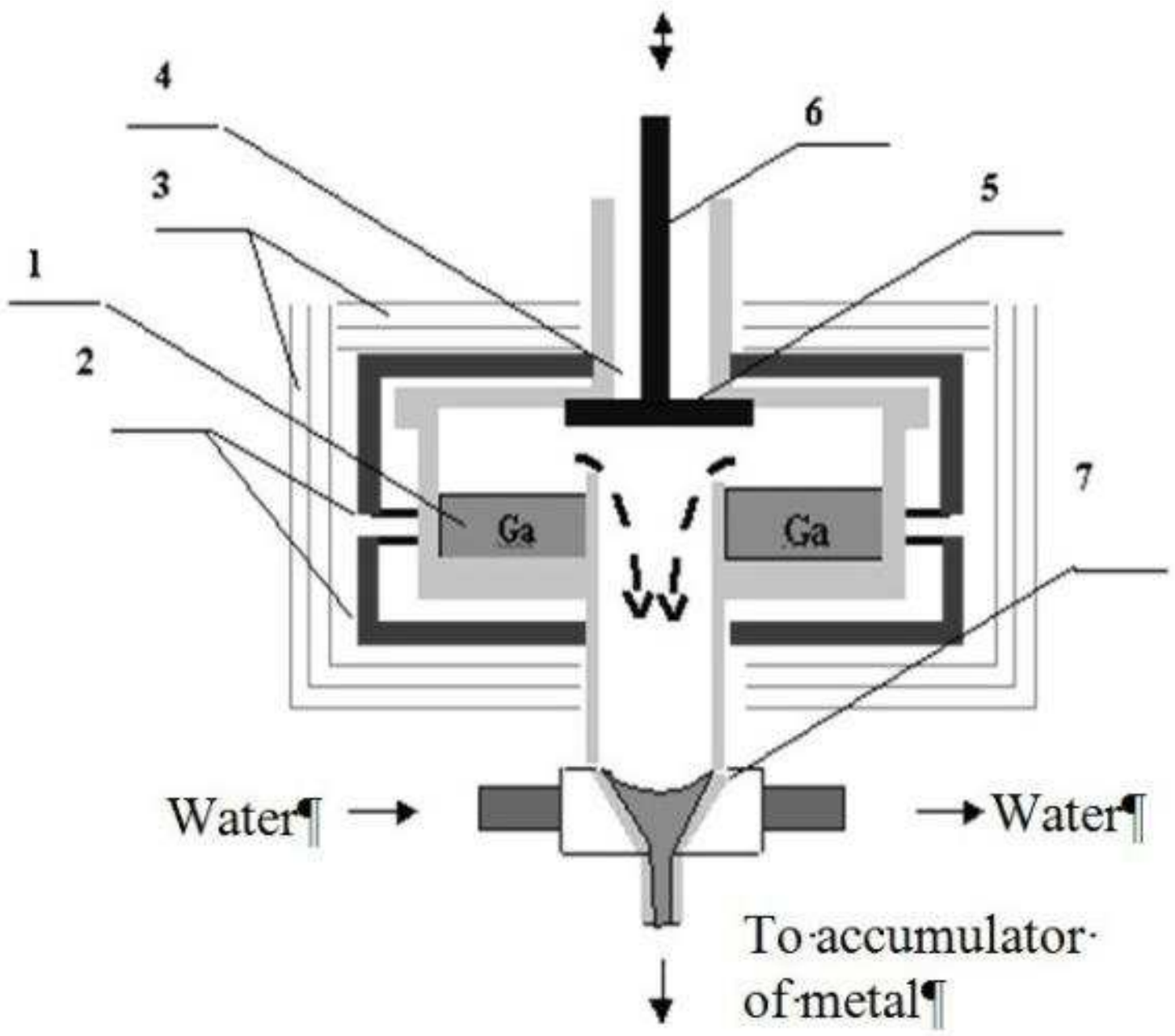

Fig. 2. Scheme of device for gallium refinement via heating and distillation in vacuum:
1 – crucible; 2 – heater; 3 – thermal shields; 4 – opening; 5 – shutter; 6 – rod; 7 – condenser.

In particular, for gallium of the purity 99.94% and $R_{res}$ ~ 100 the heating and double distillation allow for production of metal with $R_{res}$ ~ 50,000 and content of the base component about 99.9999%, the metal yield being high (up to 80%) (Table 2).

Table 2. Values of $R_{res}$ for gallium after different stages of refinement.

| Refinement stages | $R_{res}$ |
|---|---|
| Initial gallium (99.94 wt. %) | 100 |
| Heating (removal of Cd, Zn, Mg, Pb, In, Mn, gases, etc.) | 10000 |
| Distillation (purification from Al, Cu, Cr, Fe, Co, Ni, Si, etc.) | 25000 |
| Repeated distillation (additional purification from Al, Cu, Cr, Fe, Co, Ni, Si) | 50000 |

To refine cadmium, zinc and tellurium via combination of heating and distillation in vacuum, a device was developed, the scheme and operation principles of which are shown in Fig. 3 [6]. The initial metal is placed on the plate 6 with an opening in the condenser 3. The metal is heated up somewhat higher than its melting temperature to provide for its spill into the crucible. With that, the gaseous impurities are abstracted away through the opening 8, the highly volatile impurities and oxides being condensed on the condenser surface 3, while the low-volatile oxides and slags remain as a film 11 on the surface of the plate 6.

Then the condenser containing the highly volatile impurities is replaced with the condenser 5 over the crucible (the air-tightness of the chamber 1 is not violated), and the second stage of the process comes into play: removal of the low-volatile impurities where the fraction of metal abstraction is 90 to 95% at each distillation.

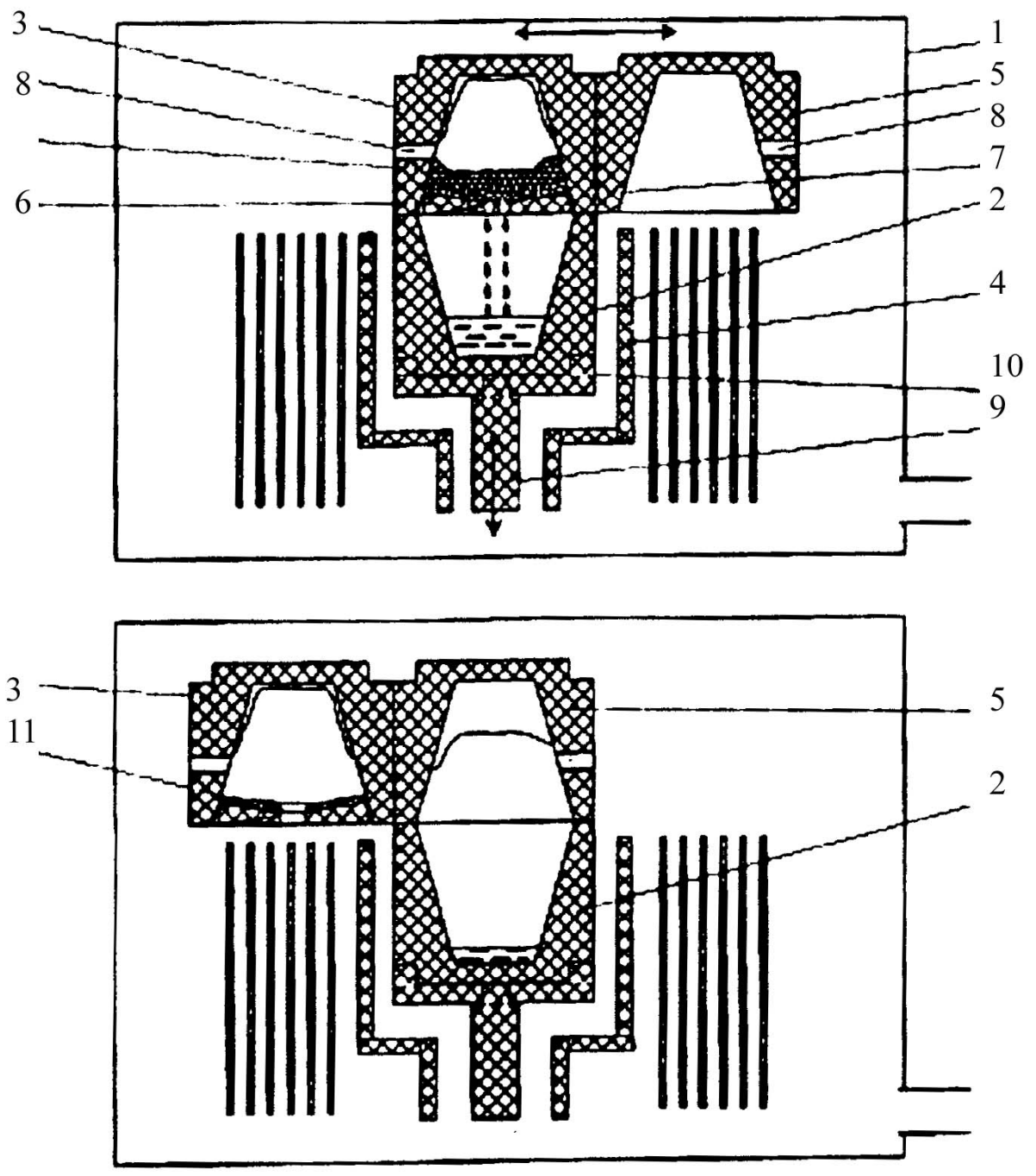

Fig. 3. Scheme of device for refinement of cadmium and tellurium via heating and distillation in vacuum: 1 – working chamber; 2 – crucible; 3 – main condenser; 4 – heater; 5 – supplementary condenser; 6 – plate; 7 – initial metal; 8 – opening; 9 – movable rod; 10 – mount; 11 – oxide film.

Table 3 gives the total impurity content (without taking into account carbon and gas-forming component) in cadmium and tellurium after various stages of refinement. As one can see, the combination of heating and subsequent distillation is an effective purification method for cadmium and tellurium even after one pass of the refinement cycle. The subsequent 2nd and 3rd distillations additionally enhance the purity of metals, although not considerably.

Table 3. Total impurity content in Cd, Zn and Te in initial states and after refinement via different techniques, %.

| Metal | Initial | 1st distillation after abstraction of volatile impurities and "filtration" | 3rd distillation | Czochralsky crystallization |
|---|---|---|---|---|
| Cd | $1.1\cdot 10^{-2}$ | $2.5\cdot 10^{-4}$ | $1.4\cdot 10^{-4}$ | $< 0.2\cdot 10^{-4}$ |
| Zn | $1.4\cdot 10^{-3}$ | $3.5\cdot 10^{-4}$ | $2.5\cdot 10^{-4}$ | $< 0.5\cdot 10^{-4}$ |
| Te | $2.6\cdot 10^{-2}$ | $7.3\cdot 10^{-4}$ | $4.2\cdot 10^{-4}$ | $< 0.8\cdot 10^{-4}$ |

The studies made to date [6-8] indicated that, as in the case with gallium, a deeper impurity separation (by an order of magnitude or more) in cadmium, zinc and tellurium is achieved using the above crystallization methods (see Table 3).

## References

1. V.A. Pazukhin, A.Ya. Fisher, *Separation and Refinement of Metals in Vacuum*. Moscow, Metallurgy, 1969, 203 p. (in Russian).
2. G.G. Devyatikh, Yu.E. Eliev, *Introduction to Theory of Deep Purification of Substances*. Moscow, Nauka, 1982, 320 p. (in Russian).

3. V.E. Ivanov, I.I. Papirov, G.F. Tikhinsky, V.M. Amonenko, *Pure and Ultra-Pure Metals*. Moscow, Metallurgy, 1965, 263 p. (in Russian).
4. G.F. Tikhinsky, G.P. Kovtun, V.M. Azhazha, *Production of Ultra-Pure Rare Metals*. Moscow, Metallurgy, 1986, 160 p. (in Russian).
5. G.P. Kovtun, A.I. Kravchenko, A.P. Shcherban', Inorg. Mat. 34 (1998) 819.
6. G.P. Kovtun, A.I. Kravchenko, A.P. Shcherban', Engineering and Production Design in Electronic Devices 3 (2001) 6.
7. V.M. Azhazha, G.P. Kovtun, I.M. Neklyudov, Engineering and Production Design in Electronic Devices 6 (2002) 3.
8. G.P. Kovtun, A.P. Shcherban', V.D. Vyrich, Visnyk of Kharkiv Nat. Univ., ser. phys. "Nuclei, Particles, Fields" 619/1/23 (2004) 95 (in Russian).

# Purification of cadmium and lead for low-background scintillators

**G.P. Kovtun[a*], A.P. Shcherban'[a], D.A. Solopikhin[a], V.D. Virich[a], V.G. Glebovsky[b]**

[a] *National Science Center "Kharkov Institute of Physics and Technology" (NSC KIPT), 1, Academicheskaya str., Kharkov 61108, Ukraine*
[b] *Institute of Solid State Physics, Russian Academy of Sciences, Chernogolovka, Russia*

The method for deep purification of Cd, $^{106}$Cd and Pb, consisting in the filtration-distillation combination is offered. It is shown that these procedures are very efficient for deep purification of natural Cd and enriched $^{106}$Cd, and archeological Pb (Black Sea, near Crimea, dated to I century B.C.). A reached level of the content of the most harmful elements, such as Ni, Cu, Fe, Mg, Mn, Cr, V, Co, Th, U, Ra, K, Rb, In, La, Lu, Sm is < 1 ppm. The pilot batches of high-purity Cd, $^{106}$Cd and Pb, applicable for production of scintillation monocrystals (Cd,$^{106}$Cd,Pb)$WO_4$ and (Cd,$^{106}$Cd,Pb)$MoO_4$ are obtained.

## 1. Introduction

To produce high-quality scintillators (Cd,$^{106}$Cd,Pb)$WO_4$ and (Cd,$^{106}$Cd,Pb)$MoO_4$, the level of initial component contamination should not exceed several ppm. The most harmful elements are the following: Ni, Cd, Fe, Mg, Mn, Cr, V, Cd and radioactive elements. An increased content of Ni, Cu > 0.2 ppm and Fe, Mg, Mn, Cr, V, Co > 2 ppm leads to the coloring of crystals and to the deterioration of their scintillation properties. The concentration of radioactive elements, e.g. Th, U, Ra, K, Rb, In, La, Lu, Sm should be << 1 ppm, as their presence in crystals increases the background of a detector. Analysis of the contamination level in the available natural Cd, enriched $^{106}$Cd and archeological lead has shown that the impurity content in them exceeds by tens and hundreds times the requirements to the initial materials for growing single-crystals of tungstates and molybdates of cadmium and lead.

The purification procedure should provide both the minimal losses of initial materials and the total waste conservation in the process of refining, in particular for $^{106}$Cd, taking into account a high cost of this material. One of the methods of deep purification of these metals is a vacuum distillation [1-3]. The interest in distillation is due to the fact that this method enables reaching a high degree of purification with a high yield of an ecologically pure product. The goal of investigations is to develop the methods and devices for purification and production of a pilot batch of high-purity Cd, $^{106}$Cd and Pb, as applied to the problems of designing of low-background scintillation detectors with the use of tungstates and molybdates of cadmium and lead. The paper presents the results of experimental investigations on the behavior of impurity elements in the natural Cd, enriched $^{106}$Cd and archeological lead.

To account for the above-mentioned requirements to the final product purity and to provide minimal losses of initial materials, the procedure developed for deep purification of Cd and $^{106}$Cd was based on our own earlier investigations on the cadmium purification by vacuum distillation [2-4]. In this connection, for the given case we have chosen an approach consisting in the combination of prefiltration and 2…5 distillations. Experimental investigations on the optimization of temperature and time regimes were carried out with archeological lead. For this purpose special devices for lead filtration and distillation were developed.

[*] Corresponding author, *E-mail address*: gkovtun@kipt.kharkov.ua

## 2. Analysis of the behavior of impurity elements in cadmium and lead

The basis for the method of metal purification is the difference in the compositions of a separated mixed liquid and a vapor formed from it. This difference is characterized by the value of relative volatility $\alpha$ of the component to be separated (with reference to the purification process this value is named as a separation factor). For the case of a diluted ideal solution the expression determining the ideal impurity separation factors $\alpha_i$ upon molecular evaporation is given in [5]. The calculated values of ideal impurity separation factors $\alpha_i$ for cadmium in the temperature range from 600 to 900 K are given in [3]. For lead the same calculations were performed. For calculations the values of element vapor elasticity taken from [6] were used.

The determined values of ideal impurity separation factors $\alpha_i$ in Cd and Pb in the temperature range from the melting point to the boiling point have shown that the spectrum of impurity elements includes light volatile impurities ($\alpha_i << 1$), heavy-volatile impurities ($\alpha_i >> 1$) and several impurities with $\alpha_i \sim 1$, belonging to the difficult-to-remove impurities. For harmful impurities in the scintillation detectors, the number of purification repetition, calculated by $\alpha_i$ values, is 100 and more that implies an effective purification of cadmium and lead from these impurities by vacuum distillation.

The results of calculations on the behavior of impurity elements in the process of Cd distillation are given in [2, 3]. Investigations were carried out on the dependence of the efficiency of Cd melt purification from the light volatile impurities on the residue fraction and the dependence of the efficiency of Cd condensate purification from the heavy-volatile impurities on the distillation efficiency at a given temperature. Analysis of the dependences has shown that more than 10 purifications of the melt from light volatile impurities with $\alpha_i << 1$ will occur upon evaporation of Cd < 10 %. For the heavy-volatile impurities with $\alpha_i >> 1$ more than 10-fold purification of the condensate will occur upon Cd refining to the condensate of > 95%.

So, for the deep purification of Cd and Pb more preferable can be realization of the process of step-by-step purification from light-volatile and heavy-volatile impurities with wastes of basic metal (10-15%) that has been earlier realized for natural cadmium refining [3]. To provide minimal losses of initial materials, in particular $^{106}$Cd, for the given case we have chosen, instead of the step-by-step purification, a method consisting in the combination of prefiltration and distillation with subsequent casting of distillates in the form of measured ingots.

The obtained results of calculations of the efficiency of cadmium and lead refining were taken into account in the development of improved distillation devices and in the choice of temperature and time regimes of distillation purification processes.

## 3. Experimental investigation of the processes of purification of natural Cd, enriched $^{106}$Cd and archaeological Pb by vacuum distillation

### 3.1. Materials and methods of purity control

The initial materials for purification were granular Cd of ChDA grade (spec 6-09-3095-78), enriched $^{106}$Cd and archeological Pb.

To exclude the ingress of background impurities into the metal being refined in the purification processes we used high-purity accessory materials and equipment. A crucible and a condenser of distillating apparatuses are made of high-purity graphite of MPG-7 grade, and a heater, heater components and screens are made of the spectrally pure graphite, corresponding to spec 48-20-90-82. The content of regulated impurities in such graphite is $\leq 6.2 \cdot 10^{-4}$ mass %. An inert atmosphere for filtration of Cd and $^{106}$Cd was the highest-grade argon gas corresponding to the State Standard 10157-88 with an argon volume fraction no less than 99.995%.

Quantitative analysis of samples determining the impurity content in the initial and refined metals was performed by the ICP-MS and AAS methods (chemical laboratory of LNGS, Assergi, Italy), as well as by the LMS method (NSC KIPT, Kharkiv, Ukraine). The AAS method was used mainly to determine the iron content in cadmium, as the Fe concentration measured by the ICP-MS method was too high. It is because the iron isotopes are overlapped with the isobaric interference from the $^{58}$Ni isotope and doubly ionized $^{112}$Cd and $^{114}$Cd. The accuracy of impurity content determination by the above-mentioned methods is 15…30%.

### 3.2. Experimental procedure, results and discussion

To develop the process of deep refining of $^{106}$Cd with minimal losses and a high product yield, we carried out preliminary investigations with natural cadmium. For this purpose a distillating apparatus and a filtrating apparatus providing condensate casting in measured ingots with an initial metal charge to 250 g were developed and fabricated. The crucible and the condenser of this distillating apparatus are interchangeable that permits to repeat distillation without removing the distillate from the condenser.

The process of high-purity cadmium production consisted in the following. The initial cadmium (about 200 g) with an initial impurity content (Table 1) was preliminary filtered, to minimize losses, in the pure argon atmosphere under pressure in the apparatus chamber ~ 120 kPa. As a result of filtration, oxides of impurity metals and slag in the form of a film remained on the surface of filtrating apparatus plate. Then the chamber was evacuated and in the process of refining the pressure in it was maintained as $10^{-3}$ Pa or lower. Cadmium, preliminary refined by filtration, was subjected to two distillations with a distillation fraction more than 98% in each process. During the process cadmium was evaporated at 630…650 K and condensed at 530…550 K.

Condensation at such temperatures leads to the partial purification of the condensate from the light-volatile impurities (Na, K, S, P, As, Se etc.) by removing them into the chamber volume through the little hole in the condenser. The heavy-volatile impurities (Fe, Ni, Co, Si, Cu, Al, Au, Ag, Pb, Tl, Sb, Bi, Li, Sn, Mn etc.) were concentrated in the residue in the crucible. After refining by distillation, cadmium was cast in the measured ingots. The similar procedure was used for refining $^{106}$Cd with 2…5 distillations depending on the degree of purity of the initial material.

Table 1. Impurity composition of natural Cd and enriched $^{106}$Cd before and after purification.

| Impurity element | Concentration in natural Cd (ppm) | | Concentration in $^{106}$Cd (ppm) | |
|---|---|---|---|---|
| | before purification | after purification | before purification | after purification |
| Ni | 30* | 0.3* / ≤ 0.2** | 0.6* | 0.6* / ≤ 0.2** |
| Cu | 47* | 0.3* / ≤ 0.2** | 5* | 0.7* / 0.5** |
| Fe | 0.4*** | 0.17*** / ≤ 0.5** | 1.3*** | 0.4*** / ≤ 0.4** |
| Mg | 30* | ≤ 0.5* / ≤ 0.05** | 12* | ≤ 0.5* / ≤ 0.05** |
| Mn | 0.2* | 0.1* / ≤ 0.3** | 0.1* | 0.1* / ≤ 5** |
| Cr | 0.2* | 0.1* / ≤ 1** | 9* | ≤ 0.5* / ≤ 0.1** |
| V | <0.005* | ≤ 0.005* / ≤ 0.08** | <0.005* | ≤ 0.01* / ≤ 0.08** |
| Co | 0.3* | ≤ 0.003* / ≤ 1** | 0.02* | ≤ 0.01* / ≤ 0.1** |
| K | 8* | ≤ 5* / 0.7** | 11* | ≤ 10* / 0.04** |
| Pb | 1000* | 3* / ≤ 1** | 270* | 8* / ≤ 0.3** |
| Th | <0.001* | ≤ 0.001* | <0.001* | ≤ 0.001* |
| U | <0.001* | ≤ 0.001* | <0.001* | ≤ 0.001* |

* ICP-MS – Inductively Coupled Plasma - Mass Spectrometry analysis
** LMS – Laser Mass Spectrometry
*** AAS – Atomic Absorption Spectroscopy

For purification of archeological lead by vacuum distillation, a special apparatus was developed providing the metal vapor condensation into the liquid phase with a high efficiency and high product yield > 95%. Surface contaminations, metal oxides and slag, similarly as for cadmium, were removed by prefiltration of the initial metal. Then the metal was placed in the crucible and heated to working temperatures (~ 1220 K), and, as a result, it was evaporated and passed into the condenser. During the distillation process the lead was purified from the heavy-volatile impurities (Mn, Ni, Co, Cu, Fe, U, Th etc.), and the light-volatile impurities (Ca, Mg, Tl, K, As etc.) were removed through the special hole due to the high temperature (~ 1020 K) in the condenser. After that the pure metal was formed in ingots in the casting device.

The impurity concentration in archeological lead (Table 2) was determined by the LMS method. Analysis of the impurity composition shows a good purification of Pb from Ni, Cu, Zn, Ag, Sb and other impurities.

Table 2. Impurity composition of archaeological lead before and after purification.

| Impurity element | Before purification, ppm | After purification, ppm | Impurity element | Before purification ppm | After purification, ppm |
|---|---|---|---|---|---|
| Mg | 0.09 | < 0.04 | Cu | **6.3** | < 0.1 |
| K | 0.5 | 0.35 | Zn | **2.6** | < 0.2 |
| Ca | 0.56 | 0.3 | As | < 0.1 | < 0.1 |
| V | < 0.07 | < 0.07 | Ag | **34** | < 0.6 |
| Cr | < 0.09 | < 0.09 | Sb | **5.4** | < 0.6 |
| Mn | < 0.08 | < 0.08 | Tl | < 0.8 | < 0.8 |
| Fe | < 0.09 | < 0.09 | Bi | < 1 | < 1 |
| Co | < 0.09 | < 0.09 | Th | < 0.7 | < 0.7 |
| Ni | < 0.2 | < 0.09 | U | < 0.7 | < 0.7 |

Analysis has shown that the residue after filtration in the form of oxide film is enriched with separate impurity elements. So, in this case, similarly to the case of cadmium, filtration serves as an additional purification element.

Comparative analysis of the obtained results evidences that the refining by vacuum distillation in combination with filtration is an effective method of deep purification of cadmium and lead. The purification procedure under consideration provides for several impurities more than 100-fold decrease of their content – to the level required for a material used in fabrication of scintillators $(Cd,^{106}Cd,Pb)WO_4$ and $(Cd,^{106}Cd,Pb)MoO_4$. Also, a high product yield (> 96%) is reached and irreversible losses are decreased. It has been found that the highest efficiency of purification takes place after the first distillation. Depending on the initial material purity, the required purity level, under similar distillation conditions, is reached by repetition of the distillation process.

In Fig. 1 presented are the samples of high-purity ingots of $^{106}Cd$ and archeological lead after filtration, distillation and casting.

## 4. Conclusions

Based on the analysis of earlier calculations of the behavior of impurity elements in metals in the process of their purification with the distillation procedure, the methods and devices were developed for deep purification of Cd, $^{106}Cd$ and Pb. A combined effect of filtration and distillation on the deep purification from harmful impurities was investigated in experiments. The use of the complex process for purification of Cd, $^{106}Cd$ and Pb provides more than 100-fold purification from harmful Ni, Cu, Mg, Co, Fe and radioactive impurities. As a result of investigations, the pilot batches of high-purity Cd, $^{106}Cd$ and Pb with a product yield > 96% and irreversible losses < 1%, applicable for production of high-quality scintillators $(Cd,^{106}Cd,Pb)WO_4$ and $(Cd,^{106}Cd,Pb)MoO_4$ were obtained.

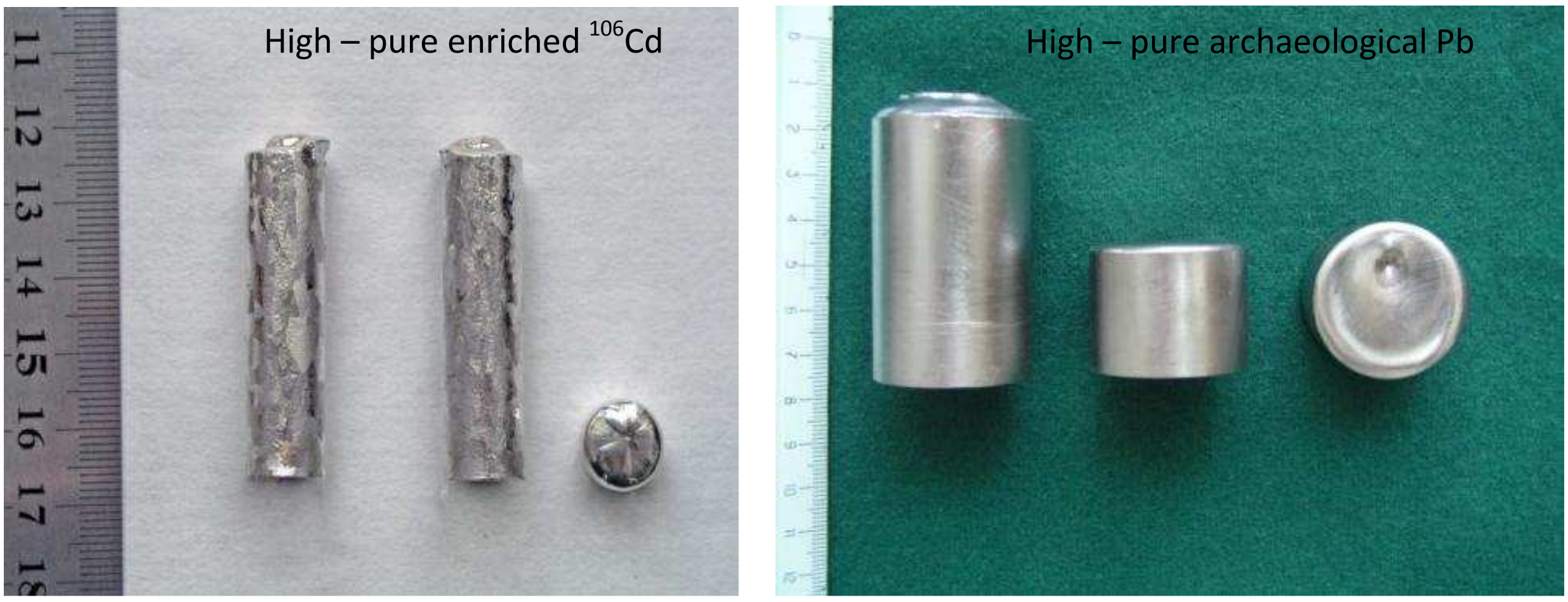

Fig. 1. Ingots of high-purity $^{106}$Cd and archaeological Pb.

## References

1. L.F. Kozin, E.O. Berezhnoj, K.L. Kozin, High-Purity Materials 5 (1996) 11.
2. G.P. Kovtun, A.I. Kravchenko, A.P. Shcherban’, Technology and Designing in Electronic Equipment 3 (2001) 6.
3. G.P. Kovtun, A.P. Shcherban’, Visnyk of Kharkiv Nat. Univ., ser. phys. “Nuclei, Particles, Fields” 642/3/25 (2004) 27 (in Russian).
4. G.P. Kovtun, A.P. Shcherban’, Patent 1246, Ukraine, C22B9/04, C22B9/187. National Science Center “Kharkov Institute of Physics and Technology” – № 2001075474; Appl. 31.07.01; Published. 15.05.02, Bulletin 5 (2002) 3.
5. V.A. Pasukhin, A.Ya. Fisher, *Separation and Refinement of Metals in Vacuum*. Moscow, Metallurgiya, 1969, 203 p. (in Russian).
6. A.N. Nesmeyanov, *Vapor Pressure of Chemical Elements*, Moscow, Publ. of Ac. Sci. USSR, 1961, 396 p. (in Russian).

# Development of techniques for characterisation of scintillation materials for cryogenic application at the University of Oxford

**V.B. Mikhailik*, H. Kraus**

*Department of Physics, University of Oxford, Keble Road, Oxford OX1 3RH, United Kingdom*

The multi-photon counting (MPC) technique was designed to record photon emission of scintillators and, as a very powerful method of material characterisation, is enjoying increasing popularity. The technique is especially advantageous for the analysis of slow scintillation processes and the investigation of temperature-dependent scintillator properties. The paper describes the latest development of the technique aiming to improve performance and widen the scope of applications. The results from characterising $MgF_2$ are presented to illustrate the capabilities of the MPC technique.

## 1. Introduction

The search for rare events is a most vibrant research area in astro-particle physics [1]. Finding neutrinoless double beta decay and detecting weakly interacting massive particles requires detectors capable of discriminating the weak and rare signal over the dominating background of spurious events caused by natural radioactivity and cosmic rays. This can be achieved, for example, in low-temperature detectors by the simultaneous measurement of the phonon and scintillation responses [2, 3]. The technique exhibits efficient event type discrimination, providing a very important tool for the identification of radioactive background. This, in combination with other advantages of cryogenic phonon detectors, such as greatly enhanced energy resolution and low threshold, elevated cryogenic phonon-scintillation detectors (CPSD) to the category of especially promising techniques for next-generation experiments searching for WIMP Dark Matter [4 – 6], neutrinoless double beta decay [7] and radioactive decay of very long-living isotopes [8, 9]. Inorganic scintillators are a key element of CPSD and in recent years there has been a continuous increase in research activities and the development of scintillator materials capable of meeting the strict design requirements of rare event experiments [10 – 14].

Unlike most conventional applications of scintillators, rare event searches do not require scintillators with fast decay times. Therefore they have the option to use scintillators which are traditionally considered as "slow" [10]. One of the ways to gain insight into the features of the scintillation process of a material, and to identify possible improvements, is via measurements of temperature dependence of decay times and the light yield. To address this issue we developed the multi-photon counting (MPC) technique that allows measurement of these scintillator parameters over a wide range of temperatures [15].

Since the first publication on MPC, there have been substantial upgrades of the technique's hard- and software with the aim to improve overall performance and extend the capabilities of MPC, making it suitable for versatile studies that involve measurements of slow scintillation over a wide temperature range [16 – 19]. MPC is also becoming a key component of newly emerging methods, such as the Monte Carlo refraction index matching technique (MCRIM) [20]. Given such success of the technique, and addressing the growing interest of the community in practical implementations of this development, as shown elsewhere in this paper, we present the main principles, describe the practical implementation of the method, and discuss performance characteristics of the MPC in its latest version.

---

* Corresponding author. *E-mail address:* vmikhai@hotmail.com

## 2. Characterisation of scintillator at low temperatures – main issues

Characterisation of scintillation detectors has so far relied almost exclusively on the use of high-gain photomultipliers (PMT) or avalanche photodiodes (APD). However it is very difficult to carry out reliable measurements of light yield using conventional techniques if the detector and scintillator under investigation are subjected to a change of temperature. In addition to the technical difficulties of operation, the response of PMTs and APDs strongly depend on temperature [21, 22] and filtering out these contributions by the light detectors can be difficult [23 – 27]. To avoid this problem it is preferable to keep the light detector at constant temperature, i.e. outside of the cryogenic apparatus. However this inevitably reduces the solid angle, causing a reduction of the light collection efficiency. In such geometry only fast scintillators with high light yield can be studied when using a conventional method for data recording and analysis [28].

Additional problems associated with the necessity to cover a large range of decay time constants arise in their studies as function of temperature. The delayed coincidence single-photon counting (DC-SPC) technique has excellent timing resolution (ca. 0.5 ns) and it is commonly used for measuring the decay characteristics of traditional fast scintillators with decay time constants in the range of nano-seconds to micro-seconds [29]. If the decay time constant of the scintillator is in the region of several micro-seconds, pulse shape analysis (PSA) can be implemented [30, 31]. One generic source of error is inherent for these measurement techniques. Measurements of the temperature dependence of the decay time constant must allow for its variation over a large range of values (from tens of μs to tens of ms). This requires a correspondingly long recording time. As it is not practically possible to fully control the rate at which ionising radiation interacts with the scintillator, there are always events recorded in which a second (or more) scintillation event will have occurred during the measurement period of the first, a so-called multiple excitation event or "pile-up". These events contribute to the total signal resulting in a roughly even time difference distribution throughout the decay time spectrum, and if not removed can cause false attribution to an additional, extra-long decay component. Thus, when analysing such data, one might easily misinterpret this background as an additional decay component with a long time constant. This issue is becoming even more important with a decrease of temperature, when the scintillation decay time increases, which in turn leads to enhancing the probability of multiple excitations. The only practical approach to overcome this problem would be to implement appropriate statistical methods of analysis allowing recognition of single and multiple events.

## 3. Setting up MPC

There are two key considerations concerning the characterisation of scintillators over a wide temperature range that arise from the previous section. Firstly, the detector and scintillator should be spatially separated and secondly, to enable off-line analysis on event-by-event basis, the single photon counting mode should be used to record the scintillation events. Fortuitously these conditions can be reconciled: single photon counting requires a rate low enough to avoid pile-up of individual photoelectrons generated. In this case, the measurements imply detection of a few tens of photons distributed throughout an interval of 10 – 1000 microseconds duration of the slow scintillation processes, and that can be easily arranged by using ~100 MHz electronics.

The MPC method is based on recording a sequence of photoelectron pulses produced by a PMT when detecting photons from a scintillation event. Each pulse in the sequence corresponds to an individual photon impinging on the photocathode of the PMT. The output PMT signal is statistical in nature, both with regard to the time interval between photons and the total number of detected photons. The distribution of arrival times of the photons provides information on the decay characteristics of the scintillation process, while the number of photons recorded per event is proportional to the light yield of the scintillator. Thus, recording a large number of scintillation events ($10^3$ – $10^4$) one can obtain the decay time characteristics and the light yield in a single measurement.

For scintillation detection we use bi-alkali 9125BQ or multi-alkali 9124A PMTs (Electron Tube Enterprise, Ruislip, UK) with a single electron pulse width of 7.5 and 5 nsec, respectively. To detect a sequence of single photoelectron pulses, a data acquisition chain with a ~10 nanosecond resolution is adequate. The charge signals of the PMT are converted into voltage pulses, using an integrating amplifier with a time constant of ~10 ns. After that, the signal is transmitted into comparatively long coaxial cables. Part of the signal is fed into a transient recorder while another is passed to the slow part of the electronics to derive the trigger.

To record the signal produced by PMT and pre-amplifier we use LeCroy CAMAC-based transient recorders TR8828D or TR8818A. The model TR8828D permits recording with a sampling interval of 5 ns; and that is useful when higher resolution is required to resolve scintillation events with a high rate of PMT pulses, i.e. while studying the scintillation processes at room temperature under conditions when the scintillator is close to the PMT [18 – 20]. The characteristic time resolution of the setup is defined by the FWHM of the individual pulses formed by the preamplifier, which accounts for ~15 ns and therefore a 5 ns sampling interval is adequate. Conversely, for the investigation of temperature dependences of the scintillation characteristics, the sampling interval can be longer (10 or 20 ns) but one needs a long record length [11, 15 – 17]. In this case the 100 MHz transient recorder TR8818A with 8 memory modules is the more suitable option as it permits recording up to $2^{17}$ samples per record.

In the original version of the MPC technique, called the multi-photon coincidence counting [15] the event selection for triggering is derived from the coincidence in two PMTs that face the scintillation crystal. The scheme of this setup is displayed in Fig. 1a. The PMT signal is passed to an integrating amplifier that produces a signal that is a measure of the total energy detected, and it is fed into single channel analysers (SCA). The discrimination thresholds of the SCA are set such as to reduce the number of pulses with low and very high amplitudes which are associated with electronic noise and spurious events caused by cosmic muons, respectively. The logic output pulses of both SCAs are fed into a coincidence unit which provides the trigger for the transient recorder. A gate and delay generator is operating in one channel to tune the arrival times of the logic pulses to the coincidence unit.

Another option that was implemented lately is the use of a self-trigger signal from one PMT [16, 20]. In this case the second PMT with its downstream circuitry is removed and the trigger is received directly from the logical output of the first SCA (see Fig. 1b). This essential hardware simplification became possible thanks to an advanced off-line analysis, capable of filtering out the majority of spurious events.

All scintillation events recorded by the transient recorder are stored in a binary data file by custom-made DAQ software and then subjected to data analysis using tailor-made software. A key aspect of the MPC technique is that it allows discrimination between single and multiple events via statistical analysis of the arrival times of individual scintillation photons. The algorithm used for searching for and eliminating multiple events is based on the idea that the decay time constant $\tau$ obtained for each event should be noticeably dissimilar between single events and multiple events. The statistical analysis consists of a combination of a cut on the number of photons, a Shapiro-Wilk likelihood cut and a Poisson statistics cut for the distribution of arrival times of the photons.

The MPC is a versatile and flexible technique that can be adapted to various experimental needs. It was first tested with a $^4$He-flow cryostat [15] and then intensely used in studies of many crystal scintillators [10, 11, 16, 32]. The implementation of the self-trigger option and the use of a single PMT for detection allowed improving the light collection of the setup by placing the sample near the focus of a concave mirror and directing the scintillation light to the photocathode. This in turn made possible studies of scintillators with only poor response at room temperature [33].

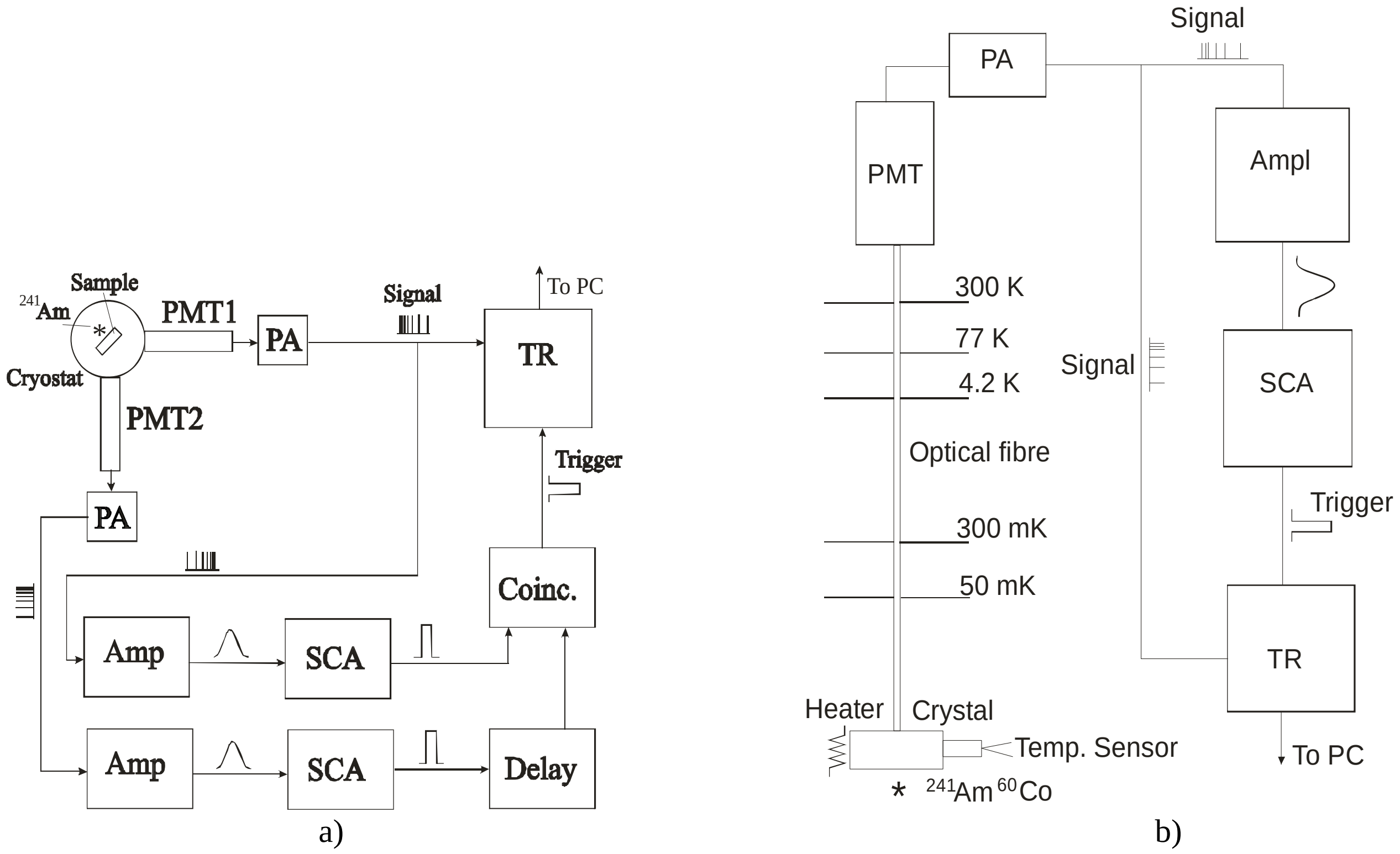

Fig. 1. Scheme of MPC setups for measurements of scintillation characteristics: a) coincidence trigger and b) self-trigger option. PMT – photomultiplier, PA – preamplifier, TR – transient recorder, Amp – linear amplifier, SCA – single channel analyser, Delay – gate and delay generator, Coinc. – coincidence unit.

As a next step we used the MPC technique for scintillation studies in a $^3He/^4He$ dilution refrigerator cryostat [17]. The crystal was attached to the mixing chamber of the cryostat and, to guide scintillation light to the PMT, we used an optical fibre. The fibre is made from quartz, having low transmittance in the far-IR region, thus providing negligible heat load onto the sample. In this experiment, for the first time the scintillation properties of $CaWO_4$ were studied down to 20 mK. The measurements confirmed that the light yield and the decay time constant do not experience a noticeable change below the temperature of liquid helium. It also demonstrated the feasibility of the MPC technique for characterisation of scintillation materials at millikelvin temperatures.

Finally it was shown that MPC can be successfully used for studies of scintillating materials, both in crystalline and powder form, when applying a conventional setup, i.e. the crystal is in close proximity to the PMT photocathode. In these studies, the condition for detection of single photoelectrons can be satisfied by choosing the energy of the excitation source. It was shown that the scintillation excited in $CaWO_4$ and $ZnWO_4$ by 60 keV $\gamma$-rays from a $^{241}Am$ source are clearly defined; the effect of pile-up of individual photoelectron pulses is negligible and the experimental data yield adequate decay time characteristics [18 – 20, 34].

## 4. Performance of the MPC technique

The key advantage of the MPC technique lies in its ability to identify and eliminate spurious contributions from multiple scintillation events. If not removed, these contributions can lead to the false attribution of an extra long decay component. The probability of multiple event excitation increases with the event rate. To visualise the effect, we carried out an experiment in which the excitation rate was varied by changing the distance between the source and the scintillator. The normalised decay curves displayed in Fig. 2a (raw data) indicate that the major effect observed is the rise in the intensity of background. Should it be a slow component, its intensity should not change after normalization. Finally, when the data were subjected to the analysis following the procedure described in the paper, the background was eliminated and all decay spectra produced virtually identical results (Fig. 2b). It is clearly seen that when multiple events are rejected, the decay spectrum shows virtually no background. This observation proves that this part of the decay spectrum is caused by multiple excitations.

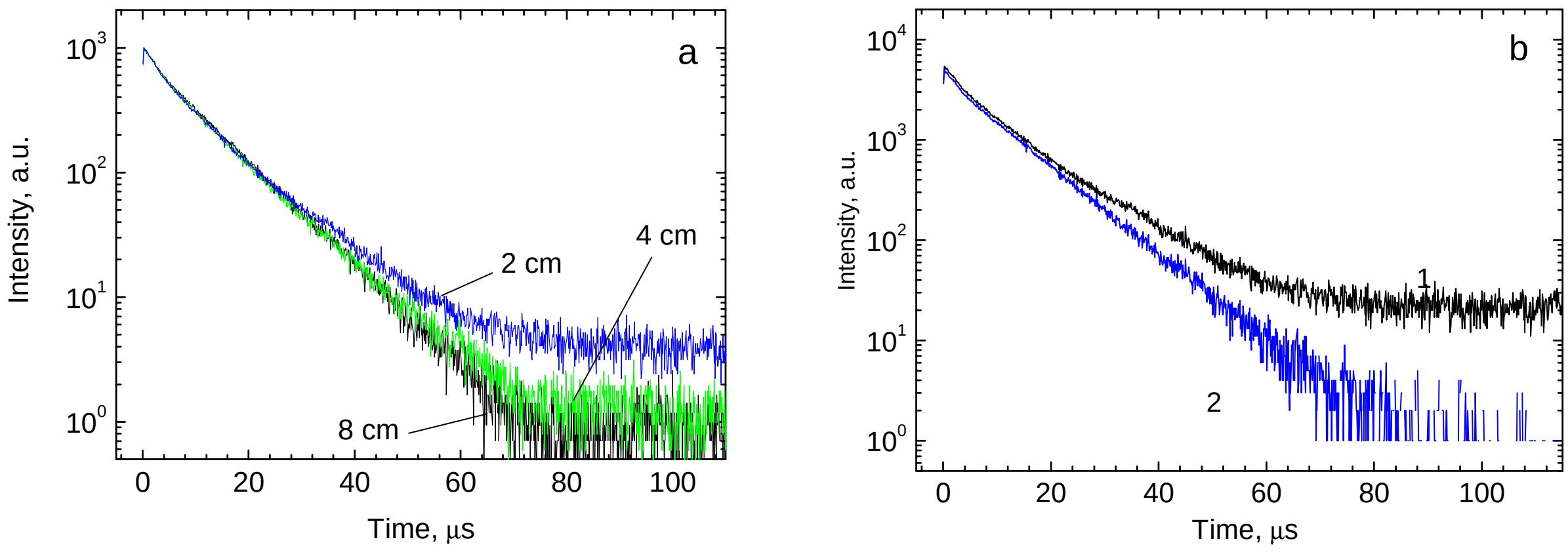

Fig. 2. (a) Normalised decay time curves of $CaWO_4$ recorded for different distances between the source ($^{241}Am$) and the scintillator at T=295 K. (b) The decay time curves of $CaWO_4$ as recorded (1) and after elimination of spurious events (2) at T=295 K.

The example presented here demonstrates that the procedures implemented give a much clearer decay time spectrum of the scintillation process, which in turn permits the decay time parameters to be determined with better accuracy. This shows the great benefit of using the MPC technique in characterising "slow" scintillators; and even more so when investigating "slow" scintillators at low temperatures, where their scintillation is even slower.

Another advantage of the MPC technique is the possibility to derive information on the light response of a scintillator by analysing the amplitude characteristics of the scintillation event from the recorded experimental data. Fig. 3a displays a typical distribution of the total area under the PMT pulse trace, which resulted from the scintillation events excited by $^{241}Am$ source (raw data). The scintillations of $CaWO_4$, placed in a $^4He$ cryostat, were recorded with a low trigger threshold at room temperature. The histogram shows all principal features of the pulse height distribution of the PMT signal: a prominent noise shoulder, the peak due to 60 keV $\gamma$-rays, and an indication of a band due to 5.5 MeV $\alpha$-particles. The histogram displayed in Fig. 3b is obtained from the same set of data after statistical analysis that removed spurious events. The graph shows that the pulse height distribution is free from noise and the 60 keV peak exhibits a Gaussian shape. The importance of this result is twofold. First, it confirms the successful performance of the implemented procedure of statistical analysis. Second, it also manifests that the position of such a clearly identified peak can be used to monitor the relative change in the light response of the scintillator with temperature. It is also worth remarking that increasing the trigger threshold reduces the contribution of the noise and permits detection of a high-energy $\alpha$-band from a $^{241}Am$ radioactive source (see Fig. 4). The latter can be useful for investigation of scintillators with poor light yield.

As a result, the MPC technique has been extensively used for the characterisation over a wide temperature range of scintillation properties of various scintillation materials. The list of investigated materials includes tungstates ($CaWO_4$, $ZnWO_4$, $CdWO_4$, $MgWO_4$, $PbWO_4$), molybdates ($CaMoO_4$, $ZnMoO_4$, $PbMoO_4$, $CdMoO_4$) fluorides ($MgF_2$, $CaF_2$) as well as other compounds ($Bi_4Ge_3O_{12}$, ZnSe, $Al_2O_3$-Ti). Here we illustrate this by the example of studies of the temperature dependence of decay times and light yields of $MgF_2$.

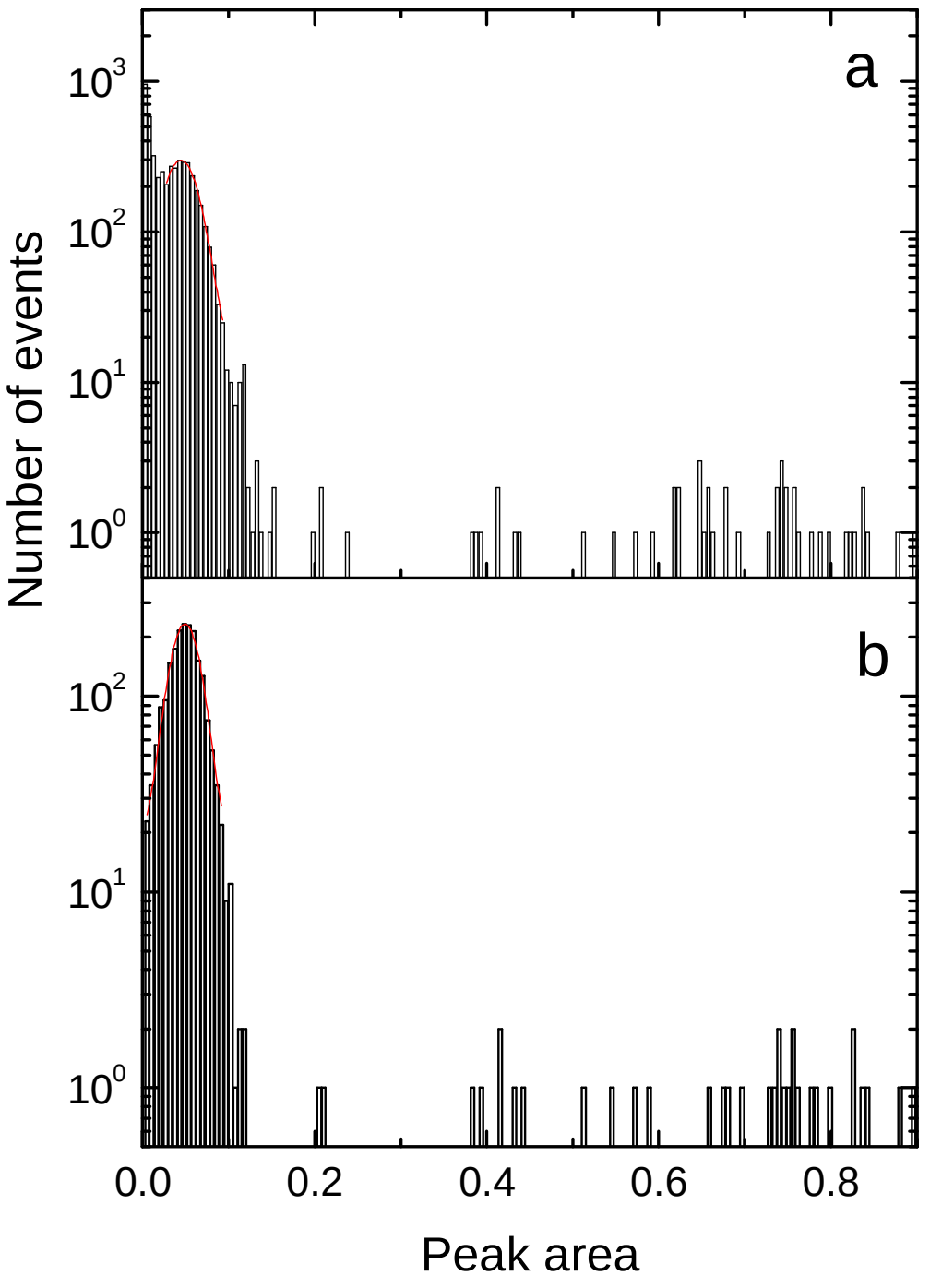

Fig. 3. Pulse height distribution of scintillation signals from a $CaWO_4$ crystal excited by $^{241}Am$ at room temperature (a). The scintillations were recorded with a low trigger threshold (0.4 V) that resulted in a prominent noise shoulder. The data analysis procedure eliminated the spurious events yielding a clear 60 keV γ-peak of $^{241}Am$ (b). The events in the bottom right part of the plot correspond to 5.5 MeV α-particles.

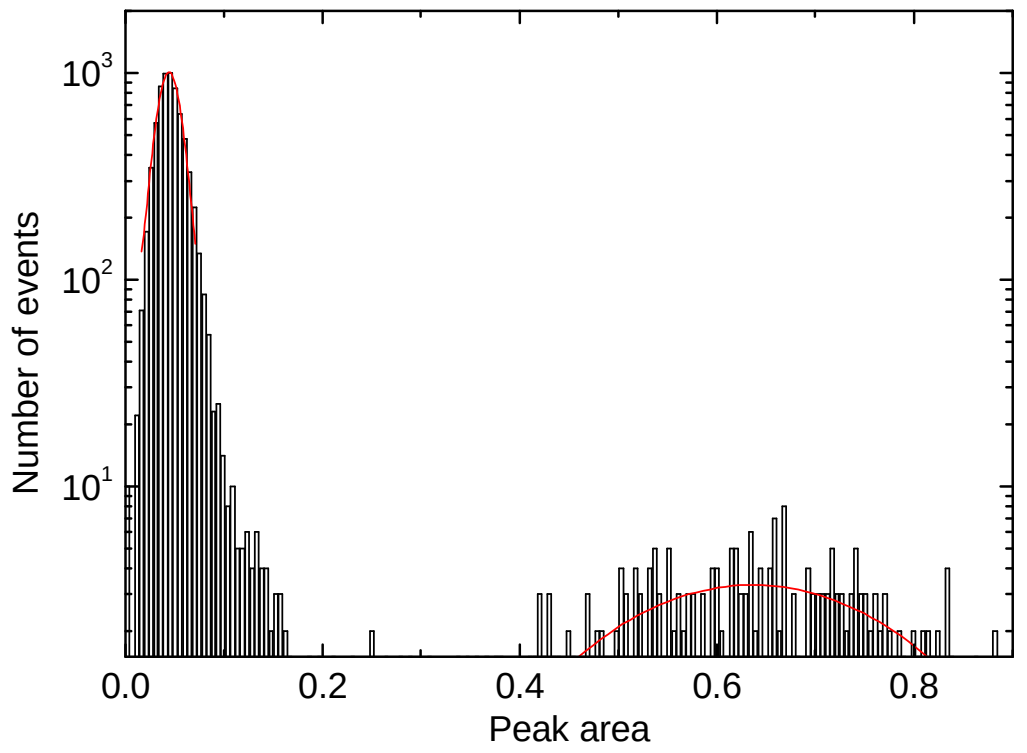

Fig. 4. Pulse height distribution of scintillation signals from $CaWO_4$ crystal excited by $^{241}Am$ at room temperature. The data were recorded with a slightly increased trigger threshold (1.0 V) to remove the PMT noise without compromising the 60 keV γ-peak of $^{241}Am$, thereby improving statistics for α-events. The 60 keV γ-peak and 5.5 MeV α-peak of $^{241}Am$ are fitted by Gaussians.

## 5. Scintillation characterisation of $MgF_2$

Fluoride compounds are thought to be attractive targets in the search for spin-dependent interaction of dark matter particles with nuclei [11]. Preliminary tests, conducted a few years ago, by Keeling [35] have shown that $MgF_2$ scintillates very efficiently at low temperatures. It is generally agreed that the intrinsic emission band of $MgF_2$ observed at 385 nm is due to the radiative decay of triplet self-trapped excitons [36, 37]. The transitions are parity forbidden and exhibit a very slow (~$10^{-3}$ s) decay time constant. Because of such a long decay, this emission is a good example to test the performance of MPC.

$MgF_2$ shows no scintillation at room temperature. The emission appears after cooling the crystal below 70 K, where it rises steeply, reaching the maximum at ca. 40 K (see Fig. 5a). Between 40 and 10 K the light output of the crystals remains fairly constant. Figure 6 shows the scintillation decay curves of $MgF_2$ measured at 61 and 10 K. The decay of $MgF_2$ at 4.2 K has been reported to be characterized as a single exponential decay with decay time constant of 6.4 ms [36]. Fitting a single exponential function yields a decay time constant $\tau$ = 5.8±0.1 ms and, given the difference in excitation conditions, we consider this result as fairly good agreement with the published data. Nevertheless, using the sum of two exponential functions gives a better fit of the experimental results (see Fig. 6). The values for decay time constants at 10 K are found to be 1.5±0.1 and 8.9±0.4 ms. The temperature evolution of the decay time constant of $MgF_2$ shown in Fig. 5b agrees well with published results [37]. The variations of the scintillation light output and decay time constant with temperature observed in $MgF_2$ represent the generic dependences of emission characteristics typical for solids that exhibit thermal quenching.

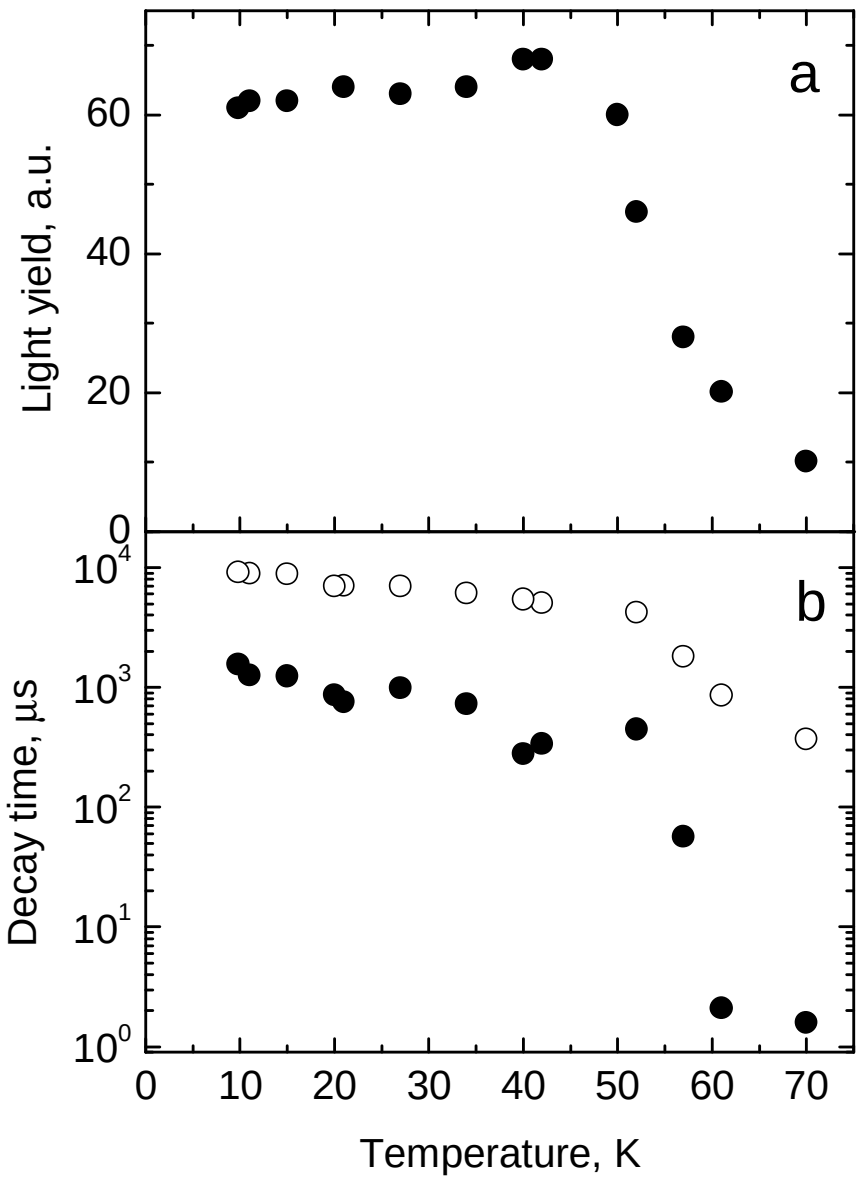

Fig. 5. Temperature dependence of scintillation light output (a) and decay time constants (b) of $MgF_2$ ($^{241}$Am α-source). Open and solid symbols in (b) represent the slow and fast decay time constants obtained from a two-exponential fit.

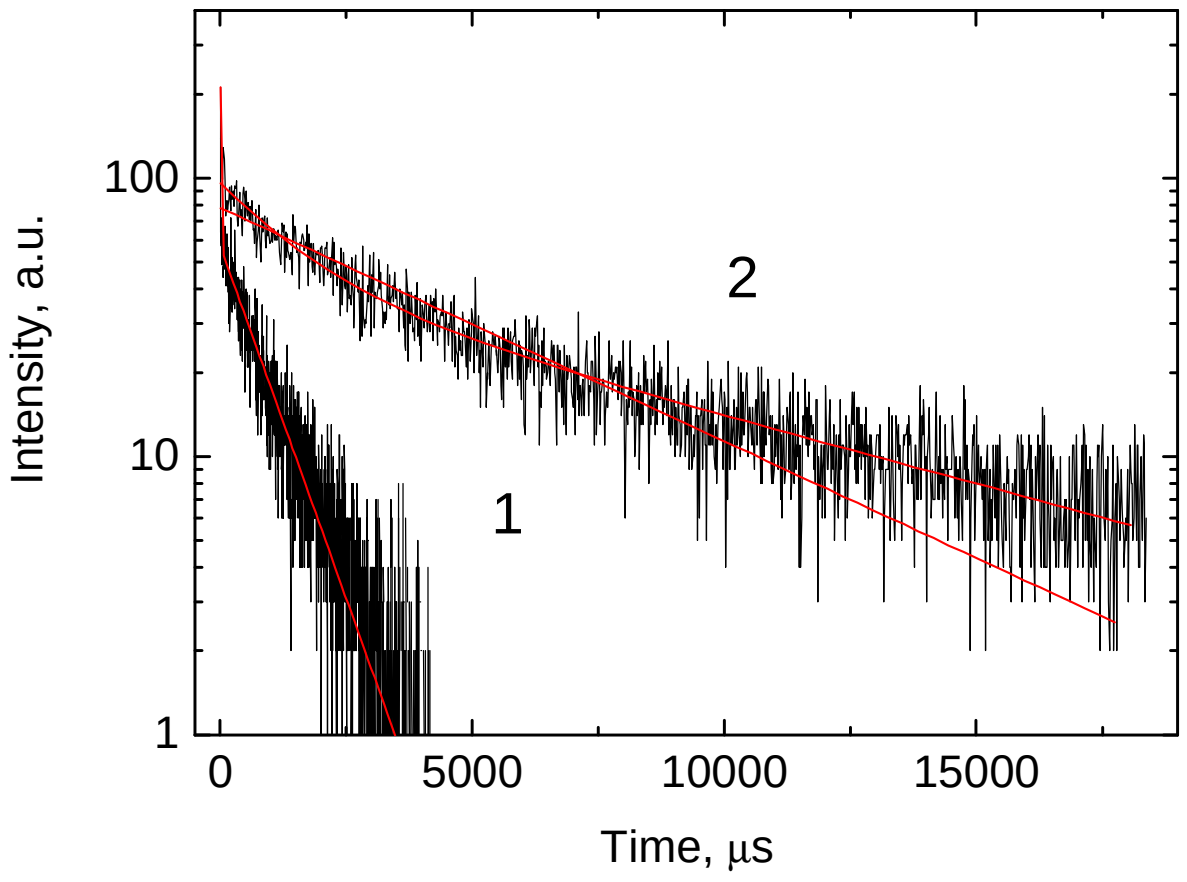

Fig. 6. Decay curves of scintillation in $MgF_2$ measured at 61 K (1) and 10 K (2). The lines show the fits of a single and of two exponential functions to the experimental data.

Having measured the light output of the crystal scintillator in a fixed geometry we can compare the light yield of the material under test with that of a $CaWO_4$ reference sample. Taking the light output of $CaWO_4$ as 100%, after correction for spectral response of our setup, we estimate the low-temperature light output of $MgF_2$ to be 68±20%. This finding is very encouraging; it proves that magnesium fluoride is potentially a good cryogenic scintillator.

## 6. Conclusion

The design of cryogenic scintillation detectors for ionising radiation in present and future experimental searches for rare events requires knowledge of the scintillation properties of various materials at low temperatures. To address this need we continued development of the multi-photon counting technique (MPC) and demonstrated that it can be successfully used for the investigation of major scintillation characteristics (decay time and light yield) over a wide temperature range. The example of characterisation of $MgF_2$, used for demonstration of the technique's performance, clearly illustrated that the MPC technique with its associated data analysis is most beneficial for characterising "slow" scintillators and it is definitely needed for the characterisation of such scintillators at low temperature, where their scintillation is even slower.

## References

1. *Status and Perspectives of Astroparicle Physics in Europe:* http://www.aspera-eu.org/images/stories/files/Roadmap.pdf
2. A. Alessandrello et al., Phys. Lett. B 420 (1998) 109.
3. P. Meunier et al., Appl. Phys. Lett. 75 (1999) 1335.
4. S. Cebrian et al., Astropart. Phys. 21 (2004) 23.
5. G. Angloher et al., Astropart. Phys., submitted.
6. H. Kraus et al., J. Phys. Conf. Ser. 39 (2006) 139.
7. S. Pirro, C. Arnaboldi, J.W. Beeman, G. Pessina, Nucl. Instr. Meth. A 559 (2006) 361.
8. P. de Marcillac et al., Nature 422 (2003) 876.
9. C. Cozzini et al., Phys. Rev. C 70 (2004) 064606.
10. V.B. Mikhailik, H. Kraus, J. Phys. D: Appl. Phys. 39 (2006) 1181.
11. V.B. Mikhailik, H. Kraus, J. Imber, D. Wahl, Nucl. Instr. Meth. A 566 (2006) 522.
12. A.N. Annenkov et al., Nucl. Instr. Meth. A 584 (2008) 334.
13. L.L. Nagornaya et al., IEEE Trans. Nucl. Sci. 55 (2008) 1469.
14. H. Kraus et al., Nucl. Instr. Meth. A 600 (2009) 594.
15. H. Kraus, V.B. Mikhailik, D. Wahl, Nucl. Instr. Meth. A 553 (2005) 522.
16. H. Kraus, V.B. Mikhailik, D. Wahl, Radiat. Meas. 42 (2007) 921.
17. V.B. Mikhailik, H. Kraus, S. Henry, A.J.B. Tolhurst, Phys. Rev. B 75 (2007) 184308.
18. H. Kraus et al., Phys. Stat. Sol. A 204 (2007) 730.
19. F.A. Danevich et al., Phys. Stat. Sol. A 205 (2007) 335.
20. D. Wahl, V.B. Mikhailik, H. Kraus, Nucl. Instr. Meth. A 570 (2007) 529.
21. J.A. Nikkel, W.H. Lippincott, D.N. McKinsey, J. Instrum. 2 (2007) P11004.
22. M. Moszynski et al., Nucl. Instr. Meth. A 504 (2003) 307.
23. H.V. Piltingsrud, J. Nucl. Med. 20 (1979) 1279.
24. J. Valentine et al., Nucl. Instr. Meth. A 325 (1993) 147.
25. P. Antonini et al., Nucl. Instr. Meth. A 488 (2002) 591.
26. L. Yang et al., Nucl. Instr. Meth. A 508 (2003) 388.
27. T. Ikagawa, Nucl. Instr. Meth. A 538 (2005) 640.
28. G. Bizarri et al., Phys. Stat. Sol. A 203 (2006) R41.
29. M. Moszynski, B. Bengtson, Nucl. Instr. Meth. A 142 (1977) 417.
30. G.F. Knoll, *Radiation Detection and Measurements*. John Wiley & Sons Inc., New York, 1999.
31. Yu.G. Zdesenko et al, Nucl. Instr. Meth. A 538 (2005) 657.
32. J. Gironnet et al., Nucl. Instr. Meth. A 594 (2008) 358.
33. V.B. Mikhailik, S. Henry, H. Kraus, I. Solskii, Nucl. Instr. Meth. A 583 (2007) 350.
34. V.B. Mikhailik et al., J. Phys. Cond. Matt. B 20 (2008) 365219.
35. R.O. Keeling, D. Phil. Thesis, Oxford, 2002.
36. R.T. Williams, C.L. Morquart, J.W. Williams, M.N. Kabler, Phys. Rev. B 15 (1977) 5003.
37. N.G. Zakharov, T.I. Nikitinskaya, P.A. Rodnyi, Sov. Phys. Sol. State 24 (1982) 709.

# R&D of radiopure crystal scintillators for low counting experiments

**F.A. Danevich***

*Institute for Nuclear Research, MSP 03680 Kyiv, Ukraine*

A programme to develop radiopure crystal scintillators for low counting experiments is discussed briefly.

## 1. Introduction

Experiments to search for rare events, i.e. dark matter particles, double beta decay, investigations of rare $\alpha$- and $\beta$-activity, measurements of neutrino fluxes, require ultra low-level background of the detector. Counting rate of a few counts per kg per day in the energy interval 2–20 keV is typical for present scintillator-based dark matter experiments [1, 2, 3]. To elaborate region of WIMP-nucleon scattering cross sections predicted by different models, a counting rate of a detector should be further decreased a few orders of magnitude. For instance, the EURECA dark matter project[1] calls for a background counting rate lower than a few events per keV per 100 kg per year at energies of a few keV [4]. Radioactive contamination of target materials (germanium and scintillation crystals) will play a key role to decrease background in the experiment.

Double beta decay projects require low as much as possible – in ideal case zero – background of a detector in a region of interest[2]. The most dangerous radionuclides for $2\beta$ experiments are $^{226}$Ra and $^{228}$Th having daughters ($^{214}$Bi and $^{208}$Tl) with large energies of $\beta$ decay. Presence of cosmogenic radioactivity should be also controlled and decreased as much as possible. A reasonable (and *measurable* with present instrumentation) level of a few $\mu$Bq/kg is discussed now (see, for instance [5, 6, 7, 8, 9]). However, further progress in the searching for $2\beta$ processes will be possible only with detectors with much better radiopurity.

At present the most promising scintillators with high light output for cryogenic dark matter search are $ZnWO_4$, $CaWO_4$, and $CaMoO_4$ [10]. The most interesting scintillators for double beta decay experiments are $CdWO_4$ ($^{106}$Cd, $^{116}$Cd), $CaMoO_4$, $PbMoO_4$ ($^{100}$Mo), ZnSe ($^{82}$Se). R&D on the scintillation materials $Li_2MoO_4$, $Li_2Zn_2(MoO_4)_3$, $ZnMoO_4$, $MgWO_4$, that can be applied both for dark matter and double beta decay search, is in progress, as well is in progress the further optimization and improvement of $ZnWO_4$, $CaWO_4$, $CaMoO_4$, $CdWO_4$, $CaF_2$, BGO, $Al_2O_3$, LiF, ZnSe, $PbWO_4$, $PbMoO_4$. Zinc and cadmium tungstates are good examples of radiopure scintillators (~0.2–1 mBq/kg level) [11, 12, 13]. Nevertheless at least a ~20-fold improvement of $ZnWO_4$ radiopurity is still needed for the EURECA experiment; and that represents a significant challenge.

## 2. Requirements to radiopurity of crystal scintillators

Let us estimate a level of radiopurity requested by the EURECA experiment. A Monte Carlo simulated background of a detector with mass of 100 kg for 1 yr of measurements is presented in Fig. 1. Assumed internal contaminations of the detector material by $^{40}$K, $^{60}$Co, $^{87}$Rb, $^{90}$Sr-$^{90}$Y, $^{137}$Cs, $^{232}$Th, $^{238}$U correspond to the activity of 0.1 mBq/kg. The energy resolution FWHM = 10% for 662 keV $\gamma$ line of $^{137}$Cs, a typical for a scintillation spectrometer, was taken for the calculations.

* Corresponding author. *E-mail address:* danevich@kinr.kiev.ua

[1] European Underground Rare Event Calorimeter Array; www.eureca.ox.ac.uk

[2] Typically, the energy release in $2\beta$ decay, when reasonable half-lives are expected, is in the range of 1–4 MeV. However, the region of energies 10–100 keV should be explored to search for double electron capture processes.

Supposing a suppression factor ≈ $10^3$, which can be reached thanks to the active background rejection with a combination of phonon and scintillation responses [14], radioactive contamination of crystal scintillators should not exceed a level of ~10 μBq/kg, which will require significant improvement compared to the present level of $ZnWO_4$, $CaWO_4$, and $CaMoO_4$ crystal scintillators (0.1 – 100 mBq/kg) [15]. Therefore, the level of ~10 μBq/kg can be accepted as a goal of a first step in R&D programme. In addition one should keep in mind also cosmogenic radionuclides. For instance, accumulation of radioactive $^{14}C$ ($Q_\beta$ = 156.475 keV, $T_{1/2}$ = 5730 yr [16]), highly undesirable for dark matter experiments, was considered in [17]. The radioactive $^{14}C$ can be produced by hadronic component of cosmic rays in any materials composed of elements heavier than carbon.

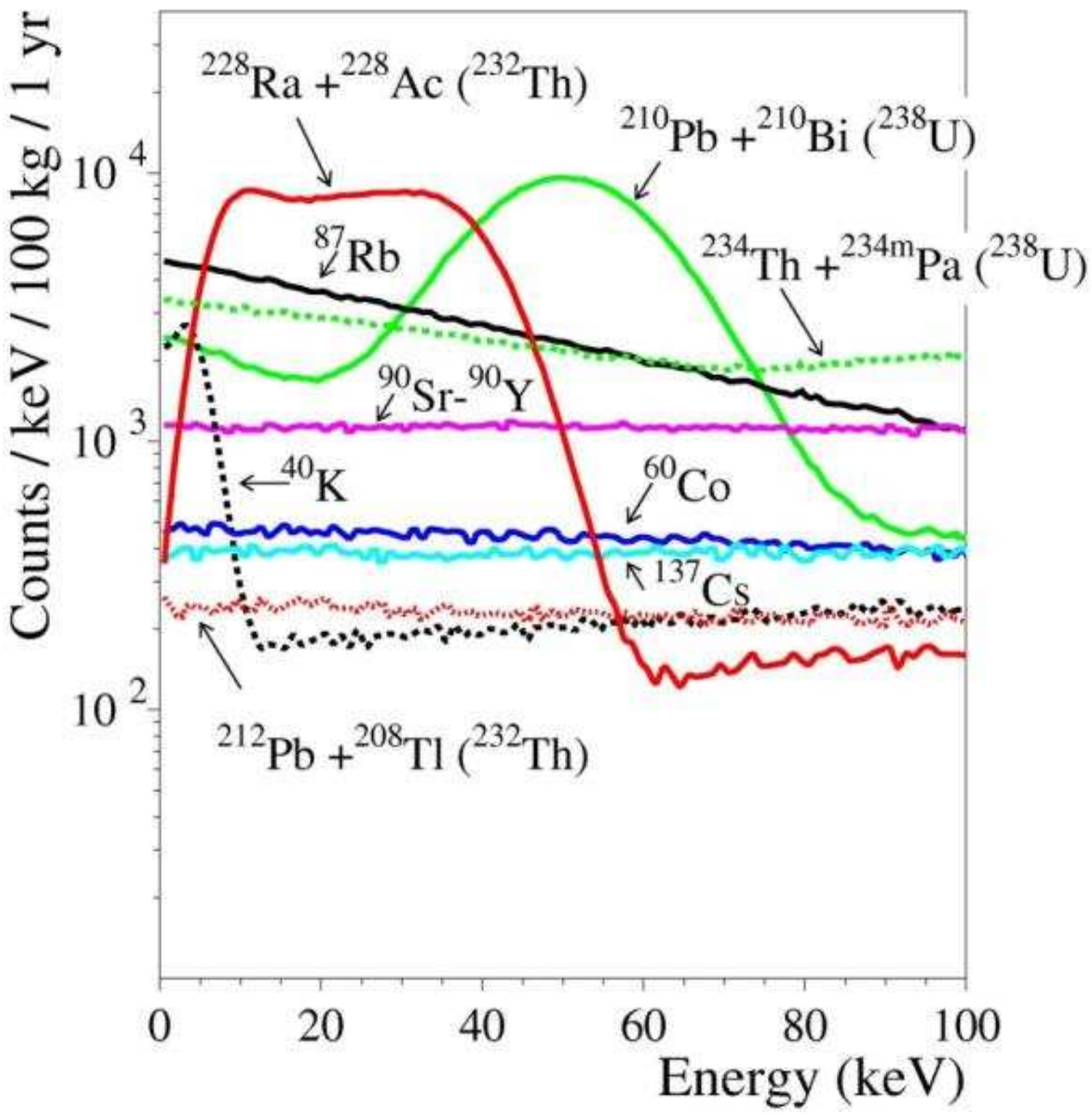

Fig. 1. Contributions to background of a scintillation detector from internal contamination by $^{40}K$, $^{60}Co$, $^{87}Rb$, $^{90}Sr$-$^{90}Y$, $^{137}Cs$, U/Th (see text for details).

## 3. A programme to develop radiopure scintillation materials for low counting experiments

A programme to develop scintillation materials for dark matter and double beta decay experiments with improved radiopurity could comprise 3 directions:

1. Development of new and improvement of existing cryogenic scintillators;
2. Improvement of radiopurity;
3. Test of radiopure scintillators.

*1. Development of new and improvement of existing cryogenic scintillators.*

1.1. Synthesis of compounds (poly-crystalline) containing certain elements and their test at low temperature. The most promising compounds should be used for crystal growth and subjected to further investigation and optimization.

1.2. Study of luminescence and scintillation properties down to very low temperatures provides important insight, beneficial towards improving cryogenic scintillators.

1.3. Analytical instrumentation is necessary to control chemical impurities in raw materials with sensitivity on the level of ~0.1 ppm. This is especially important for transition metals.

1.4. Study of deep purification of raw materials on light output of scintillators at low temperatures. It should be emphasized that scintillator performance could benefit from the deep purification required to address the radiopurity issue.

*2. Improvement of radiopurity.*

2.1. Deep purification of raw materials is supposed to be the most important issue that needs addressing. Metal purification by vacuum distillation, zone melting, and filtering are very promising approaches [18, 19], while further study is necessary for the purification of Ca, Li, Se in order to achieve the required low levels of radioactive contamination.

2.2. Two to four step re-crystallization[1], involving inspection and assessment of the produced scintillators after each step.

2.3. Screening at all stages through ultra-low background γ-, α-, β-spectroscopy is needed in the production of compounds for crystal growing (choice of raw materials, quality control of purified elements and compounds).

2.4. All work should be done using highly pure reagents, lab-ware and water. Careful protection from radon should be foreseen. All chemistry should be done in a clean room, and, as far as possible, in nitrogen atmosphere.

*3. Test of radiopure scintillators.*

3.1. The low-background scintillation measurements are currently the most appropriate methods of examining the performance of scintillators.

3.2. Final testing of scintillators involves their operation as low-background cryogenic detectors.

3.3. R&D of ultra-low-radioactive instrumentation with the sensitivity at the level of 10 μBq/kg (able to operate at low, at least liquid nitrogen, temperatures) is necessary. In itself the problem of measurements of crystal scintillators radiopurity on the level of 10 μBq/kg is rather complicated task, demanding an extended R&D (a low-background set-up to measure crystal scintillators in a temperature region down to liquid nitrogen was proposed in [21]). It should be stressed, measurements of low energy β-active radionuclides were never realized at such a level of sensitivity.

## 4. Conclusions

Next generation low background scintillation (cryogenic) experiments call for extremely low (μBq/kg) level of radioactive contamination of scintillation materials. The most radiopure existing crystal scintillators [$ZnWO_4$, $CdWO_4$, as well as developed for dark matter experiments NaI(Tl) and CsI(Tl)] have one-three order of magnitude level of radioactive impurities worse. An extended programme is necessary to develop scintillators satisfying requirements of high sensitivity dark matter and double beta decay experiments. The programme could consist of the following main directions: development of new scintillators, improvement of radiopurity of crystal scintillators, testing of scintillators on a required level of 0.1–0.01 mBq/kg. The principal components of the programme are: i) deep purification of raw materials, ii) a few step re-crystallization, iii) careful screening at all stages by ultra-low background α-, β-, and γ-spectrometry, and iv) use of radiopure reagents, lab-ware, equipment and installations, production of raw materials, crystal growing and their storage in radon free atmosphere. Special efforts are necessary to prevent cosmogenic and neutron activation of materials.

[1] As it was demonstrated recently, recrystallization can decrease radioactive contamination of $CaWO_4$ crystals one order of magnitude [20].

## Acknowledgments

Author would like to thank Prof. H. Kraus (University of Oxford, UK) for valuable comments and corrections. It is a pleasure to express gratitude to Mr. A. Dossovitskiy and Mr. A. Mikhlin (NeoChem Company, Moscow, Russia), Prof. B.V. Grinyov and Dr. L.L. Nagornaya (Institute for Scintillation Materials, Kharkiv, Ukraine), Prof. M. Korjik (Institute for Nuclear Problems, Minsk, Belarus), Prof. G.P. Kovtun and Dr. A.P. Shcherban (Institute of Physics and Technology, Kharkiv, Ukraine), Dr. P. de Marcillac (CNRS-IAS Orsay, France), Dr. V. Mikhailik (University of Oxford, UK), Dr. S. Pirro (INFN - Sezione di Milano Bicocca, Italy), Dr. V. Shlegel (Institute of Inorganic Chemistry of SD RAS, Russia), Dr. I. Solsky (SRC "CARAT", Lviv, Ukraine) for fruitful discussions, and Dr. V.V. Kobychev (INR, Kyiv, Ukraine) for Monte Carlo simulation of the background components presented in Fig. 1. The support from the project “Kosmomikrofizyka” (Astroparticle Physics) of the National Academy of Sciences of Ukraine is gratefully acknowledged.

## References

1. R. Bernabei et al., Eur. Phys. J. C 56 (2008) 333.
2. H.S. Lee et al., Phys. Let. B 633 (2006) 201.
3. W. Westphal et al., Czech. J. Phys. 56 (2006) 535.
4. H. Kraus et al., J. Phys. Conf. Series 39 (2006) 139.
5. G. Bellini et al., Eur. Phys. J. C 19 (2001) 43.
6. F.A. Danevich et al., Nucl. Instr. Meth. A 556 (2006) 259.
7. Yu.G. Zdesenko et al., Astropart. Phys. 23 (2005) 249.
8. A.N. Annenkov et al., Nucl. Instr. Meth. A 584 (2008) 334.
9. M. Bongrand (on behalf of the SuperNEMO collaboration, AIP Conf. Proc. 897 (2007) 14.
10. H. Kraus et al. (EURECA collaboration), these Proceedings, p. 7.
11. P. Belli et al., “Search for 2β decay of Zinc and Tungsten with the help of low-background $ZnWO_4$ crystal scintillators”, Preprint ROM2F/2008/22 – Roma 2 University, 2008, 20 p. (arXiv:0811.2348 [nucl-ex]); submitted to Phys. Rev. C;
D.V. Poda, these Proceedings, p. 50.
12. F.A. Danevich et al., Phys. Rev. C 68 (2003) 035501.
13. P. Belli et al., Phys. Rev. C 76 (2007) 064603.
14. V.B. Mikhailik, H. Kraus, J. Phys. D: Appl. Phys. 39 (2006) 1181.
15. F.A. Danevich, these Proceedings, p. 28.
16. R.B. Firestone et al., *Table of Isotopes*, 8th ed., John Wiley & Sons, New York, 1996 and CD update, 1998.
17. Yu.G. Zdesenko, O.A. Ponkratenko, V.I. Tretyak, J. Phys. G 27 (2001) 2129.
18. G.P. Kovtun et al., these Proceedings, p. 54.
19. P. Belli et al., “Development of enriched cadmium tungstate crystal scintillators to search for double beta decay processes in $^{106}$Cd”, Preprint ROM2F/2008/17 – Roma 2 University, 2008, 9 p. (to be published in Nucl. Instr. Meth. A).
20. F.A. Danevich et al., these Proceedings, p. 37;
F.A. Danevich et al., “Effect of recrystallization on the radioactive contamination of $CaWO_4$ crystal scintillators” (in preparation).
21. F.A. Danevich, H. Kraus, V.B. Mikhailik, “Ultra-low background scintillation spectrometer for measurements of the radioactive contamination of materials and scintillation crystals for cryogenic rare events experiments”, LPD KINR note 5/2008 (unpublished).

# List of the RPScint'2008 participants

| | |
|---|---|
| 1. Danevich Fedor | Institute for Nuclear Research, Kyiv, Ukraine |
| 2. Degoda Vladimir | Kyiv Taras Shevchenko National University, Kyiv, Ukraine |
| 3. de Marcillac Pierre | CNRS Institut d'Astrophysique Spatiale, Orsay, France |
| 4. Dossovitskiy Alexey | JSC NeoChem, Moscow, Russia |
| 5. Grigoriev Dmitry | Institute of Inorganic Chemistry SD RAS, Novosibirsk, Russia |
| 6. Kobychev Vladislav | Institute for Nuclear Research, Kyiv, Ukraine |
| 7. Korzhik Mikhail | Institute of Nuclear Problems, Minsk, Belarus |
| 8. Kraus Hans | University of Oxford, Oxford, United Kingdom |
| 9. Mikhailik Vitalii | University of Oxford, Oxford, United Kingdom |
| 10. Mokina Valentyna | Institute for Nuclear Research, Kyiv, Ukraine |
| 11. Nagornaya Liudmila | Institute for Scintillation Materials, Kharkiv, Ukraine |
| 12. Nikolaiko Andrew | Institute for Nuclear Research, Kyiv, Ukraine |
| 13. Pirro Stefano | INFN - Sezione di Milano Bicocca, Milano, Italy |
| 14. Poda Denys | Institute for Nuclear Research, Kyiv, Ukraine |
| 15. Podviyanuk Ruslan | Institute for Nuclear Research, Kyiv, Ukraine |
| 16. Polischuk Oksana | Institute for Nuclear Research, Kyiv, Ukraine |
| 17. Raina P.K. | I.I.T., Kharagpur, India |
| 18. Shcherban Alexey | NSC "Kharkiv Institute of Physics and Technology", Kharkiv, Ukraine |
| 19. Shekhovtsov Alexey | Institute for Single Crystals, Kharkiv, Ukraine |
| 20. Shlegel Vladimir | Institute of Inorganic Chemistry SD RAS, Novosibirsk, Russia |
| 21. Solopikhin Dmitry | NSC "Kharkiv Institute of Physics and Technology", Kharkiv, Ukraine |
| 22. Solsky Ivan | Institute of Materials, SRC Carat, Lviv, Ukraine |
| 23. Sofroniuk Andriy | Kyiv Taras Shevchenko National University, Kyiv, Ukraine |
| 24. Spassky Dmitry | Skobeltsyn Institute of Nuclear Physics, M.V. Lomonosov Moscow State University, Moscow, Russia |
| 25. Stryganyuk Grygoriy | Institute of Materials, SRC Carat, Lviv, Ukraine |
| 26. Verdier Marc-Antoine | IPN, Lyon, France |

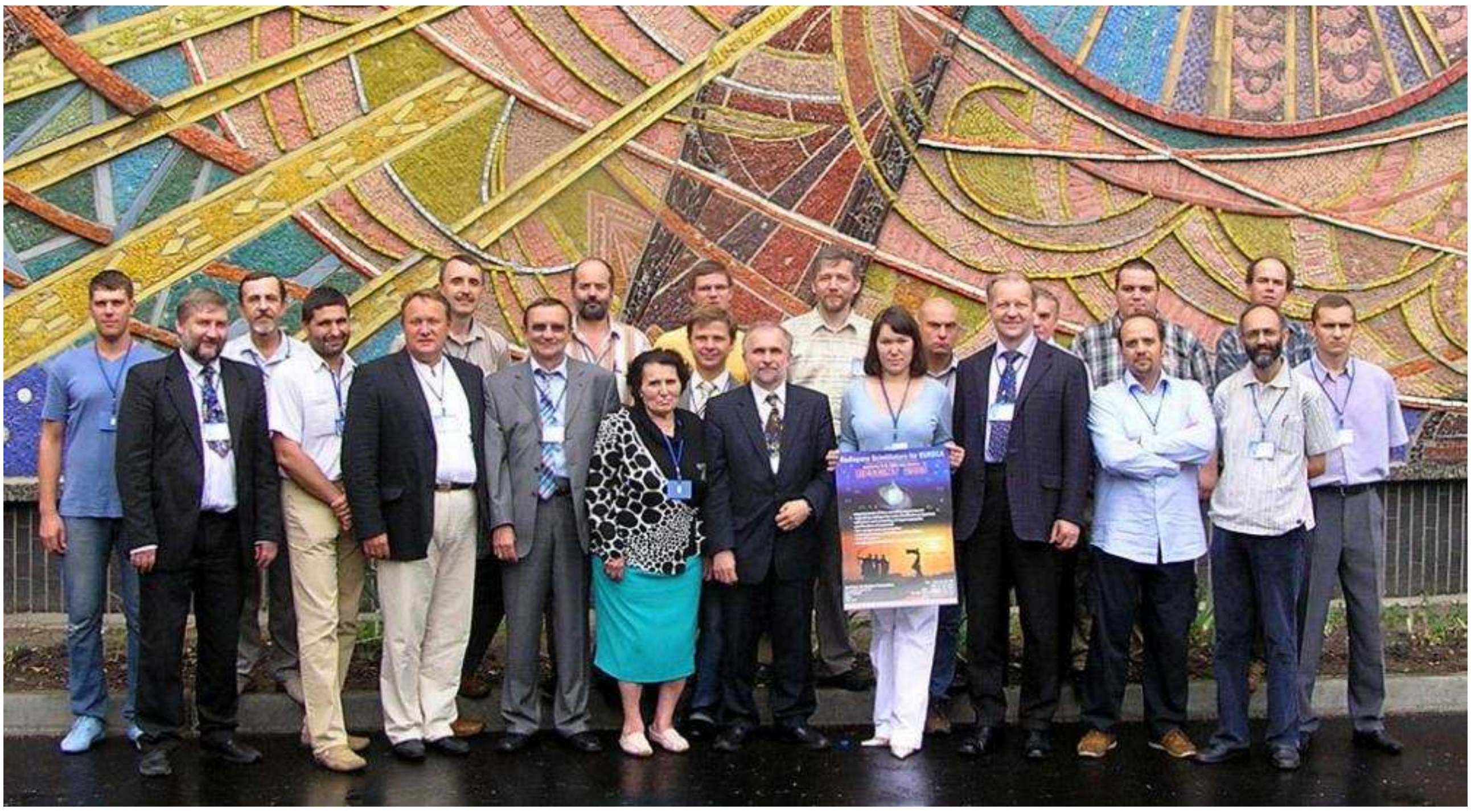

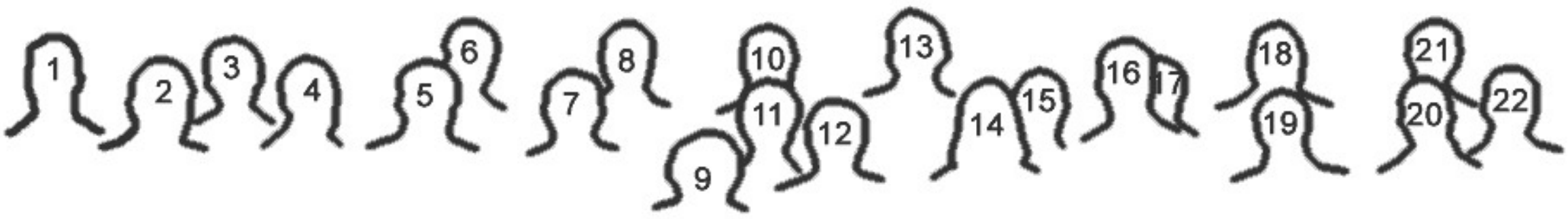

**Participants of the RPScint'2008 Workshop:**

1 - Denys Poda; 2 - Dmitry Grigoriev; 3 - Andrew Nikolaiko; 4 - Alexey Dossovitskiy; 5 - Mikhail Korjik; 6 - Vitalii Mikhailik; 7 - Ivan Solsky; 8 - Vladimir Shlegel; 9 - Liudmila Nagornaya; 10 - Dmitry Spassky; 11 - Grygoriy Stryganyuk; 12 - Fedor Danevich; 13 - Vladislav Kobychev; 14 - Valentyna Mokina; 15 - Alexey Shekhovtsov; 16 - Hans Kraus; 17 - Alexey Shcherban; 18 - Marc-Antoine Verdier; 19 - Stefano Pirro; 20 - P.K. Raina; 21 - Pierre de Marcillac; 22 - Dmitry Solopikhin

---